\documentclass[%
pre,
twocolumn,
amsmath,amssymb,f
]{revtex4-2}

\usepackage{amsmath}
\usepackage{gensymb}
\usepackage{graphicx}%
\usepackage{dcolumn}%
\usepackage{bm}%
\usepackage{multirow}
\usepackage[utf8]{inputenc}
\usepackage[T1]{fontenc}
\usepackage{mathptmx}
\usepackage{subfigure}
\usepackage[normalem]{ulem}
\draft %
\usepackage{xcolor}
\definecolor{ForestGreen}{RGB}{34,139,34}

\begin{document}
	
	\title{Lattice Boltzmann method for fluid-structure interaction in compressible flow}

	\author{Abhimanyu Bhadauria}
	\author{Benedikt Dorschner}
	\email[Corresponding author ]{bdorschn@ethz.ch}
	\author{Ilya Karlin}
	\affiliation{Department of Mechanical and Process Engineering, ETH Zurich, 8092 Zurich, Switzerland}

	\date{\today}
	
	\begin{abstract}
	We present a two-way coupled fluid-structure interaction scheme for rigid bodies  using a two-population lattice Boltzmann formulation for compressible flows.
	 Arbitrary Lagrangian-Eulerian  formulation of the discrete Boltzmann equation on body-fitted meshes is used in a combination with polynomial blending functions.
	The blending function approach localizes mesh deformation and allows treating multiple moving bodies with a minimal computational overhead. We validate the model with several test cases of vortex induced vibrations of single and tandem cylinders and show that it can accurately describe dynamic behavior of these systems. 
	Finally, in the fully compressible regime, we demonstrate that the proposed model accurately captures complex phenomena such as  transonic flutter over an airfoil.

	\end{abstract}
	
	
	\maketitle %

	\section{\label{sec:Intro}Introduction }
 	
	Interaction between fluid and deformable or moving bodies is important area of research in computational mechanics. Behavior of solid structures under the influence of fluid flow, and vice versa, is of special interest in the fields of bio-mechanics, mechanical engineering and aerospace engineering with applications ranging from blood flow inside the heart and blood vessels \cite{Anand_2020}, material erosion due to bubble cavitation \cite{Hsiao_2015,Chahine_2015}, effect of shock or blast waves on solid structures \cite{Subramaniam_2009,Aune_2021}, and vibration or deformation of aircraft wings at high speeds \cite{Thomas_2002,Liu_2001,Yuan2021}.
	For aerospace vehicle design, the study of fluid structure interactions (FSI) at transonic and supersonic speeds is especially important in order to detect shock induced vibrations and deformations such as aeroelastic flutter. A robust and efficient model to accurately predict fluid loads on structures is therefore crucial.
	
	The lattice Boltzmann method (LBM) has steadily been gaining prominence in the domain of computational fluid dynamics (CFD). 
 	Promising results have been shown in a wide array of flows ranging from turbulence \cite{FRAPOLLI20182}, multiphase flows\cite{A.Mazloomi2015} and multi-component flows \cite{Sawant2020} to rarefied gas flows \cite{Staso2016} and relativistic hydrodynamics \cite{Mendoza2010}, to name a few.
	LBM arrives at the macroscopic equations of fluid dynamics from a mesoscopic perspective, where the flow is described by discretized particle distribution functions (populations) $f_i(\bm{x},t)$, which are associated with a set of discrete velocities  $ \cal{C} $ $= \verb|{| \bm{c_i}, i = 0,..., Q-1 \verb|}|$, forming the links of a space-filling lattice. 
	The LBM algorithm eventually reduces to a simple and highly efficient stream-and-collide procedure, where the populations are advected along the discrete velocities and relaxed towards local equilibrium distribution functions, which are designed to recover the Navier-Stokes equations in the hydrodynamic limit. 
	
	However, despite a lot of progress in the development of various extensions of LBM, the classical LBM is limited to the incompressible flow regime.
	This is mainly due to the geometrical restrictions of standard lattices, such as $D2Q9$ or $D3Q27$ in $D=2$ or $D=3$ dimensions with $Q=9$ and $Q=27$ discrete velocities, respectively, which induce errors in the fluid stress tensor and break Galilean invariance. 
	While systematically increasing the number of velocities, leading to high-order lattices \cite{Chikatamarla2006,XIAOWEN2006,Chikatamarla2009,Siebert2008,CHIKATAMARLA2010}, 
	was thought to be a path towards compressible flows, severe restrictions due to  
	the added computational costs and tight bounds on the temperature range remain.
	Another approach is to retain standard lattices but introduce 
	appropriate correction terms to counteract the anomalous terms in the stress tensor. 
	In the literature, different proposals for the implementation of such correction terms from a variety of authors exist \cite{Parsianakis2007,Prasianakis2008,Feng_2015,Huang_2019,Hosseini_2020}.  
	Recently, we have extended this approach within the two population setting, which has shown promising results for compressible flows on standard lattices \cite{Saadat2019,saadat2021extendedGas}. 
	
	Another intrinsic property of LBM is that it is based on uniform Cartesian grids. While this allows for a highly efficient and exact advection scheme, non-uniform body-fitted grids can be 
	advantageous to resolve the thin boundary layers of complex geometries. Moreover, on non-uniform meshes, a variable Courant-Friedrichs-Levy (CFL) number can be achieved easily, while it is fixed to unity for the classical LBM. 
	The class of so-called off-lattice Boltzmann methods were developed to overcome these limitations and extend LBM to non-uniform grids. 
	A majority of these methods turn to  purely Eulerian formulations such as finite-volumes  \cite{Nannelli1992,Xi1999,Patil2009}, finite-differences \cite{Fakhari2015,Hejranfar2014}, or finite-elements \cite{Duester2006,Li2005} in order to solve the discrete Boltzmann equation. While this does allow for non-uniform grids and a flexible CFL number, a partial differential equation for each discrete velocity needs to be solved in order to advect the populations. This typically requires small times steps and repeated non-local evaluation of spatial gradients, which leads to a significant computational overhead and can result in prohibitively high costs \cite{Kraemer2017}. Another avenue are semi-Lagrangian methods \cite{shu2002taylor,cheng2004lattice,bardow2006general,Kraemer2017,wilde2020semi}, which start from a characteristic equation to resolve these issues. Promising results have been shown for wall-bounded turbulent flows \cite{DIILIO_2018} and even compressible flows \cite{Saadat2020,Saadat2020a,wilde2020semi}.
	
	Here, we build on these results and continue our development of semi-Lagrangian LBM for fluid-structure interaction problems of compressible flows.
	In the literature, there exist a variety of FSI schemes also within the realm of LBM.
	However, the majority are based on Cartesian meshes (see, e.g., \cite{Dorschner2018,Dorschner2018a})  using Cartesian boundary conditions or a form of the immersed boundary method (see, e.g., \cite{Feng_2004,Zhu_2011,Nagar_2015}). 
	For these schemes, the added costs of FSI coupling is low, since even for large displacements, no re-meshing or alike is required. 
	However, as stated above, the uniformity of the mesh limits the attainable resolution at the boundary.
	
	In order to keep the body-fitted nature of the grid while also allowing for mesh motion, we turn to the Arbitrary Lagrangian-Eulerian (ALE) scheme. The ALE approach allows for some nodes of the mesh to move with the boundary, as in a Lagrangian description, while nodes away from the boundary stay fixed, as in an Eulerian description. Using ALE, the need to re-mesh the entire domain after displacement of the body is reduced, and the model can accommodate much larger distortions of the domain than afforded by a purely Lagrangian description, with a higher resolution near the wall than possible with a purely Eulerian description \cite{Donea2004}. 
	
	Within the ALE description, care must be taken to accommodate multiple bodies in the domain, and bodies whose motion is not a rigid body motion. In such cases, it is necessary to limit the mesh deformation due to the moving body (or bodies) to its own vicinity in order to maintain mesh conformity throughout the domain. In the literature, this problem is solved by using either of two methods. One approach is the use of polynomial blending functions that smoothly distribute the motion of the mesh spatially, such that mesh nodes close to and attached to the body move with it, while the mesh nodes that are far away from the moving body remain stationary \cite{Persson2009,Donea2004,Akhtar2007}. 
	Another approach is to solve a differential equation, such as the diffusion or linear elasticity equation in order to generate the mesh deformation field	\cite{WICK20111456,Wick2013}. 
	While the latter is more robust, solving an additional differential equation adds a significant amount of computational costs compared to solving a polynomial equation. 
	
	In our previous works \cite{Saadat2020,Saadat2020a}, it was demonstrated that by using an arbitrary Eulerian-Lagrangian (ALE) formulation in combination with a dual population LB formulation, high-speed compressible flows with moving geometries can be captured accurately and efficiently. 
	However, in those works, only a single body was considered and the motion was prescribed.
	Here, we aim at extending this framework towards genuine two-way coupled FSI, including multiple bodies as well as un-prescribed motion. To that end, we start with a rigid body FSI using an ALE formulation, where solids do not deform but are coupled with the fluid through a spring-mass system. Multiple bodies with independent displacements are accounted for by using a blended mesh formulation. Finally, we demonstrate that the proposed scheme is able to accurately capture complex phenomena such as vortex induced vibrations (VIV) and aeroelastic flutter, which is of crucial importance to aerospace applications. To the best of our knowledge there does not exist a LBM model for fluid-structure interaction in the compressible flow regime. 
	
	The paper is organized as follows: In section Sec.~\ref{sec:LBModel} we review the ALE description of LB and its extension to the semi-Lagrangian framework. We also introduce concept of blended meshes, which is crucial for the treatment of multiple moving bodies with conforming meshes. Sec.~\ref{sec:FSI} describes the fluid-structure coupling, the force computation, and the boundary conditions necessary for the LB model. The ordinary differential equations that define the rigid body motion of the bodies 
	are defined in Sec.~\ref{sec:StrModel}. Validation of the coupled system is presented in Sec.~\ref{sec:Res}, where we consider one degree of freedom (DOF) and two DOF systems 
	for flows past oscillating cylinders as well as airfoil flutter.
	Simulations over a range of Reynolds number and Mach number regimes are presented. It is shown that the LB model together with a simple yet robust mesh deformation scheme ensures mesh conformity even under highly irregular motion. 
	Finally, conclusions are summarized in Sec.~\ref{sec:Concl}. 
	
	\section{\label{sec:LBModel} Lattice Boltzmann Model }
	\subsection{\label{sec2:l4}Arbitrary Lagrangian-Eulerian Description of LB }
\begin{figure}[h!]
		\includegraphics[width=\linewidth]{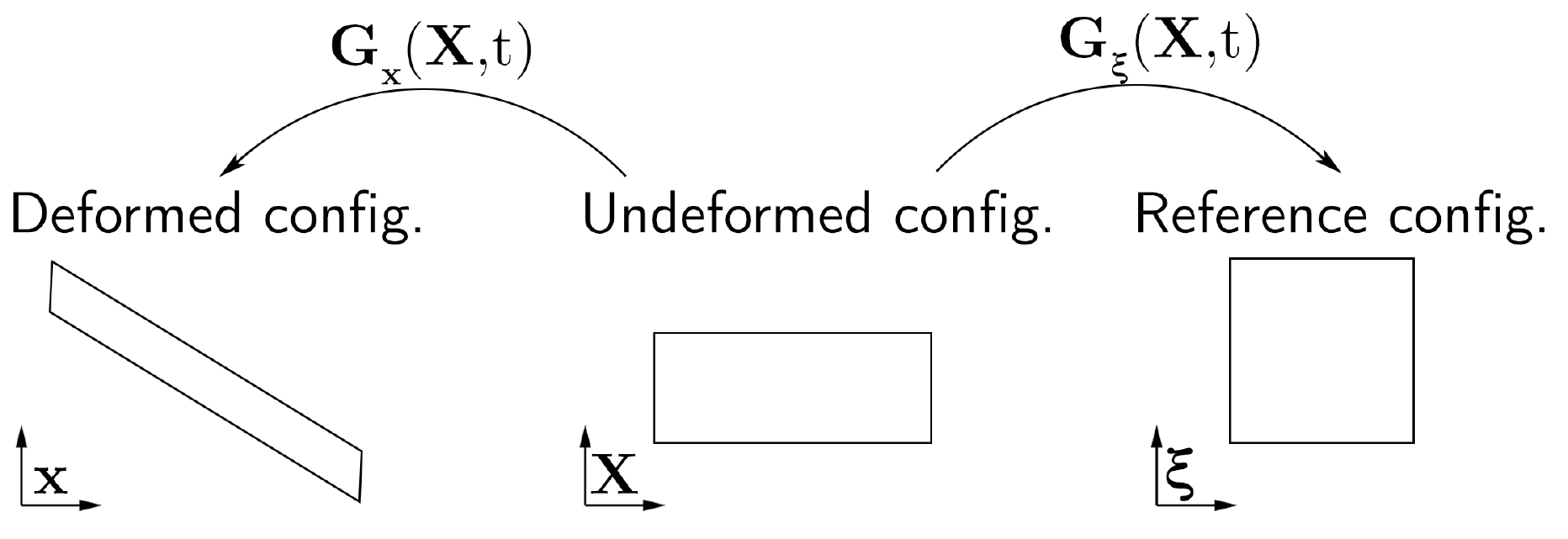}
		\caption{\label{fig:Frames} Deformed, undeformed, and reference configurations. }
	\end{figure}

	In this section we briefly review the ALE formulation for LBM as proposed in \cite{Saadat2020a}. To that end, we consider the Boltzmann equation, 
	\begin{align} \label{eqn:DboltzmannEqn}
	\frac{\partial f_i}{\partial t}+ \bm{c}_i\cdot{\nabla_{\bm{x}} f_i} = \Omega_i,
	\end{align}
	where the populations $ f_i(\bm{x},t) $ are defined on the deformed physical domain, $ (\bm{x},t)$ and 
	$\Omega_i$ is the collision operator. In the ALE description, all computations take place on the undeformed domain (see Fig.~\ref{fig:Frames}). {Hence, we wish to obtain a mapping $ \bm{G}_{\bm{x}}(\bm{X},t) $ between the undeformed domain, $ (\bm{X},t_0)$ and the deformed domain, $ ( \bm{x},t)$. We also introduce a unit reference frame, $\bm{\xi}$ which is fixed, and the transformation $\bm{G}_{\xi}(\bm{X},\rm t)$ between the undeformed domain and the unit reference domain. 
	For now, we concern ourselves with the deformed and the undeformed domains, we will return to the unit reference domain in Sec.~\ref{sec2:l3} }.
 	
	In the simplest case of prescribed motion of a rigid body, the mapping from the deformed domain to the undeformed domain is a rigid transformation, consisting of a combination of translation and rotation. For coupled problems, where such a mapping is not known a-priori, it can be constructed at each time step from the motion of the body. Further, in case of multiple moving bodies, the mapping will not be a rigid transformation. This scenario is addressed in Sec.~\ref{sec2:l5}.
	
	Assuming the existence of such a continuous mapping between the deformed and the undeformed initial domain $ \bm{x} =  \bm{G}_{x}(\bm{X},t)  $, that maps a location $ \bm{X} $ in the undeformed domain to a location $ \bm{x} $ in the deformed domain, we have for the temporal derivative of this mapping, the mapping velocity $ \bm{V}_G $,
	\begin{align}
\left.	\bm{V}_G  = \dfrac{\partial \bm{G}_x }{\partial t} \right|_{\bm{X}}.
	\end{align}
	For the time derivative of the populations $f_i$ in the deformed domain we can write,
	\begin{align} \label{eqn:dt_ref}
	\frac{\partial f_{i}}{\partial t}=\frac{d f_{i}}{d t}-\bm{V}_{G} \cdot \bm{\nabla}_{\bm{X}} f_{i}=\left.\frac{\partial f_{i}}{\partial t}\right|_{\bm{X}}-\bm{V}_{G} \cdot \bm{\nabla}_{\bm{X}} f_{i}.
	\end{align}
	The space derivative can similarly be transformed to the undeformed domain,
	\begin{align} \label{eqn:dx_ref}
	\nabla_{\bm{x}} f_i = \bm{g}^{-1} \nabla_{\bm{X}} f_i.
	\end{align}
	Here $ \bm{g} $ is  the Jacobian of the ALE mapping and its inverse is given by 
	\begin{align} 
	\bm{g}^{-1}=\frac{1}{\operatorname{det} \bm{g}}\left[\begin{array}{cc}
	\partial_{Y}y & -\partial_{X}y \\  
	-\partial_{Y}x & \partial_{X}x
	\end{array}\right],
	\end{align}
	where the mapping metrics $\partial_{X}x, \partial_{Y}x, \partial_{X}y$ and $\partial_{Y}y$ are the partial derivatives $  \dfrac{ \bm{G}_{x}(\bm{X},t)} {\partial \bm{X}} $, and
	\begin{align}
	\operatorname{det} \bm{g}=x_{X} y_{Y}-y_{X} x_{Y}.
	\end{align}

	Finally, from Eq.~\eqref{eqn:DboltzmannEqn}, \eqref{eqn:dt_ref}, and \eqref{eqn:dx_ref}, we have the ALE description of the Boltzmann equation in the undeformed domain,
	\begin{align} \label{eqn:DboltzmannEqn_ALE}
	\left.\frac{\partial f_{i}}{\partial t}\right|_{\bm{X}}+\hat{\bm{c}}_{i}\cdot \bm{\nabla}_{\bm{X}} f_{i}=\Omega_{i},
	\end{align}
	where we can see that the discrete velocity $ \bm{c}_i $ has been replaced by $ \hat{\bm{c}}_{i} $, which is simply the transformation of the discrete velocity set to the initial undeformed domain,
	\begin{align}
	\hat{\bm{c}}_{i}=\left(\bm{g}^{-T} \bm{c}_{i}-\bm{V}_{G}\right).
	\end{align}

	The resulting equation is in the same form as the discrete Boltzmann equation in the deformed domain, with only the discrete velocities different. These transformed discrete velocities cannot be assumed to be integers anymore and an off-lattice propagation scheme is now necessary.

	\subsection{\label{sec2:l3}Semi-Lagrangian Framework}
	
\begin{figure}[t]
		\includegraphics[width=1\linewidth]{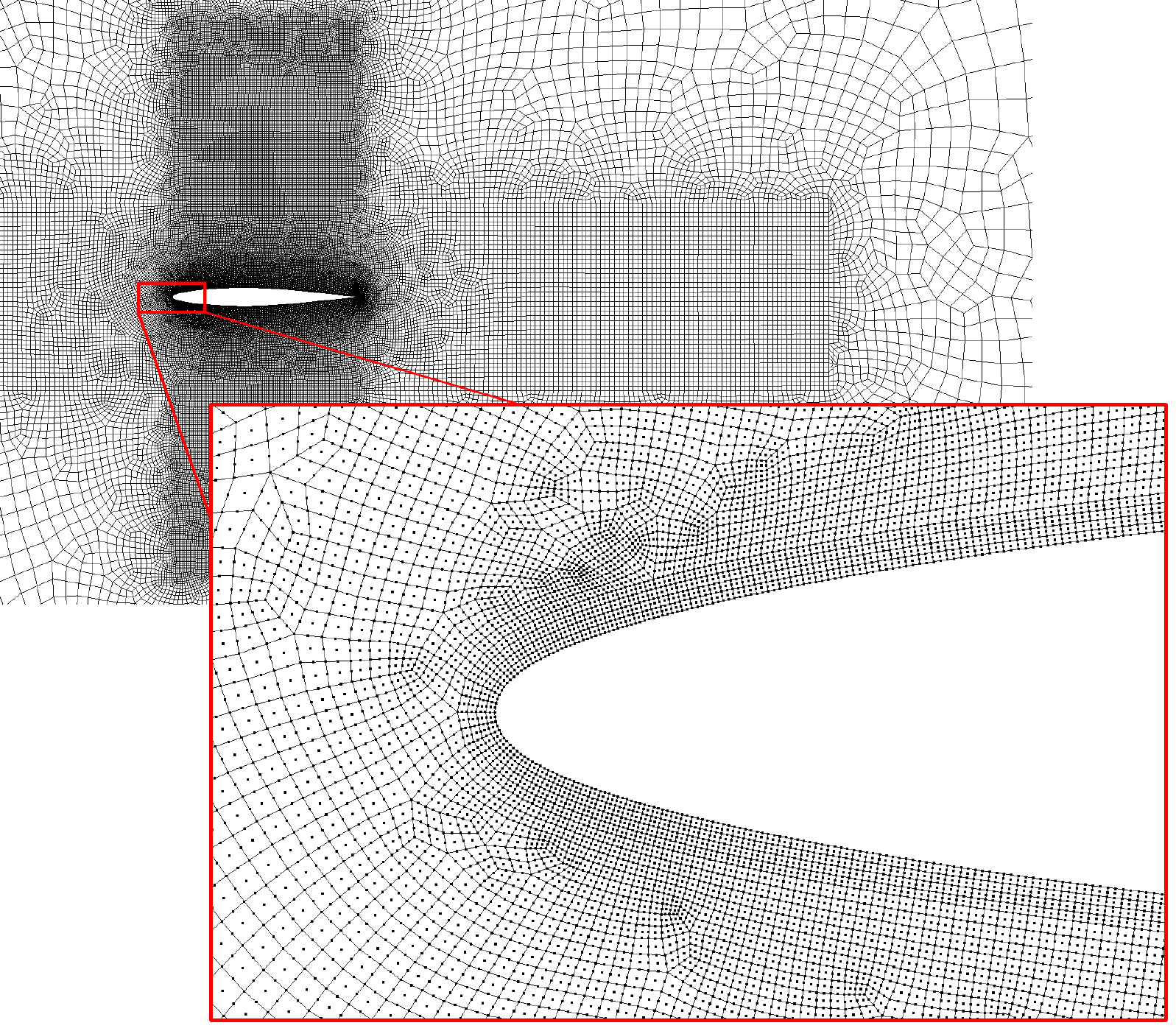}
		\caption{\label{fig:unsMesh} Unstructured mesh around the  NACA64A010 airfoil.}
	\end{figure}

	In this paper, we turn towards semi-Lagrangian propagation as our choice of off-lattice scheme, which was recently 
	extended to compressible flows using a two-population LBM on standard lattice and body-fitted meshes \cite{Saadat2020}. This leads to a simpler and more efficient treatment of complex boundaries while keeping the $D2Q9$ lattice structure, even at high Reynolds and Mach numbers. An example of a body fitted mesh over a NACA64A010 airfoil, with refinement zones to adequately resolve shock waves, is shown in Fig.~\ref{fig:unsMesh}.
	
\begin{figure}[h!]
		\includegraphics[width=\linewidth]{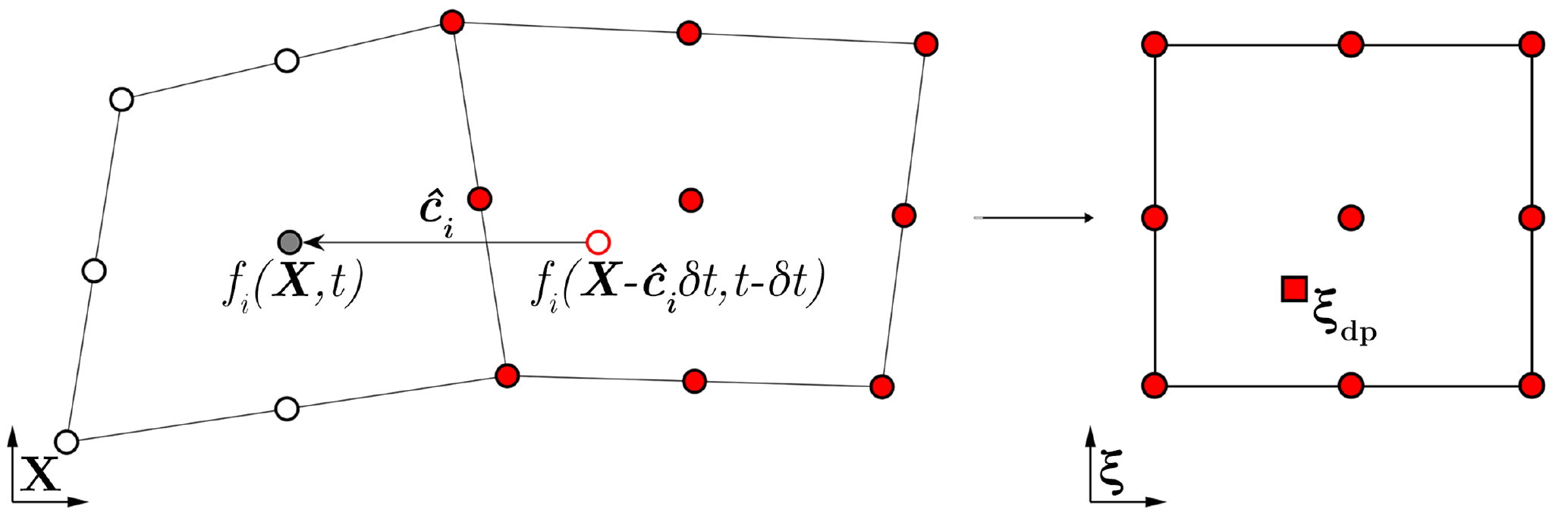}
		\caption{\label{fig:SL_Inter} Schematic of a second-order finite-element mesh, the semi-Lagrangian advection along the discrete velocity $\hat{\bm{c}}$ and the mapping to reference cell \cite{Saadat2020}.}
	\end{figure}

	The off-lattice nature of the model implies that the  departure points of the characteristic lines do not 
	necessarily coincide with a lattice node anymore
	and therefore, the reconstruction of the populations at these points require interpolation.
	Fig.~\ref{fig:SL_Inter} describes such a scenario that results from a non-uniform grid.  
	
	In order to obtain the values of the populations $ f_i $ at the off-lattice departure points, we need to interpolate based on the known values of the populations. 
	Here, this is done using a second-order finite-element reconstruction,
	\begin{align} \label{eq:SemiLagrProp}
	f_{i}(\bm{X}, t)=f_{i}\left(\bm{X}-\hat{\bm{c}_{i}} \delta t, t-\delta t\right)=\sum_{s=1}^{9} f_{i}\left(\bm{\xi}_{s}, t-\delta t\right) N_{s}\left(\bm{\xi}_{\mathrm{dp}}\right), 
	\end{align}
	where the summation is carried out over all collocation points $\bm \xi_{s}$, $s=1,\dots, 9$ of a reference cell with $ \bm{\xi}  =  (\xi,\eta), (-1 \leq \xi,\eta \leq 1) $.
	$ N_s $ are standard second-order Lagrangian finite-element shape functions and $\bm \xi_{dp}$ is the departure point on the reference cell.
	Here, we use the total Lagrangian formulation which means that all computations take place on the finite-element reference cell, and hence, as a first step we need to compute the mapping, $\bm{G}_{\xi}( \bm{X}_{\mathrm{dp}} , t )$ of the departure point in the undeformed frame, $ \bm{X}_{\mathrm{dp}} = \bm{X}-\hat{\bm{c}_{i}}{\delta t}$, to the departure point in unit reference frame, $\bm{\xi_{dp}}$. This involves solving the following system for $\bm{\xi_{dp}}$: 
	\begin{align} \label{eq:depPnt}
	\bm{X}_{\mathrm{dp}}=\sum_{s=1}^{4} \bm{X}_{s} N^{\prime}_{s}\left(\bm\xi_{\mathrm{dp}}\right),
	\end{align}
    {	where $ N_{s}^{\prime} $ are standard first-order Lagrangian finite-element shape functions. The finite-element interpolation here is computed over a first-order element, instead of second-order element in order to save computational time. Reducing the interpolation order for the departure point computation has negligible effect on the accuracy of the model (see also \cite{Saadat2020}).  
    }
	
	The computation of spatial gradients, is done using the finite-element approximation of the first derivative. For a variable $ C $ we have, 
 
	\begin{align} \label{eqn:ddx+fem}
	\partial_{X} C =J^{-1} \sum_{s} C_{s} \partial_{\xi} N_{s},
	\end{align}
	where $C_{s}$ are the values of the variable at collocation points and $J^{-1}$ is the inverse of the Jacobian of the transformation matrix computed as,
	
	\begin{align}
	J^{-1}=\frac{1}{\operatorname{det} J}\left[\begin{array}{cc}
	\partial_{\eta} y & -\partial_{\xi} y \\
	-\partial_{\eta} x & \partial_{\xi} x
	\end{array}\right],
	\end{align}
	
	and
	\begin{align}
	\operatorname{det} J=\partial_{\xi} x \partial_{\eta} y-\partial_{\xi} y \partial_{\eta} x,
	\end{align}
	
	is the determinant of the Jacobian.
	The metrics of transformation $\partial_{\xi} x$, $\partial_{\eta} x$, $\partial_{\xi} y$, $\partial_{\eta} y$ are  computed with 
	\begin{align}
	\partial_{\xi} x=\sum_{s} x_{s} \partial_{\xi} N_{s}.
	\end{align}

	To summarize, the semi-Lagrangian propagation on moving grids can therefore be split into two steps, (i) solution of Eq.~\eqref{eq:depPnt} to determine the departure point in the reference frame, and (ii) finite-element interpolation of the population at the departure point using Eq.~\eqref{eq:SemiLagrProp}.
	
	\subsection{\label{sec2:l2}  Kinetic Equations }

    Here, we briefly describe the two-population model with correction terms as proposed in \cite{Saadat2019}, where populations $ f $  are used to conserve mass and momentum, and populations $ g $ are responsible for energy conservation. We limit ourselves to two dimensions and the standard D2Q9 lattice with the discrete velocities $\bm{c}_i=(c_{ix},c_{iy}),\ c_{i\alpha}\in\{-1,0,1\}$. The kinetic equations read: 
	\begin{align} 
    	f_{i}(\bm{X},t) - f_i \left(\bm{X}-\hat{\bm{c}_{i}} \delta t, t - \delta t\right)&=\omega\left(f_{i}^{\mathrm{eq}}-f_{i}\right) + \delta t \phi_i ,		\label{eqn:f_boltz}
	    \\
    	g_{i}(\bm{X}, t)-g_i \left(\bm{X}-\hat{\bm{c}_{i}} \delta t, t - \delta t\right)&=\omega\left(g_{i}^{\mathrm{eq}}-g_{i}\right) \nonumber \\
    	&+ \left(\omega_{1}-\omega\right)\left(g_{i}^{*}-g_{i}\right),
	\end{align}
	where $\delta t$ is the time-step. The local equilibria $ {f_i}^{\rm eq}$  and ${g_i}^{\rm eq} $, and the quasi-equilibrium $ {g_i}^* $ satisfy the local conservation laws for the macroscopic quantities: density $\rho$, velocity $\bm{u}$, and total energy $ E $,
	\begin{align}
	&\sum_{i = 0}^{Q-1} f_i^{\rm eq} = \rho, \\
	&\sum_{i = 0}^{Q-1} f_i^{\rm eq} \bm{c}_{i} =\rho \bm{u}, \\
	&\sum_{i = 0}^{Q-1} g_i^{\rm eq} = 2 \rho E.
	\end{align}
	The temperature $T$ is defined as, 
	\begin{align}
	T = \frac{1}{C_v}\left(E - \frac{u^2}{2}\right),
	\end{align}
	where $C_v \text{ and } C_p$ are the specific heat at constant volume and pressure, respectively, related by $C_v = C_p - R$, and $R$ is the adiabatic gas constant, which here is set to 1. The adiabatic exponent is denoted by $ \gamma = C_p/C_v $. The Mach number is defined as ratio of velocity to the speed of sound ${\rm{Ma}} = u/c_s$, where $c_s = \sqrt{\gamma R T}$ is the speed of sound.

    The equilibrium populations must also satisfy the Maxwell-Boltzmann (MB) relations in order to recover the full Navier-Stokes-Fourier equations in the hydrodynamic limit.

	The equilibrium populations, $ {f_i}^{\rm eq} $,  can be written in product form as, 
	\begin{equation} \label{eqn_feq}
	f_i^{\rm eq} = \rho \Phi_{c_{ix}} \Phi_{c_{iy}}, %
	\end{equation}		
	where,  
 		\begin{align} 
	&	\Phi_{-1} = \frac{{-u_\alpha}+{u_{\alpha}^{2}}+T}{2} ,
		\\
	&	\Phi_{0}  = 1-\left({u_{\alpha}^{2}}+T\right) ,
		\\
	&	\Phi_{+1} = \frac{{u_\alpha}+{u_{\alpha}^{2}}+T}{2},
		\end{align}
    where $ \alpha \in \{ x,y \}$.
	Equilibrium populations $ {g_i}^{\rm eq} $ and  quasi-equilibrium $ {g_i}^{*}$ are constructed to ensure that the higher order moments of energy are reproduced  \cite{KARLIN2013} and read,
\begin{align}
	&g_i^{\rm eq}= W_i \left(2\rho E +  \frac{ \bm{q}^{\rm eq}\cdot \bm{c}_{i}}{T} \right.\nonumber\\
&	+ \left. \frac{(\bm{R}^{\rm eq}-\rho E T \bm{I}): (\bm{c}_{i}\otimes\bm{c}_{i} - T \bm{I})}{2 T^2} \right) + \Psi_i,
	\label{eq:geq}\\
&	g_i^{*}= W_i \left(2\rho E +  \frac{ \bm{q}^{*}\cdot \bm{c}_{i}}{T} \right.\nonumber \\
&\left.+\frac{(\bm{R}^{\rm eq}-\rho E T \bm{I}): (\bm{c}_{i}\otimes\bm{c}_{i} - T \bm{I})}{2 T^2}\right) + \Psi_i, 	
	\label{eq:gstar}
\end{align}	    
	with the weights  
	\begin{equation}
	    W_i = W_{c_{ix}} W_{c_{iy}}, %
	\end{equation}
	and
	\begin{subequations}
		\begin{align}
        	&	W_{-1} =\frac{T}{2}, \\
        	&	W_{0}  =1 - T, \\
        	&	W_{+1} = \frac{T}{2}.
		\end{align}
	\end{subequations}	
	
    $\bm{q}^{\mathrm{eq}} $ is the equilibrium heat flux vector, %
		\begin{equation}
    			\bm{q}^{{\rm eq}}
    			=\sum_{i=0}^{Q-1} g_{i}^{\mathrm{eq}} c_{i } =2 \rho \bm{u} H,
		\end{equation}
where the enthalpy is defined as $ H = (C_v+1)/T + \bm{u{^2}}/2$. $	\bm{R}^{\mathrm{eq}} $ is the 4th order Maxwell-Boltzmann moment defined as,
		\begin{equation}
			\bm{R}^{\mathrm{eq}} 
    		  \begin{aligned}[t]
	    		&= \sum_{i=0}^{Q-1} g_{i}^{\mathrm{eq}} \bm{c_{i }} \otimes \bm{c_{i }} 
		    	\\
			    & =2 \rho E\left(T \bm{I} + \bm{u} \otimes \bm{u}\right)
			    +2 \rho T\left(T \bm{I} + 2 \bm{u} \otimes \bm{u}\right).
	    	\end{aligned}
		\end{equation}    
	The term $ \phi_i $ in Eq.~\eqref{eqn:f_boltz} is the correction term for the momentum equation, which is defined as 
		
	\begin{equation}
	    \begin{aligned}
	     \phi_i = -\frac{\bm{c}_{i}}{2} \cdot \nabla  \odot \left[\left(\frac{1}{\omega}-\frac{1}{2}\right) \nabla \odot ( \rho \bm{u}(1-3T) -\rho u \odot u \odot u  )\right].
	    \end{aligned}
	\end{equation}

	Another correction term, $ \Psi_i  $, for the energy equation, appears in the definition of the equilibrium and quasi-equilibrium populations, $ {g_i}^* \text{ and } {g_i}^{\rm eq} $ (Eq.~\eqref{eq:geq} and \eqref{eq:gstar}), 
	which reads,
	\begin{equation}
	    \begin{aligned}
	     \Psi_i = \bm{B}_{i} \cdot \left(\frac{\rho(1-3 T)\left(T^{2}+ 2(\bm{u} \odot \bm{u})T+E (\bm{u} \odot \bm{u})\right)}{T} \right) ,
	    \end{aligned}
	\end{equation}

    Here, $\odot$ is the symbol for the Hadamard product or the component-wise product. 
    
	These correction terms are designed to cancel out the errors in the fluid pressure tensor that are originating from the restriction of the $D2Q9$ lattice.  
	While we only present the resulting expressions here, a detailed derivation can be found in \cite{Saadat2019}. 

	From the Chapman-Enskog analysis of the kinetic equations together with the equilibrium populations, we are able to recover the full Navier-Stokes-Fourier equations:  
\begin{align}
&\partial_t \rho + \nabla\cdot (\rho \bm{u})=0,
\label{eq:dtrho}\\
&\partial_t (\rho\bm{u}) +  \nabla\cdot ({\rho\bm{u}\otimes\bm{u} })+ \nabla\cdot \bm{P}=0,
\label{eq:dtu}\\
&\partial_t (\rho E)+\nabla\cdot(\rho E\bm{u})+\nabla\cdot\bm{q}+\nabla\cdot(\bm{P}\cdot\bm{u})=0,
\label{eq:dtE}
\end{align}

with the heat flux $\bm{q} = -\kappa \nabla T$ 
and 
the viscous stress tensor, 
	\begin{equation} 
	\bm{P} = p\bm{I}  - {\mu}\left[\nabla\bm{u}  + \nabla\bm{u}^{\dagger} -\frac{1}{C_{v}}(\nabla\cdot\bm{u})\bm{I} \right],
	\end{equation}
	where the pressure is given by $p = \rho T $. 
The viscosity $\mu$ and the thermal conductivity $\kappa$ are related to the relaxation parameters $\omega$ and $ \omega_1 $ by
	\begin{equation}
		\begin{aligned}
		\mu = \left( \dfrac{1}{\omega} - \dfrac{1}{2}   \right) p\delta t ,  \qquad
		\kappa = C_p \left( \dfrac{1}{\omega_1} - \dfrac{1}{2}   \right) p \delta t.
		\end{aligned}
	\end{equation}
 
	\subsection{\label{sec2:l5} Concept of Blended Mesh} 
	
	In the case of a single body undergoing a prescribed motion, 
	the mapping  $ \bm{x} =  \bm{G}_x(\bm{X},t) $ is known a-priori. In such cases, the ALE transformation is obtained trivially by a combined translation and rotation transformation of the entire domain. The entire domain assumes a rigid motion that mirrors the motion of the moving body. For coupled motion, this mapping is not available analytically and is instead constructed at each time step from the values of the body displacement. The time derivatives in this case are then calculated using first-order finite-differences.%
	
	{Here, we focus  on the polynomial blended function approach for maintaining mesh conformity, for reasons explained below.}
	\begin{figure} [h!]
		\centering
		\includegraphics[width=\linewidth]{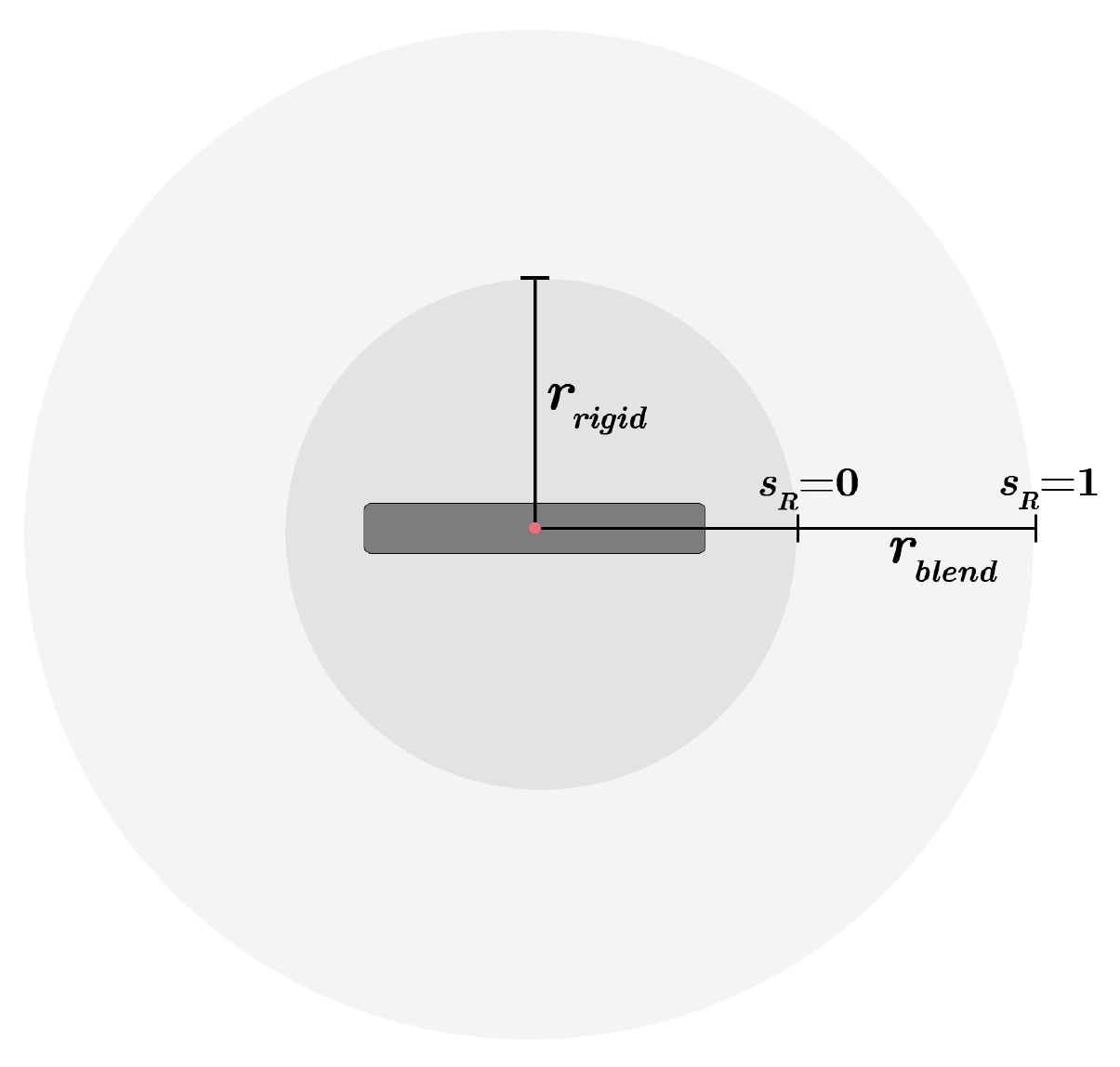}
    		\caption[]{Schematic of the blended mesh showing the region of influence around a moving body.  }
		\label{fig:blended}
	\end{figure}
	
	In this approach, a moving body is given a region of influence, inside of which the motion of the mesh is influenced by the motion of the body. This region is divided into two parts: the rigid region and the blended region (see Fig.~\ref{fig:blended}). Inside the rigid region (dark gray in Fig.~\ref{fig:blended}), the mesh motion coincides with the motion of the body. This ensures mesh conformity and quality close to the walls. Inside the blended region, the influence of the moving body on mesh deformation is gradually reduced, and vanishes at the boundary of the blended region (light gray in Fig.~\ref{fig:blended}). Beyond the blended region, the mesh is stationary and no special treatment of the nodes is necessary. The choice of how big to make the blended and rigid regions are dependent on the size of the body in question and the amplitude of the motion.
	
	Inside the blended region, the distribution of mesh motion is done using blending functions. 
	In this work, we use a 5th order polynomial function:
	\begin{align}\label{eqn:blendPol}
	r_5(s_R) = 10s_R{^3} - 15s_R{^4} + 6 s_R{^5} , \qquad s_R \in [0,1],
	\end{align}
	where $ s_R $ is  the normalized position of the mesh node inside the blended region.
 
	With Eq.~\eqref{eqn:blendPol}, the new position of a node in the physical domain, $ \bm{x} $, can be computed from the rigid mapping, $\bm{G}_{x}^{\prime}(\bm{X},t) $, and position of the node in the undeformed domain, $ \bm{X} $, as:  
	\begin{equation}
	\bm{x}= r_{blend} \cdot \bm{X} + (1-r_{blend}) \cdot \bm{G}_{x}^{\prime}(\bm{X},t), 
	\end{equation}
	where $r_{blend}$ is the blending factor, 
	\begin{equation}
	r_{blend} =\left\{
	\begin{array}{ll}
	0, & \text { if } s_R<0, \\
	1, & \text { if } s_R>1, \\
	r_5(s_R), & \text { otherwise, }
	\end{array}
	\right.\end{equation}
	and $ \bm{G}_{x}^{\prime}(\bm{X}) $ is the discrete map from physical to moving domain, computed at each time-step from the output of the structural solver.   
	
	In this first approach to moving bodies, we use radial distances from the center of the body for defining the region of influence. While this approach works well, it may not be optimal for slender bodies. In that case, an approach based on the distance field from each point on the moving body might be more efficient.
	
	The blended function approach can be extended easily to multiple bodies by assigning a region of influence to each individual moving body. 
	Given the approach as described above, it can be seen that it is not suitable for multiple moving bodies that are placed in very close proximity to each other, since the regions of influence cannot overlap.
	
	{For the simulations in this paper, we only consider motion of one or more rigid bodies. In case of multiple bodies, there is sufficient distance between them to ensure that each can be assigned an exclusive region of influence. 
	As such we do not stand to benefit from the expensive solution of differential equations, at each time-step, to obtain the ALE map. However, for cases involving relative motion of closely placed bodies, or any arbitrary non-rigid motion of bodies, solving an additional differential equation might be a robust and attractive alternative.  }

	\section{\label{sec:FSI} Fluid Structure Coupling}
 
In this work, we use a weak coupling between solid and fluid with a staggered time-stepping approach, where
fluid and solid time steps are carried out sequentially and information is exchanged in between. 
In particular, after a fluid time-step the LB solver computes the forces acting on the body and passes it to the structural solver, which uses the forces to compute the displacement, before passing the displacement back to the fluid solver. While this coupling is first-order accurate in time and does not enforce fluid-solid interface conditions exactly,
it is computationally efficient. 
It is further known that such loose coupling induces artificial energy at the interface due to its staggered nature  \cite{Piperno_2001}. This can lead to severe instabilities for small solid-fluid density ratios, a phenomenon known as the added-mass effect.
Alternatively, strong coupling approaches using sub-iterations to converge to the solid-fluid interface condition can be used to mitigate these issues.
However, the added-mass effect is proportional to the time step size for compressible flows
and has negligible influence for the cases considered here and no stability issues have been observed \cite{Kollmannsberger_2009}. 
Moreover, the LB time step size is well within the stability limits that are imposed by structural solver that is used here. 

The force $\bm{F}(\bm{x}_j,t) = [F_{x},F_{y}]^T $ acting on the $j$th element on the wall is given by
\begin{align}
	\bm{F}(\bm{x}_j,t) = \bm{\sigma}_f \cdot (\bm{n} \mathrm{ds} ),
\end{align}
where $\mathrm{ds}$ is the area of the wall element, $\bm{n}$ is the outward-pointing wall normal vector, and the fluid stress tensor is given by
\begin{align}
  \bm{\sigma}_f = -p\bm{I} - (1-\frac{\omega}{2})  \bm{P}^{(1)}.
\end{align}
The non-equilibrium stress tensor can be obtained by 
 \begin{align}
   \bm{P}^{(1)} = \sum_{i}^{Q-1} f_{i}^{(1)} \bm{c}_i \otimes \bm{c}_i,
 \end{align}
where the non-equilibrium populations $ f_i^{(1)}$ are computed by
\begin{equation}
    \begin{aligned} \label{eq:fnoneq1}
      f_i^{(1)} \approx  f_i - f_i^{eq}.
    \end{aligned}
\end{equation}
Finally, the total force acting on the body can then be obtained via a summation of the element-wise forces over the wall,
 \begin{align}
   \bm{F}_{total} = \sum_{j} \bm{F}(\bm{x}_j,t).
 \end{align}

\subsection{\label{sec3:l1} Fluid Boundary Conditions}
	\subsubsection{Wall Boundary Conditions	} \label{sec:wallBC}
	\begin{figure}[h]
		\centering
		\includegraphics[width=1\linewidth]{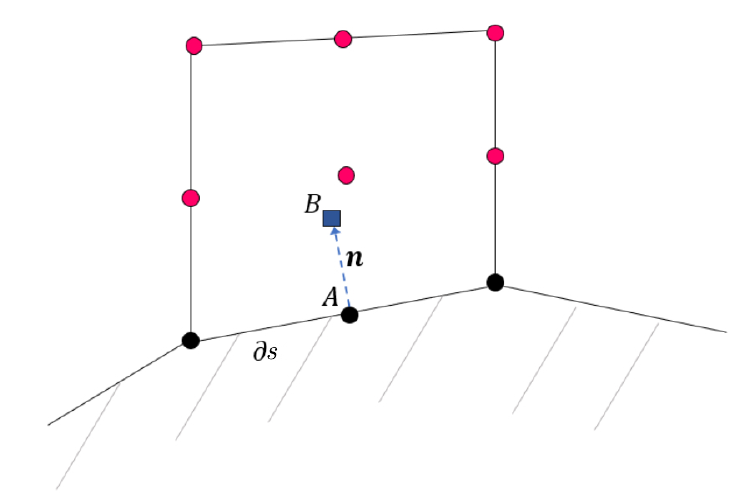}
		\caption[]{ Schematic of a boundary cell with the unit normal $\bm{n}$ at a collocation point A on the wall boundary $\partial s$. The point B is a unit distance away and is used to impose zero-gradient conditions at the wall \cite{Saadat2020}. }
		\label{fig:wallBC}
	\end{figure}
	To impose wall boundary conditions in the fluid solver we follow the approach of \cite{dorschner2015_grad,Saadat2020} and replace the missing populations at the wall nodes with Grad's approximation (see Fig.~\ref{fig:wallBC}),
	\begin{align}\begin{aligned}
	f_{i}^{\text {miss }}=& f_{i}^{\mathrm{eq}}\left(\rho_{\text {tgt }}, \bm{u}_{\text {tgt }}, T_{\mathrm{tgt}}\right) \\
	&+\delta t f_{i}^{(1)}\left(\rho_{\text {tgt }}, \bm{u}_{\text {tgt }}, T_{\text {tgt }}, \nabla \bm{u}_{\text {tgt }}, \nabla T_{\mathrm{tgt}}\right),
	\end{aligned}\\
	\begin{aligned}
	g_{i}^{\text {miss }}=& g_{i}^{\mathrm{eq}}\left(\rho_{\text {tgt }}, \bm{u}_{\text {tgt }}, T_{\mathrm{tgt}}\right) \\
	&+\delta t g_{i}^{(1)}\left(\rho_{\text {tgt }}, \bm{u}_{\text {tgt }}, g_{\text {tgt }}, \nabla \bm{u}_{\text {tgt }}, \nabla T_{\mathrm{tgt}}\right),
	\end{aligned}\end{align}
	where we use the target values $\rho_{\rm tgt}$, $\bm{u}_{\rm tgt}$, and $T_{\rm tgt}$
	that are left to be specified in order to impose the boundary conditions. 
	The equilibrium populations $f_{i}^{\mathrm{eq}}$ and $g_{i}^{\mathrm{eq}}$ are obtained from Eq.~\eqref{eqn_feq} and \eqref{eq:geq}. 
	The non-equilibrium population are defined using non-equilibrium values of the higher order moments. $ f_{i}^{\mathrm{(1)}} $ is defined as, 
	\begin{equation}
         \begin{aligned} \label{eq:fnoneq}
           f_i^{(1)} = W_i \left( \frac{\bm{P}^{(1)}(\bm{c}_{i} \otimes \bm{c}_{i} - T\bm{I}) }{2T^2},  \right),
         \end{aligned}
    \end{equation}
    with the non-equilibrium pressure tensor, $\bm{P}^{(1)}$,
    \begin{equation}
         \begin{aligned}
         \bm{P}^{(1)} = -\frac{1}{\omega}\rho T 
         \left( \bm{S} - \frac{1}{C_v} \right) (\nabla \cdot  \bm{u} ) \bm{I}, 
         \end{aligned}
    \end{equation}
	
	and the population $ g_{i}^{\mathrm{(1)}} $ is given as, 
	\\
 	\begin{equation}
     \begin{aligned} \label{eq:gnoneq}
       g_i^{(1)} = W_i \left( \frac{\bm{q}^{(1)}\cdot \bm{c}_{i}}{T} + \frac{\bm{R^{(1)}}(\bm{c}_{i}\otimes\bm{c}_{i}  - TI) }{2T^2}  \right),
    \end{aligned}
    \end{equation}
	with
	\begin{equation}
	\bm{q}^{(1)}=-\frac{2}{\omega_{1}} \rho C_{p} T \nabla T+2 \bm{u}\cdot  \bm{P^{(1)}},
	\end{equation}
	and 
	\begin{equation}
	\bm{R}^{(1)}=-\frac{2}{\omega_{1}} \rho T\left[\bm{S}(E+2 T) + \bm{u} (\nabla E)^\dagger + \nabla E \bm{u}^\dagger \right].  
	\end{equation}
	
	Here, $\bm{S}$ is the strain rate tensor defined as,  
    \begin{align} \label{eq:strainrate}
     \bm{S} = \nabla\bm{u} + \nabla^\dagger{\bm{ u}}.
    \end{align}
	Spatial gradients are obtained using a first order finite-element approximation of the derivative as given in Eq.~\eqref{eqn:ddx+fem}.

	To enforce no slip conditions at the wall ( point A in Fig.~\ref{fig:wallBC}), the target values of the velocity is set to zero, $\bm{u}_{tgt} = 0$. The target values for density and temperature at the wall are obtained by applying a zero-gradient condition normal to the wall,
	\begin{align}
	&\dfrac{\partial{\rho}}{\partial{\bm{n}} } = 0,\\
	&\dfrac{\partial{T}}{{\partial\bm{n}}} = 0.
	\end{align}
	For any scalar, this can be achieved by taking a first-order finite-difference approximation at a location $ B $, which lies a unit normal distance away from the wall (see Fig.~\ref{fig:wallBC}). In the present model with semi-Lagrangian propagation, this distance can be at maximum equal to the time-step size $\delta t$. 

	The value of the scalar at this point, $B$, is obtained by interpolation using second-order finite-elements. 
 	Using this value, together with the zero-gradient boundary conditions, we have 
	\begin{align}
&	\rho_{\rm A}  = \rho_{\rm tgt}  = \rho_B,\\
&	T_{\rm A}  = T_{\rm tgt}  = T_B.
	\end{align}
	For nodes that do not lie on the wall but lie close to the wall, and as a result also have missing populations, the target values are assigned as the value of the variable at the previous time-step.
	
	\subsubsection{Inlet and Outlet}
 
	At the inlet, we use equilibrium populations to impose the inlet target values directly.
	
    At the outlet, we also impose equilibrium conditions and the target values are obtained in the same manner as for the wall boundary condition using a zero gradient condition for density, temperature as well as for velocity.

	\section{\label{sec:StrModel}Rigid Body Structural Solver}
 	\begin{figure}[h!]
		\includegraphics[width=\linewidth]{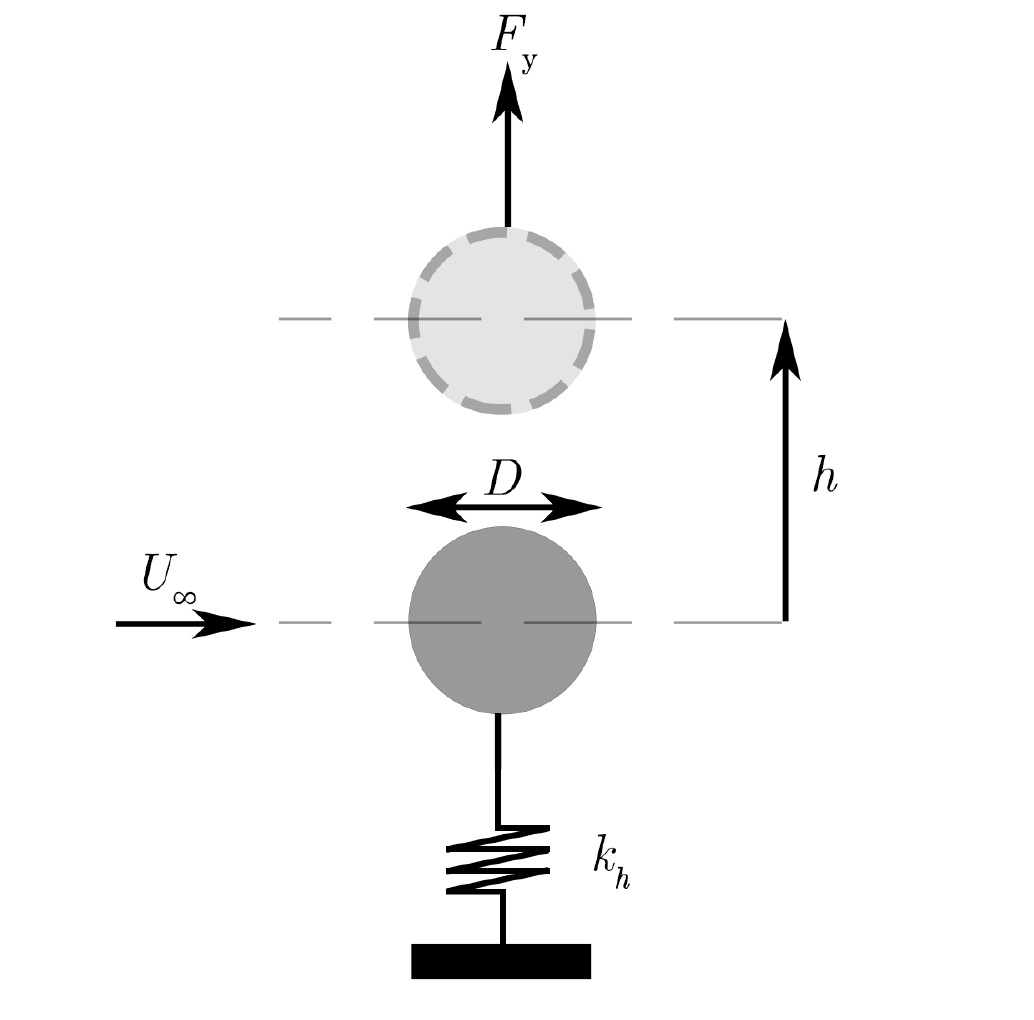}   
		\caption{\label{fig:1dofCYL} Schematic of a 1DOF system of a sphere in cross-flow, mounted on a linear spring. }
	\end{figure}
	Until now we have described the fluid part of the fluid structure interaction. We now describe the structural model, which 
	takes fluid forces on the body as an input and provides the resultant displacement of the body as an output. We present two generalized models, for 1 degree of freedom and 2 degrees of freedom motion, respectively. 
	
	\subsection{\label{sec2:l7} 1DOF Spring-Mass System }
	A single degree of freedom system, with displacement $ h $ and external driving force $ F_y $, as shown in Fig.~\ref{fig:1dofCYL}, can be modeled using the Newton's second law of motion as:
	\begin{equation}
   	\begin{aligned} 	\label{eq:1dofSM}
	m\ddot{h}  + k_h h = F_y, 
	\end{aligned} 
	\end{equation}
	where $ m $, and $ K_h $ are the mass of the system and the spring constant, respectively. Here, we omit the damping term and restrict ourselves to undamped motion.
	
	\subsection{\label{sec2:l8} 2DOF Spring-Mass System with Torsional Spring}
	\begin{figure}[h!]
		\includegraphics[width=1\linewidth]{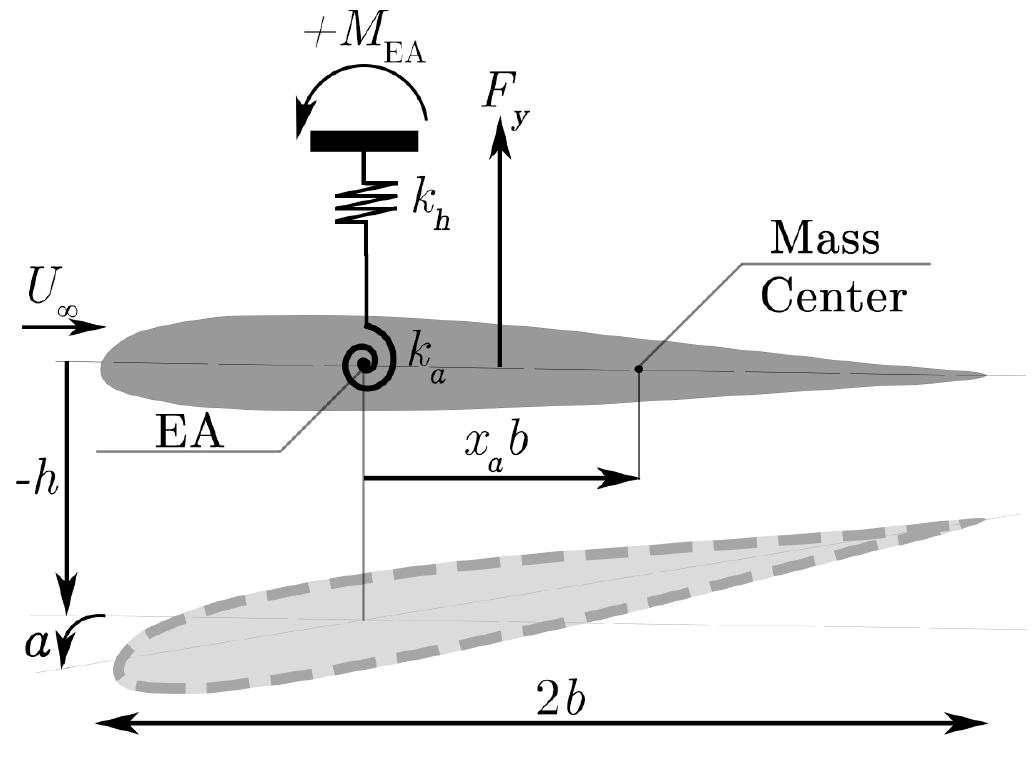} 
		\caption{\label{fig:2dofAfoil} Schematic of a 2DOF system of an airfoil, mounted on a linear and torsional spring }
	\end{figure}
	
	The two degree-of-freedom system with coupled pitching and plunging displacements, as shown in Fig.~\ref{fig:2dofAfoil}, can be described using the second law of motion as before, which yields:
	\begin{subequations} \label{eq:2dofSM}
		\begin{align} 				
		m\ddot{h} + S_\alpha \ddot{\alpha}   + k_h h  &= F_y, \\
		S_\alpha\ddot{h} + I_\alpha \ddot{\alpha}   + k_\alpha \alpha  &= M_{\rm EA}.
		\end{align}
	\end{subequations}
	
	Here, $m$ is the total mass of the system, $ h $  and $\alpha $ are the plunge and pitch displacements, $ I_\alpha $ is the moment of inertia about the elastic axis (EA), and $ S_\alpha $ represents the static moment about the EA. 
    $ K_h$  and $K_\alpha $ are the spring constants for the linear (plunge) and torsion (pitch) spring, respectively. 
    The transverse force (lift) and the moment about the EA are represented by $ F_y$  and $ M_{\rm EA}$, respectively. Once again, we only consider undamped systems in this work. The airfoil half chord $b$ is used to normalize distances, and $x_a$ is the normalized distance of the mass center from the elastic, axis which is positive if the center of mass behind the EA (towards the trailing edge) and negative if it is ahead.

	The resulting system of ordinary differential equations are solved using a 4th-order Runge-Kutta method. 
	
	\section{\label{sec:Res} Results	}
	In the following section we present several test cases that were validated against what exist in the literature in order to show the suitability of our model to capture the dynamics and non-linearity arising from the coupling between the fluid and structural solvers.   
	We build up from the well studied case of 1DOF vortex induced vibrations (VIV) of a single cylinder in incompressible flow, to the more complex dynamics of multiple bodies.
	Finally, to demonstrate a fully compressible case with two degree of freedom, we consider the case of flutter of an airfoil at transonic speeds.    
	\subsection{\label{sec4:l2} 1DOF System: Oscillating Cylinder }
	\begin{figure}[h!]  
		\centering
		\includegraphics[width=0.8\linewidth]{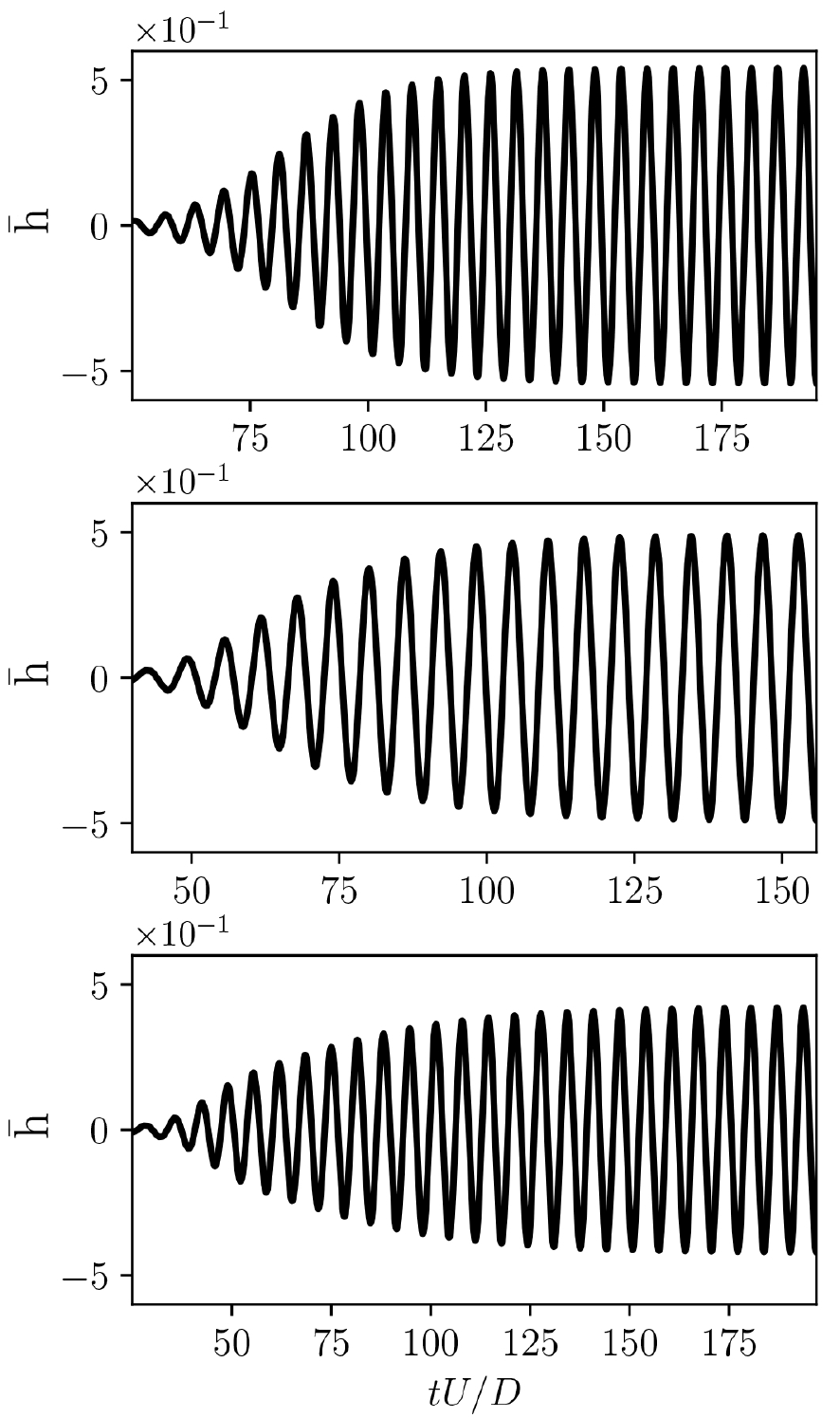}
		\caption{1DOF cylinder in cross-flow. Normalized displacement ${\rm \bar{h}} \text{ (=}h/D )$ over time. Top: $ {\rm Re} = 90 $; middle: $ {\rm Re} = 100 $; bottom: $ {\rm Re} = 110 $.    }
		\label{fig:QiuRe_Disp}  
	\end{figure}
	
	\begin{figure*}[t]
		\includegraphics[width=0.85\linewidth]{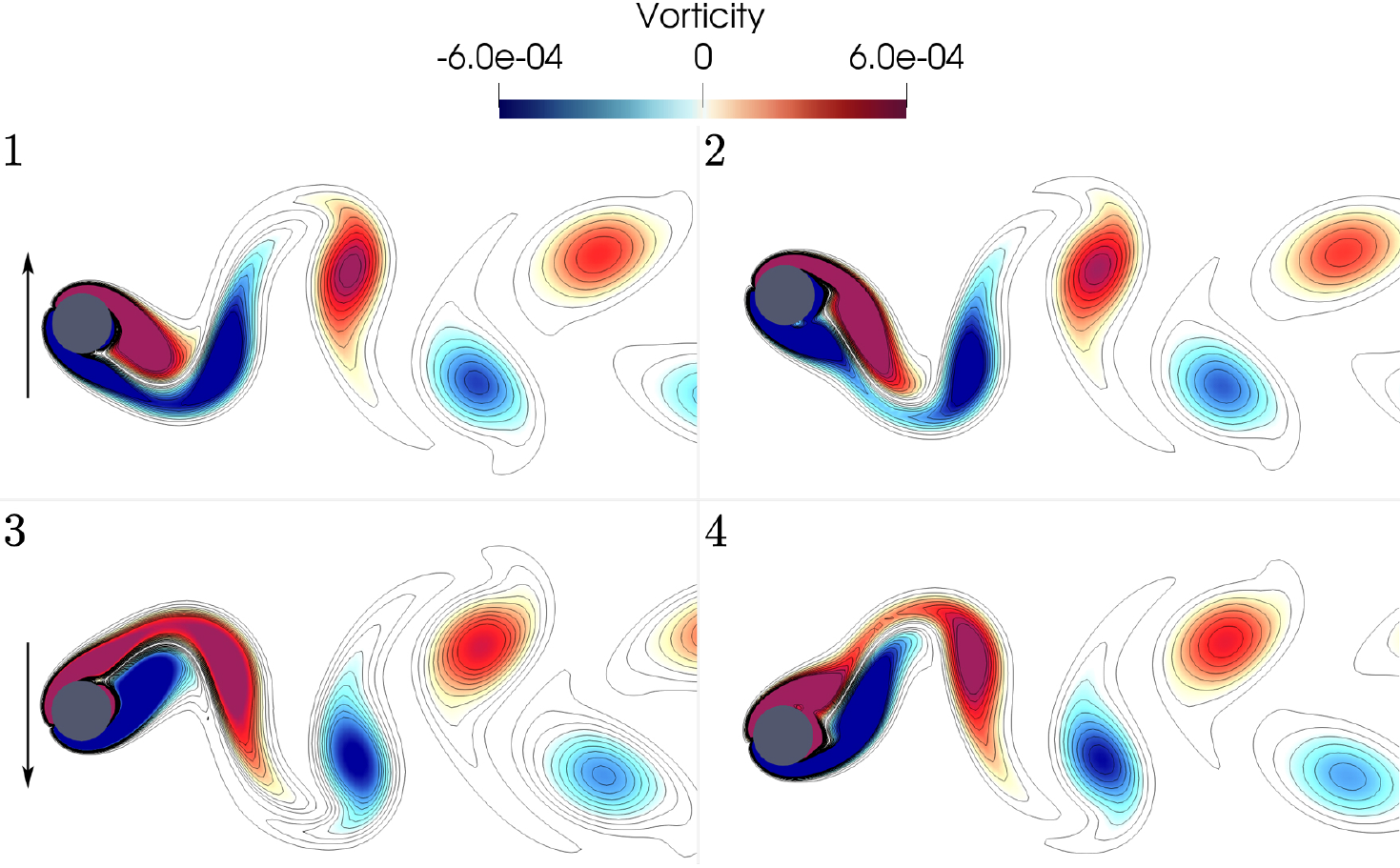}
		\caption{\label{fig:QiuRe90_Vort}1DOF cylinder in cross-flow. Vorticity contours at $ {\rm Re}=90 $.}
	\end{figure*}
	\begin{figure*}[t]
		\includegraphics[width=0.85\linewidth]{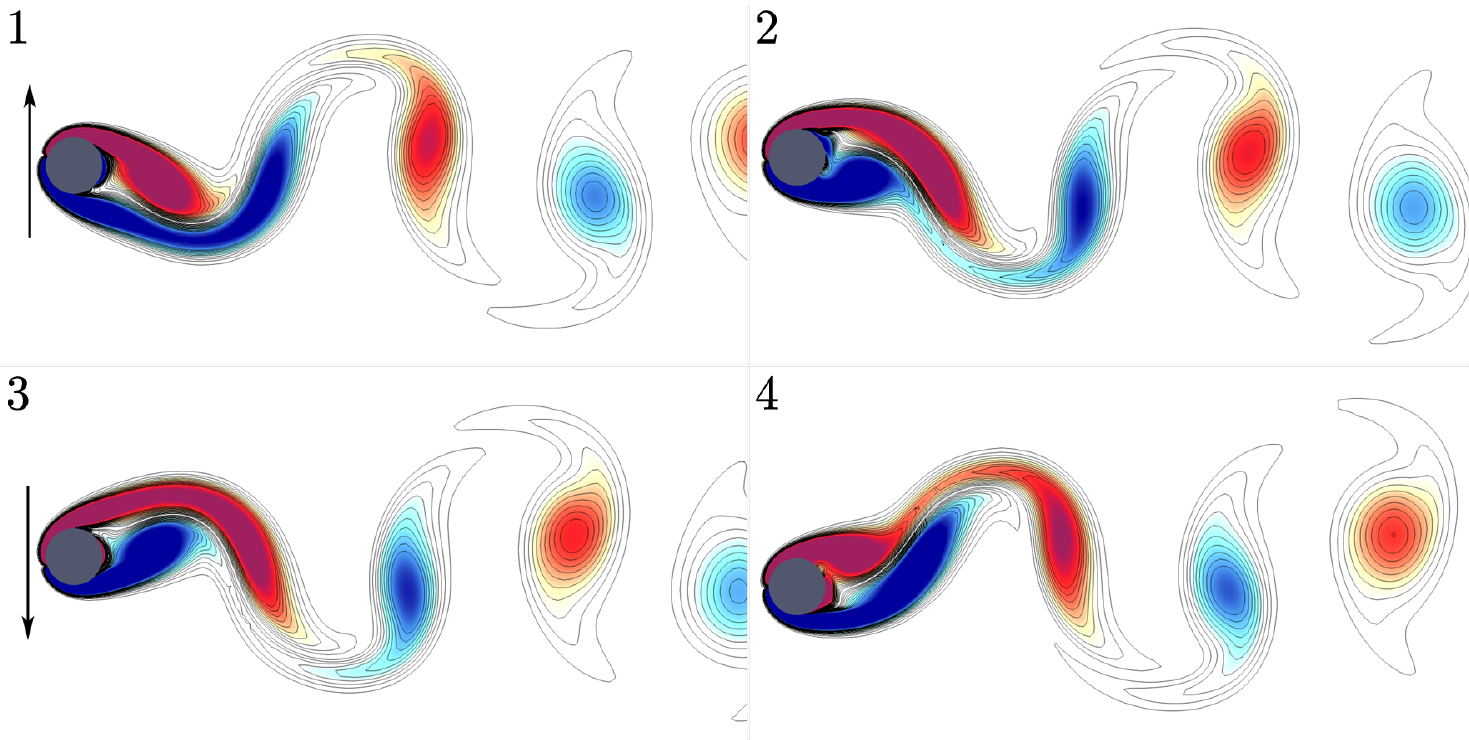}
		\caption{\label{fig:QiuRe110_Vort} 1DOF cylinder in cross-flow. Vorticity contours at $ {\rm Re}=110 $. For the color bar, please refer to Fig.~\ref{fig:QiuRe90_Vort}.}
	\end{figure*}
	
	The motion of an elastically mounted cylinder immersed in cross-flow is widely studied and is used as a model problem for the analysis of vortex induced vibrations. The schematic of the system is shown in Fig.~\ref{fig:1dofCYL}. 

	Here, we compare our results to those obtained by Qiu et al \cite{Qiu2019} where they used a fourth-order spectral difference method. A second-order finite-element mesh with 23,116 elements and 92,906 degrees of freedom was used to discretize the domain. The body had a diameter of 200 lattice units and the near wall mesh size was 2 lattice units. The time-step  was set to $ \delta t  =2/3$. 
	
	Non-dimensionalising Eq.~\eqref{eq:1dofSM} in the same way as Zhao et al\cite{Zhao2013a}, we have,
    \begin{align} 	
	\ddot{h}  + 4 \pi^2 h  = \dfrac{2 {V_R}^2 C_L}{\pi m^*}\label{eq:1dofSM_nonDim},
	\end{align}
    where the lift coefficient is given by $C_L = 2 F_y/ ( \rho U_\infty^2 D )  $
    and	 $ m^* $ is the non-dimensional mass or the mass ratio, $ V_R $ is the reduced velocity, and $ f_N $ is the natural frequency of the system.  
	
	The non-dimensional parameters are defined as: 
 	\begin{align} 
	& m^*  = \dfrac{4m}{\rho \pi D^2 },	\\
	&	V_R  = \dfrac{U_\infty}{f_n D},	\\
	&  \label{eqn:fN1} 	 f_N = \dfrac{1}{2\pi}\sqrt{\dfrac{K_h}{m}}. 
	\end{align}
 
	Fixing the mass ratio $ m^*=10 $, and the reduced velocity $ V_R=0.06{\rm Re} $, allows us to set the natural frequency $ f_N $ of the 
	structural system according to Eq.~\eqref{eqn:fN1}. The Reynolds number is defined as $ {\rm Re}=\rho U_{\infty} D / \mu $. The free-stream Mach number is set to ${\rm{Ma}}_\infty=U_\infty/\sqrt{\gamma T_\infty}=0.1$, $T_\infty=1/3$ and $\gamma=1$.
	The Strouhal number is given by ${{\rm St}} = fD/U_\infty$, where $f$ is the vortex shedding frequency.
	\begin{figure}[t]  
		\centering
	     \includegraphics[width=0.8\linewidth]{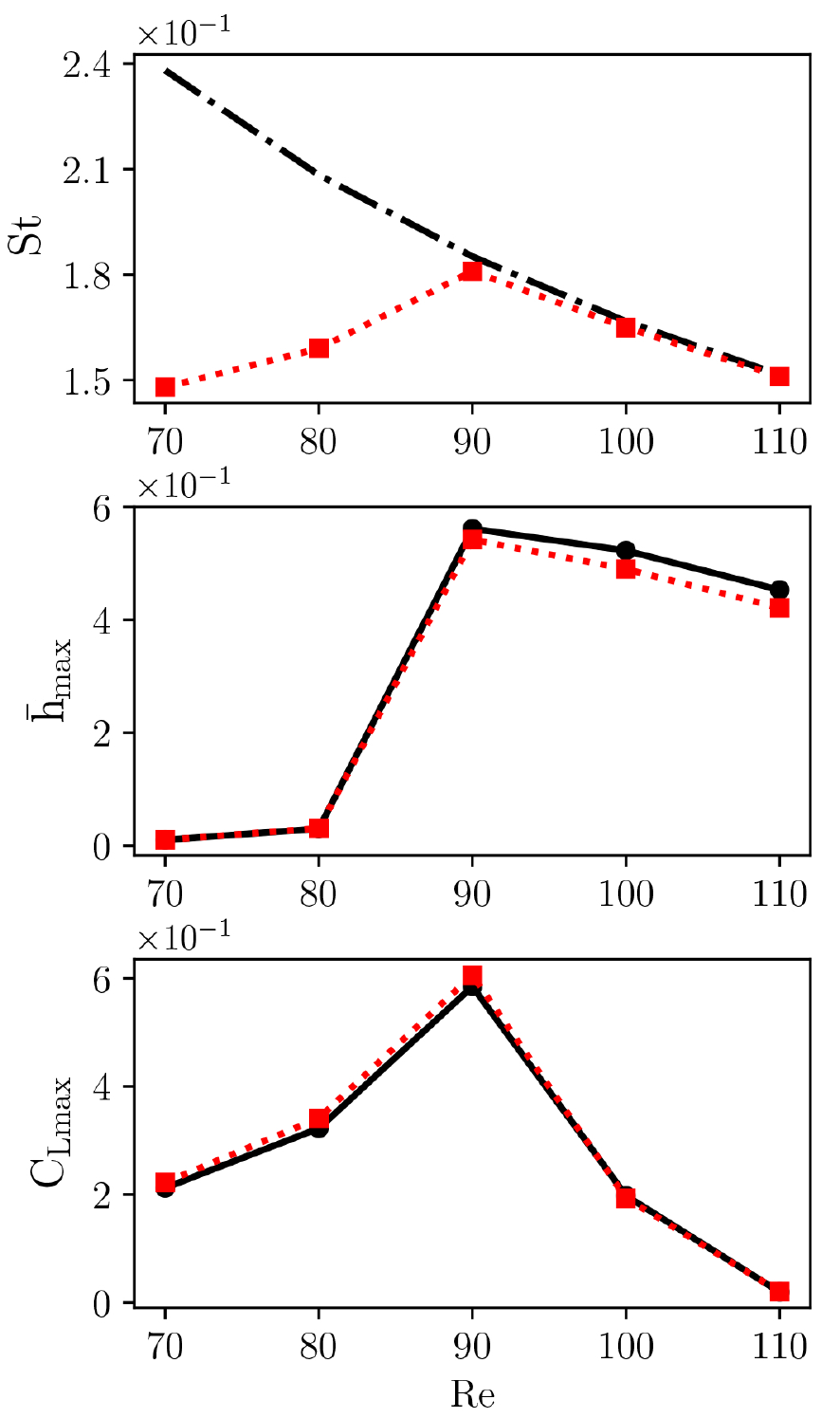}   \\
		\caption{1DOF cylinder in cross-flow. Top: non-dimensional frequency of vortex shedding (red) and non-dimensional natural frequency of structural system (black).  Middle: normalized amplitude of oscillation: present (red) and literature \cite{Qiu2019} (black). Bottom: amplitude of lift coefficient against: present (red) and literature (black). \cite{Qiu2019}}
		\label{fig:QiuRe_Comp}
	\end{figure}
    From the literature \cite{prasanth_mittal_2008}, it is well known that the maximum amplitudes of oscillation are characterized by the synchronisation of the vortex shedding frequency and the natural frequency of the oscillating system. This phenomenon, known as lock-in, presents itself physically as a sudden jump in the oscillation amplitude and the lift coefficient. 
    
	Simulations were carried out at Reynolds numbers between ${\rm Re}=70$ and ${\rm Re}=110$, in the vicinity of the expected lock-in regime. Fig~\ref{fig:QiuRe_Comp} shows the evolution of the oscillations over time, in the the lock-in regime. Snapshots of vorticity for ${\rm Re}=90$ (Fig.~\ref{fig:QiuRe90_Vort}) and ${\rm Re}=110$ (Fig.~\ref{fig:QiuRe110_Vort}) show the vortex shedding from the oscillating cylinders.
	The natural frequency of the structural system, from Eq.~\eqref{eqn:fN1}, is inversely proportional to the Reynolds number, given a fixed free-stream velocity. A clear picture of the onset of the lock-in regime can be seen in fig \ref{fig:QiuRe_Comp} (top). It can be observed that at approximately ${\rm Re}=90$, the vortex shedding frequency approaches the natural frequency of the oscillating system. A sharp jump in the amplitude of oscillation, a result of lock-in, is observed between ${\rm Re}=80$ and ${\rm Re}=90$. The corresponding jump in the maximum lift coefficient results in values that are approximately twice as large compared to the case of a stationary cylinder. The values of maximum amplitude and lift coefficient that were measured in our simulations are plotted in Fig.~\ref{fig:QiuRe_Comp} (middle and bottom). The  agreement with literature is very good, both qualitatively and quantitatively, as can be seen in Fig.~\ref{fig:QiuRe_Comp}.
	The simulations were run for a sufficiently long time to avoid accounting for the initial transient (see Fig.~\ref{fig:QiuRe_Disp}) .

	\subsection{\label{sec4:l3} 1DOF System: Tandem Linked Oscillating Cylinders } 
	\begin{figure}[h!]
	\includegraphics[width=\linewidth]{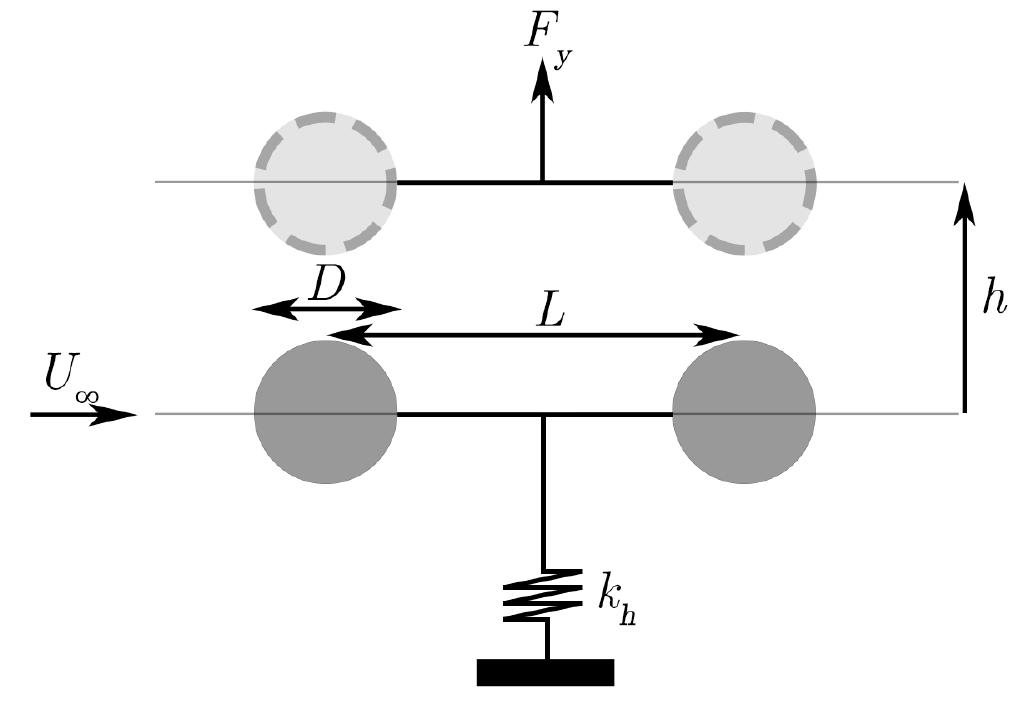} 
		\caption{\label{fig:1dof2Sph}  Schematic  of  a  1DOF  system  of  two rigidly attached cylinders  in  cross-flow, mounted on a linear spring. }
	\end{figure}
	
	The addition of another body to the system leads to the emergence of interesting dynamic behavior, even without an additional degree of freedom. Fig.~\ref{fig:1dof2Sph} represents this spring mass system, with two spherical bodies  that are rigidly connected by a rod, which causes both bodies to oscillate as one. The fluid domain was discretized with a second-order finite-element mesh with 42,068 elements and 169,271 degrees of freedom. The radius of the body was 200 lattice units and the near wall mesh size was 2 lattice units. The time-step was set to $ \delta t= 2/3 $.
   
   	\begin{figure*}[t]
		\includegraphics[width=0.77\linewidth]{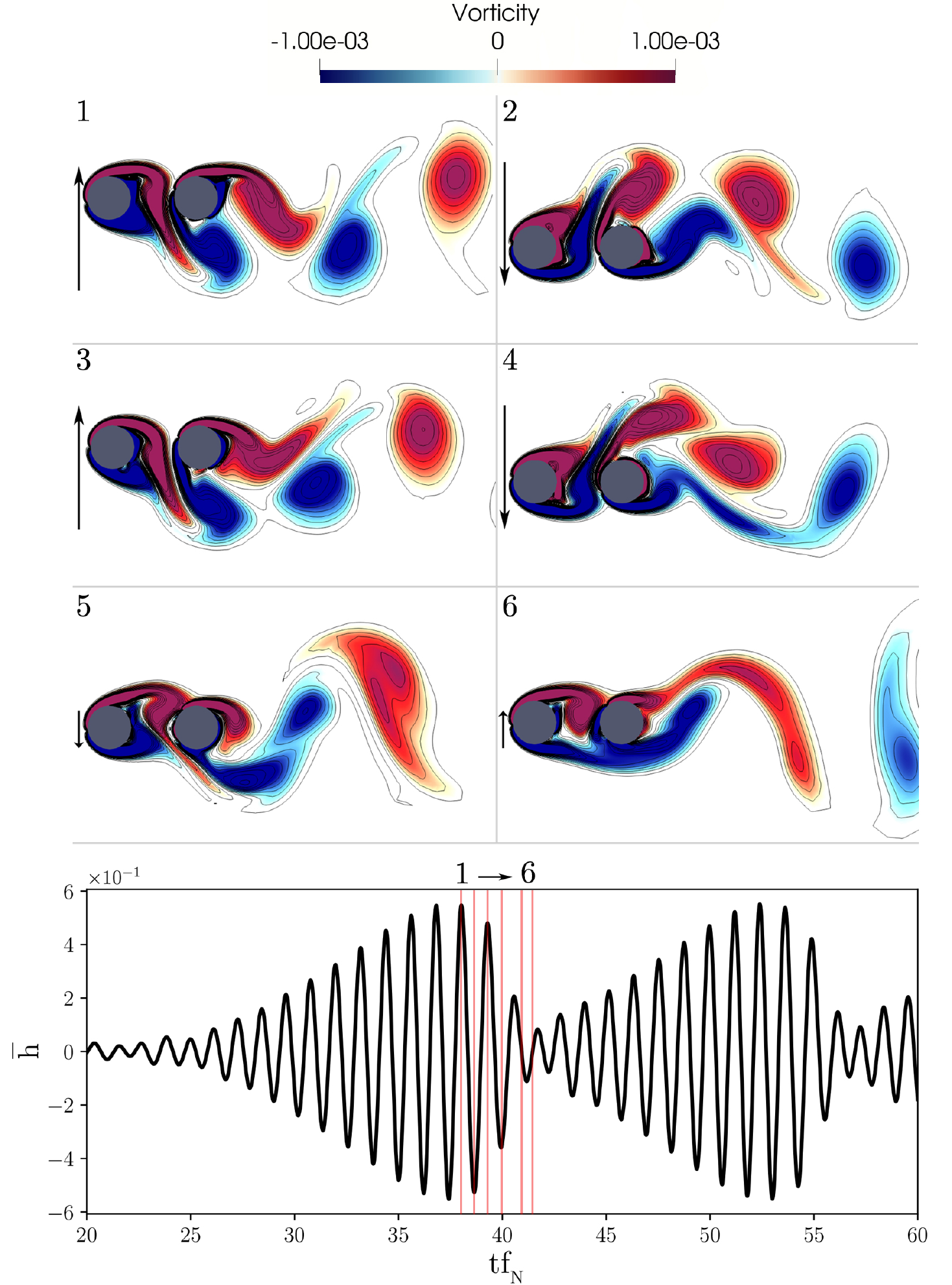}
		\caption{\label{fig:1dof2Sph+beats_vort}  1DOF linked cylinders. Top: vorticity contours. Bottom: normalized displacement $ \bar{h} (=h/D )$ over time for the case $L/D=2$, $V_R=5$.  }
	\end{figure*}
	\begin{figure*}[t]	
		\subfigure[$ \rm L/D=2,V_R=5  $]{\includegraphics[width=0.33\linewidth]{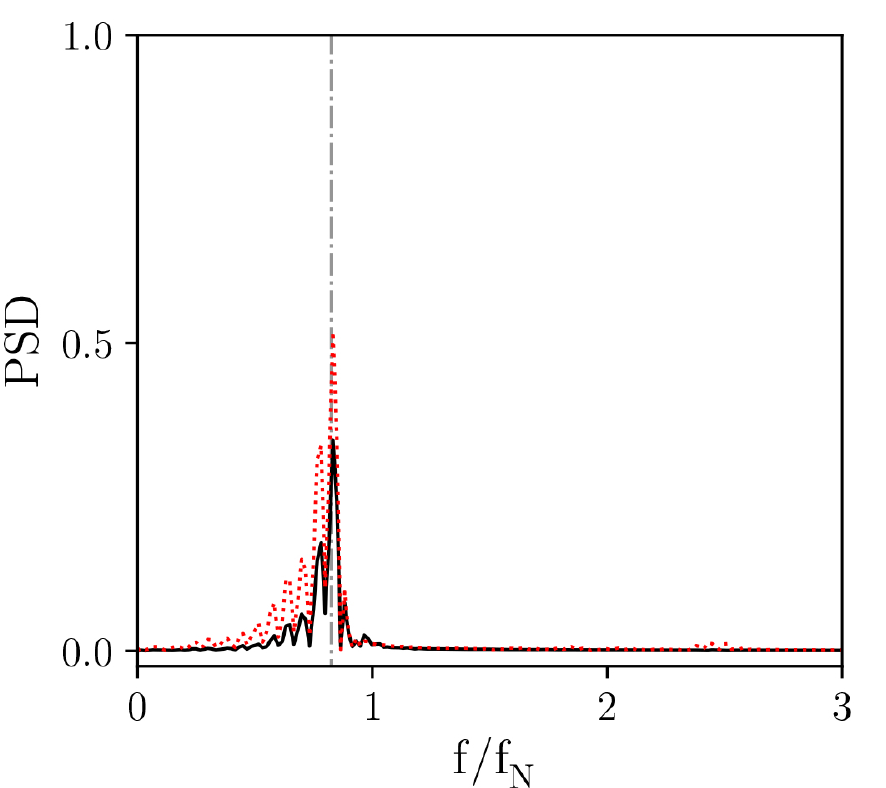}} 
		\subfigure[$ \rm L/D=6,V_R=6 $]{\includegraphics[width=0.33\linewidth]{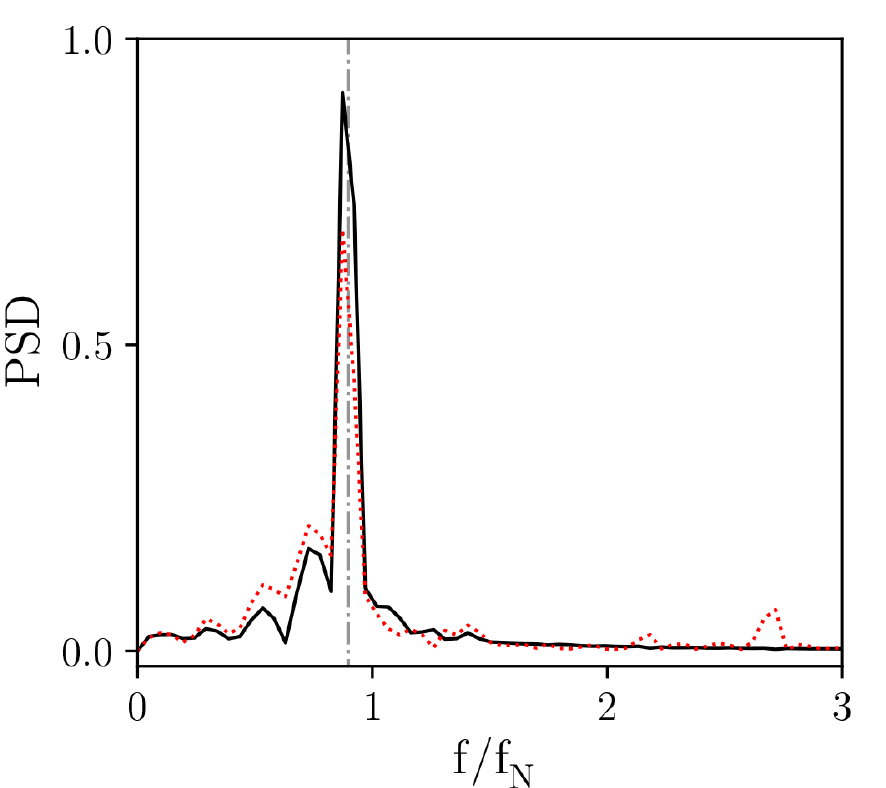}}
		\subfigure[$ \rm L/D=6,V_R=9  $]{\includegraphics[width=0.33\linewidth]{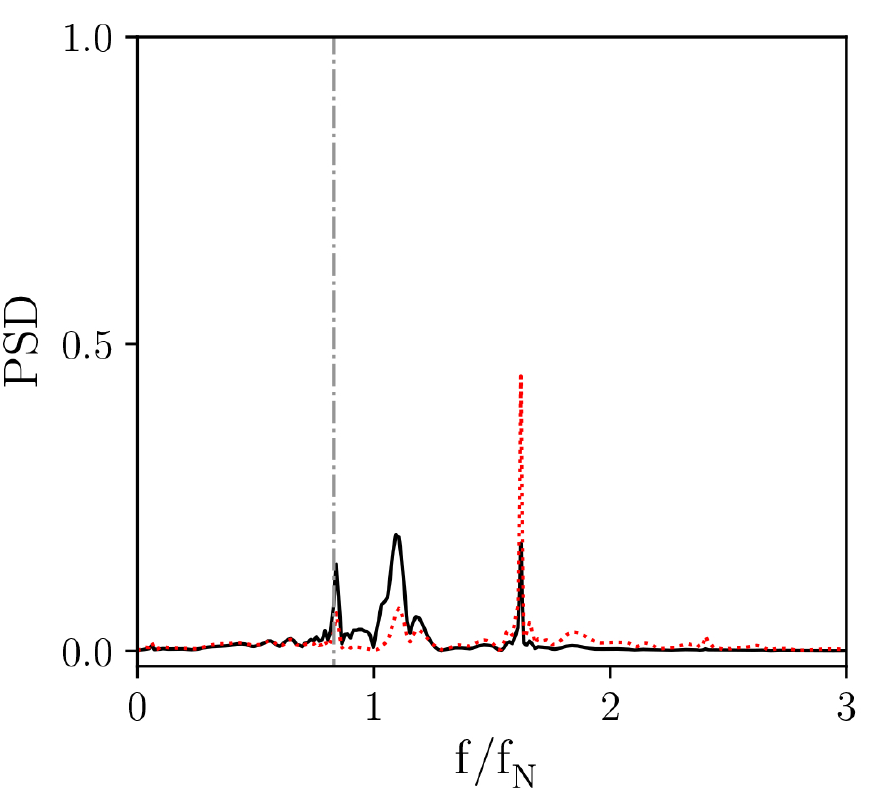}}
		\caption{\label{fig:1dof2Sph+fft} Power spectral density (PSD) of transverse displacement (black) and lift (red) for the 1DOF system of linked cylinders. Plotted on the $x$-axis is the ratio between the oscillation frequency and the natural frequency of the spring-mass system. The gray line marks the vortex shedding frequency. }
	\end{figure*}
	
	\begin{figure}[t]
		\includegraphics[width=0.8\linewidth]{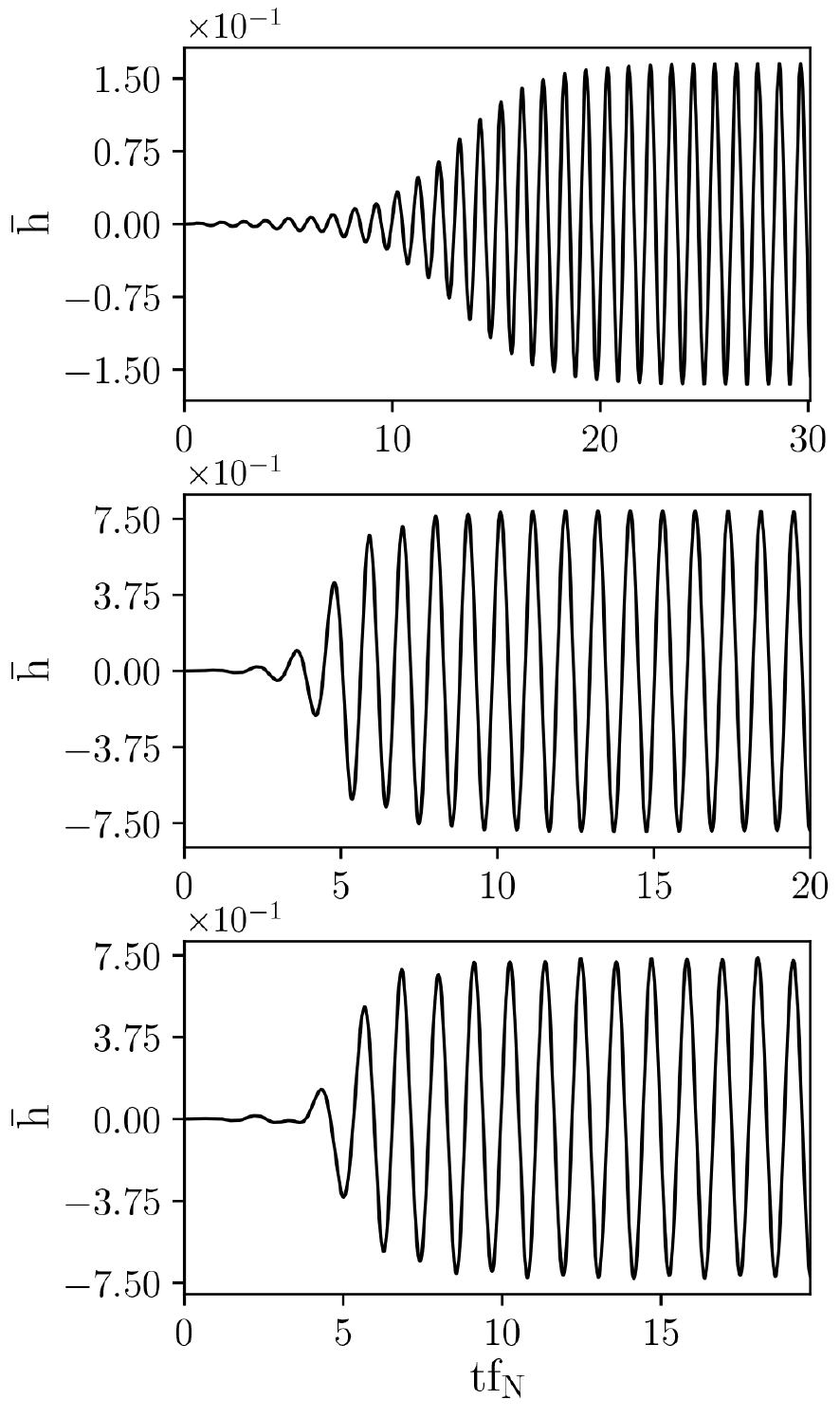}
		\caption{\label{fig:1dof2Sph+vr6}  1DOF system with linked cylinders: normalized displacement ${\rm \bar{h}} \text{ (=}h/D )$ over time for $ {\rm Re}=150 \text{ and } V_R = 6 $. Top: L/D=2. Middle: L/D=4. Bottom: L/D=6.  } 
	\end{figure}
	
	\begin{figure}[h!]
		\includegraphics[width=\linewidth]{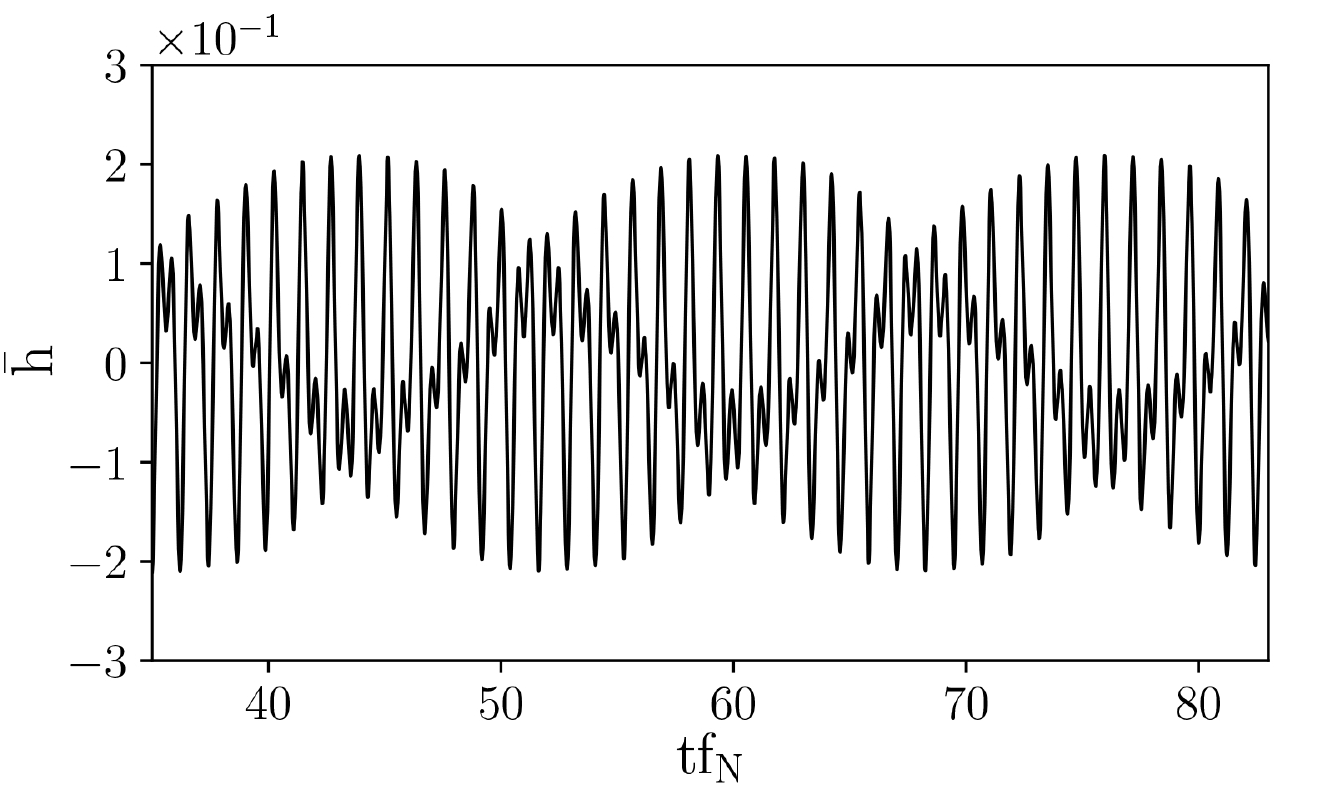} 
		\caption{\label{fig:1dof2Sph+multiMode}  1DOF system with linked cylinders: normalized displacement ${\rm \bar{h}}\text{ (=}h/D )$ over time for $ {\rm Re}=150, V_R = 9, L/D=6 $  }
	\end{figure}
	
	There is no change to the equation of motion (Eq.~\eqref{eq:1dofSM_nonDim}) that was defined in the previous section. The ratio of the body mass to the mass of the displaced fluid was set to  $m^* = 2$.  
   
	Despite the fact that the motion is limited by the linkage, the effect of the downstream body, in the wake of the upstream body, reveals interesting dynamics. 
	Zdravkovich \cite{Zdravkovich1988} studied the flow and dynamic response of multiple cylinders in VIV and classified them into different regimes based on the distance between bodies. Based on this classification, proximity interference and wake interference regimes are defined. Proximity interference occurs when the distance between the bodies is sufficiently small, while wake interference occurs when one body is submerged in the wake of another. The overlap of these regimes, proximity and wake interference, is also possible and leads to complex dynamic behavior. 
	
	Here, we present test cases of VIV of two tandem cylinders at $ {\rm Re}=150 $ exhibiting all these regimes, i.e., $L/D=2,4,6$. We also demonstrate the affect of the reduced frequency and compare our results to the work of Zhao et al \cite{Zhao2013}. In the following tests, the free-stream Mach number is set to ${\rm Ma}_\infty=0.1$ with $\gamma=1$ and $T = 1/3 $.
	
	For $L/D=2$, Zhao et. al reported beating behavior in the range of reduced velocity $ 4.5 \leq V_R \leq 6 $. Beats are caused by interference between waves which differ by a small amount in frequency, leading to a periodic oscillation where amplitude rises slowly to its peak value and then drops rapidly before repeating again. This behaviour is well captured by our model and Fig.~\ref{fig:1dof2Sph+beats_vort} shows the vorticity contours and the time evolution of the amplitude for $ V_R=5 $. According the classification by Zdravkovich, $L/D=2$ falls under the proximity and wake interference regime with intermittent vortex shedding from the upstream cylinder. 
	We can observe that the vortex shedding occurs for both the upstream and downstream cylinders. It is interesting to note that this phenomenon does not occur for stationary tandem cylinders at $L/D=2$, i.e., vortex shedding from the upstream cylinder only occurs due to the motion of the cylinders. In the stationary case, the flow is largely stagnant in the gap between the cylinders, resulting in much lower values of the lift coefficient on the upstream cylinder. This stagnation flow in the gap also presents itself momentarily in the case of the elastic cylinders, and is responsible for the beating. In Fig.~\ref{fig:1dof2Sph+beats_vort}(6), we can see such a case when the gap is mostly stagnant, leading to a drop in lift, and causing a sharp drop in oscillation amplitude. The frequency domain is shown in Fig.~\ref{fig:1dof2Sph+fft}(a) where the secondary modes causing the beats can be visualized.    
	
	At $ V_R=6 $, the natural frequency of the system is approximately equal to the vortex shedding frequency of a single stationary cylinder. The resulting synchronization leads to a regular periodic motion of the cylinders, as shown in Fig.~\ref{fig:1dof2Sph+vr6} for $L/D=2,4$, and 6. Even with dual cylinders the lock-in regime centers approximately around the vortex shedding frequency of a stationary cylinder. Due to a hard lock-in, the power spectral density as shown in Fig.~\ref{fig:1dof2Sph+fft}(b) exhibits a single peak, which is close to the natural frequency of the system, in this regime. 
	
	At higher values of reduced velocity, $ V_R\geq9 $ for $L/D=6$, Zhao et al \cite{Zhao2013} reported transverse oscillations with two or more dominant frequencies. Fig.~\ref{fig:1dof2Sph+multiMode} shows the time series of transverse displacement for $L/D=6$ and $ V_R=9 $ which is able to reproduce the data reported in the literature. The frequency domain Fig.~\ref{fig:1dof2Sph+fft}(c) shows two significant modes, which are  larger than the natural frequency and a minor mode at the vortex shedding frequency. 
	
	The combined results of all cases, over a range of reduced velocities and distances between cylinders, are reported in Table \ref{tab:table3}. 
	It shows good quantitative agreement with what is reported in the literature while the previously shown plots are able to reproduce the dynamics reported in literature qualitatively. 
	
	\begin{table}[h!]
		\caption{\label{tab:table3} Comparison of semi-Lagrangian LBM with literature \cite{Zhao2013} for two linked cylinders oscillating in cross-flow.  }
		\begin{ruledtabular}
			\begin{tabular}{cccccc}
				\multirow{3}*{$ V_R=\dfrac{Uf_N}{D} $}	&	&\multicolumn{2}{c}{Present}&\multicolumn{2}{c}{Reference \cite{Zhao2013}}\\\cline{3-4} \cline{5-6}
				\rule{0pt}{3ex} 	&	L/D & $ \rm \bar{y}_{max}$  & ${ \rm St} = fD/U$  & $ \rm \bar{y}_{max}$  & ${\rm St} = fD/U$  \\ [1ex]\hline \\ [-1.5ex]
				\multirow{4}*{6} &	  2 & 0.5 			& 0.925 	& 0.5  			& 0.92 		\\ [1ex]
				&	  4 & 0.8  			&  0.97 	& 0.82 			& 0.95  	\\ [1ex]
				&	  6 & 0.74 			& 0.897 	& 0.75 			& 0.907 	\\ [1ex] \hline \\ [-1.5ex]
				{5}				&	  2	& 0.55			& 0.83		& 0.56			& 0.849		\\[1ex]
				{9}				&	  6	& 0.2			& 0.826		& 0.2			& 0.823		\\[1ex]
			\end{tabular}
		\end{ruledtabular}
	\end{table}

	\subsection{\label{sec4:l4} 1+1DOF System: Tandem Independently Oscillating Cylinders }  
 	\begin{figure}[h!]
		{\includegraphics[width=\linewidth]{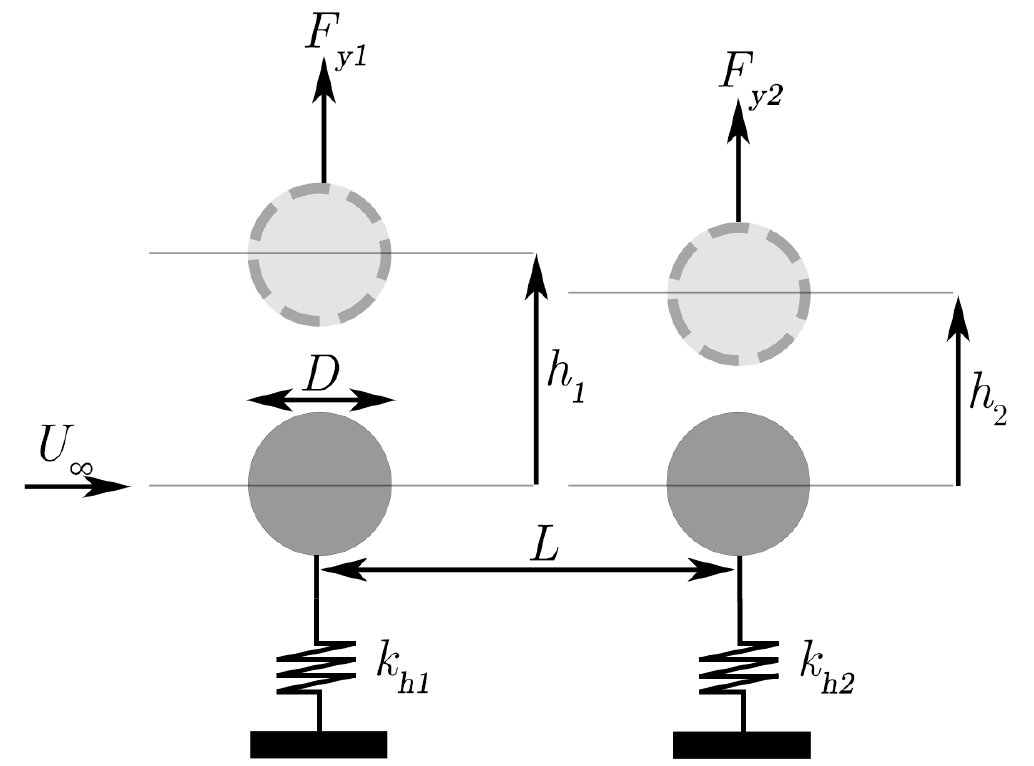}}\\
    	\caption{\label{fig:1p1dof2Sph} Schematic of 1+1DOF system with two independently oscillating cylinders in cross flow, attached to a linear spring.  }
	\end{figure}
	
    By mounting each cylinder on separate springs, we introduce another degree of freedom, transverse to the direction of flow of the fluid. This spring-mass system with two independently oscillating bodies is shown in Fig.~\ref{fig:1p1dof2Sph}. We use a second-order finite-element mesh with 34,392 elements and a total of 141,143 degrees of freedom. The body diameter was 200 lattice units with a near wall mesh size of 1 lattice unit. The time-step is set to $\delta t=1/3 $. For the implementation of the blended mesh, the radius of the rigid zone around each body, $ r_{rigid} = 50 $ lattice units, and the radius of the blended zone, $ r_{blend} = 400 $ lattice units.

	Once again, there is no change to the equation of motion Eq.~\eqref{eq:1dofSM_nonDim}, however, we now have to solve one differential equation for each body. 

	The cylinders are mounted in-line with a $L/D=5$, ${\rm Re}=1000$, and a free-stream Mach number of ${\rm Ma}_\infty = 0.1$ with $\gamma = 1$ 
	and $T_\infty = 1/3 $. We simulate the response of a wide range of reduced velocities, from 4 to 10, and compare our results to the work of Jester and Kallinderis \cite{Jester2004}. The mass ratio is set to
	$m^* = 4$. 
	\begin{figure*}[t]
		\includegraphics[width=0.75\linewidth]{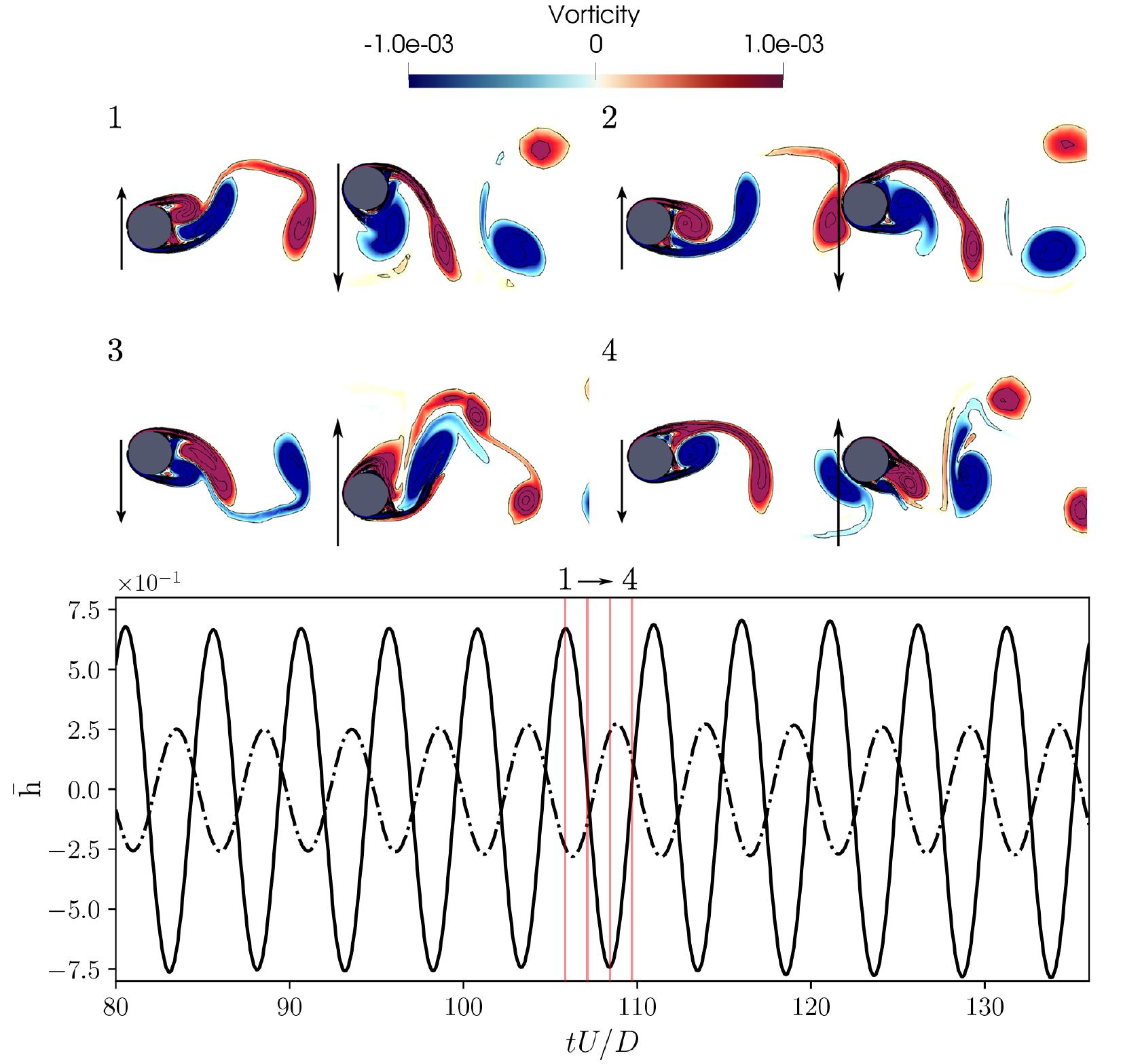}
		\caption{\label{fig:1p1dof2Sph+Vr6} 1+1DOF system with two independently oscillating cylinders: vorticity contours (top) and cylinder displacement (bottom).  }
	\end{figure*}
	\begin{figure*}[t]
		\includegraphics[width=0.9\linewidth]{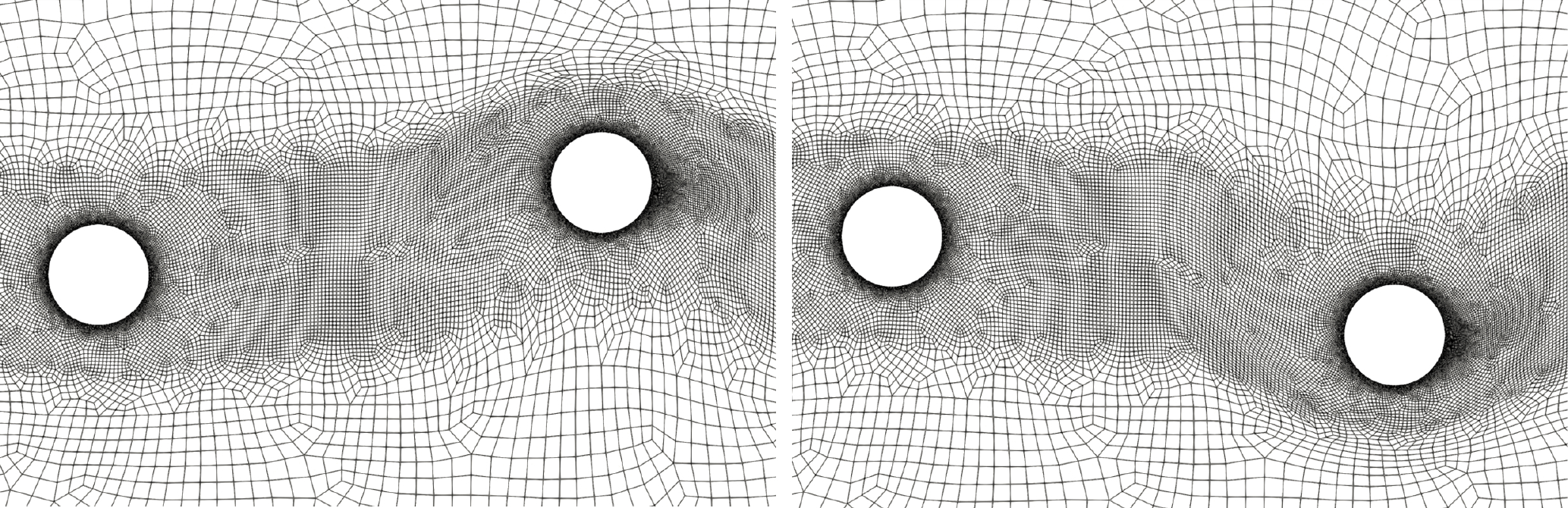}
		\caption{\label{fig:1p1dof2Sph+blend} 1+1DOF system with two independently oscillating cylinders: blended mesh at two time instances, corresponding to maximum displacements of downstream cylinder. }
	\end{figure*}
	
	Different from the previous case of linked cylinders, the forces experienced by each body act solely on it and we therefore see distinct responses for the upstream and the downstream cylinder. According to the classification by Zdravkovich, this is also a case of wake-interference. We expect to see the upstream cylinder behave like an isolated cylinder, with a maximum amplitude peak near the range of reduced velocities that synchronize with the vortex shedding frequency. Away from this lock-in region, the amplitude of oscillation quickly drops. 
	
	The downstream cylinder, in the wake of the upstream cylinder, experiences a much larger amplitudes, over a much larger range of reduced velocities. Thus the dynamic response of the downstream cylinder is very different from the response of an isolated cylinder, while the upstream cylinder behaves much in the same way. The results agree qualitatively with previous experimental findings of \cite{King1976}  as well quantitatively with numerical results obtained in \cite{Jester2004}. 

	A visualization of the wake interference phenomenon is shown in Fig.~\ref{fig:1p1dof2Sph+Vr6}, where we plot a snapshot of vorticity contours along with the displacement of both cylinders with time, for reduced velocity $ V_R=6 $. 
	It can be observed that there is a phase shift of $ \pi/2 $ between the upstream and the downstream cylinder.
	We can further see that the gap between the two cylinders is large enough for vortex shedding to occur and the vortices, generated from the upstream cylinder, impinge on the downstream cylinder.
	For instance, in subfigure 1 and 2 of Fig.~\ref{fig:1p1dof2Sph+Vr6}, the positive vortex (red) generated by the upward-traveling upstream cylinder, impinges on the downstream cylinder going downwards. Similarly, in picture 3 and 4 of Fig.~\ref{fig:1p1dof2Sph+Vr6}, the negative vortex (blue) from the upstream cylinder impinges on the downstream cylinder traveling upwards. This wake interference leads to higher transient loads on the downstream cylinder, which causes the downstream cylinder to have a higher amplitude oscillations, even at much higher values of reduced velocity. In Fig.~\ref{fig:1p1dof2Sph+blend} we visualise the grid, deformed using blended function approach, around the cylinders at two time instances which correspond to the maximum displacement of the downstream cylinder. Animations of the mesh deformation and the vorticity can be found in the supplementary material. 
	
	Fig.~\ref{fig:1p1dof2Sph+jester} shows the maximum amplitude of both cylinders plotted against the reduced velocity. The solid line represents the results of \cite{Jester2004}, which are in very close agreement with our results.
	\begin{figure}[h!]
 		{\includegraphics[width=\linewidth]{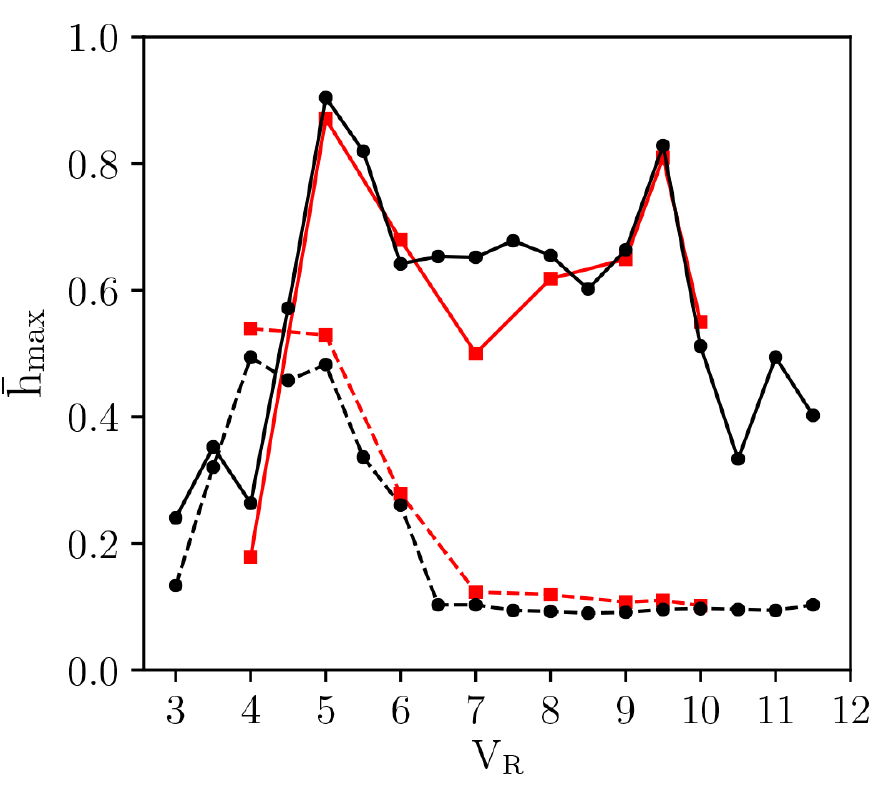}}		\caption{\label{fig:1p1dof2Sph+jester} 1+1DOF system with two independently oscillating cylinders: maximum normalized displacement for upstream (dashed) and downstream cylinder (solid) against reduced velocity. Present (black) against literature \cite{Jester2004} (red). }
	\end{figure}
	
 	\subsection{\label{sec4:l5} 2DOF System: NACA 64A010 Flutter Analysis  }   
	
	Flutter refers to an aeroelastic instability that occurs due to a coupling between aerodynamic, elastic, and inertial forces. Flutter vibrations are typically seen in slender bodies, especially for airfoils at transonic speeds due to the transient loads and shock interactions. If not controlled or actively damped, this can cause significant structural damage to the system. Below a certain threshold, the oscillations are damped out and pose no danger to the structure, but beyond, a divergent oscillating response known as flutter is observed.
    A neutrally stable response can also be observed,
    which it is known as a limit cycle oscillation (LCO).
    
	Here, we demonstrate that our model can capture these dynamic phenomena by simulating the transonic flow over a NACA64A010 airfoil with 2 DOFs and at free-stream Mach number ${\rm Ma}_\infty=0.85$ with $\gamma = 1.4$ and at $T_\infty=0.2$.
	As sketched in Fig.~\ref{fig:2dofAfoil}, the airfoil is mounted on a torsional and a linear spring, which allows a 2DOF motion of the airfoil.
 	\begin{figure*}[t]	
		\subfigure[$ {\rm Re}=1\times10^4 $]{\includegraphics[width=0.33\linewidth]{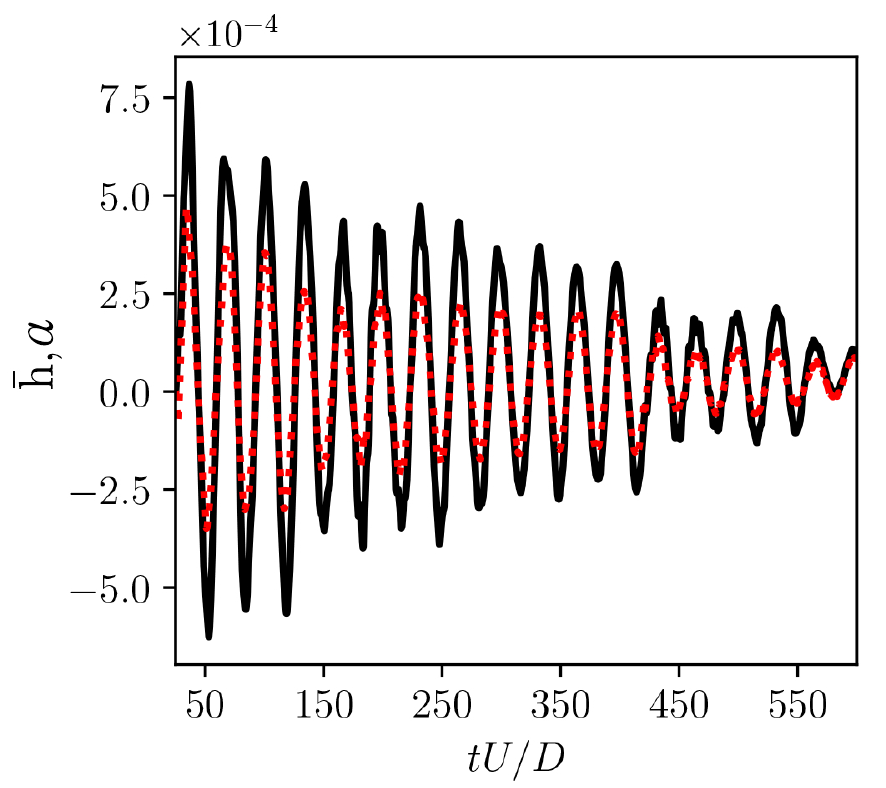}} 
		\subfigure[$ {\rm Re}=1.25\times10^4 $]{\includegraphics[width=0.33\linewidth]{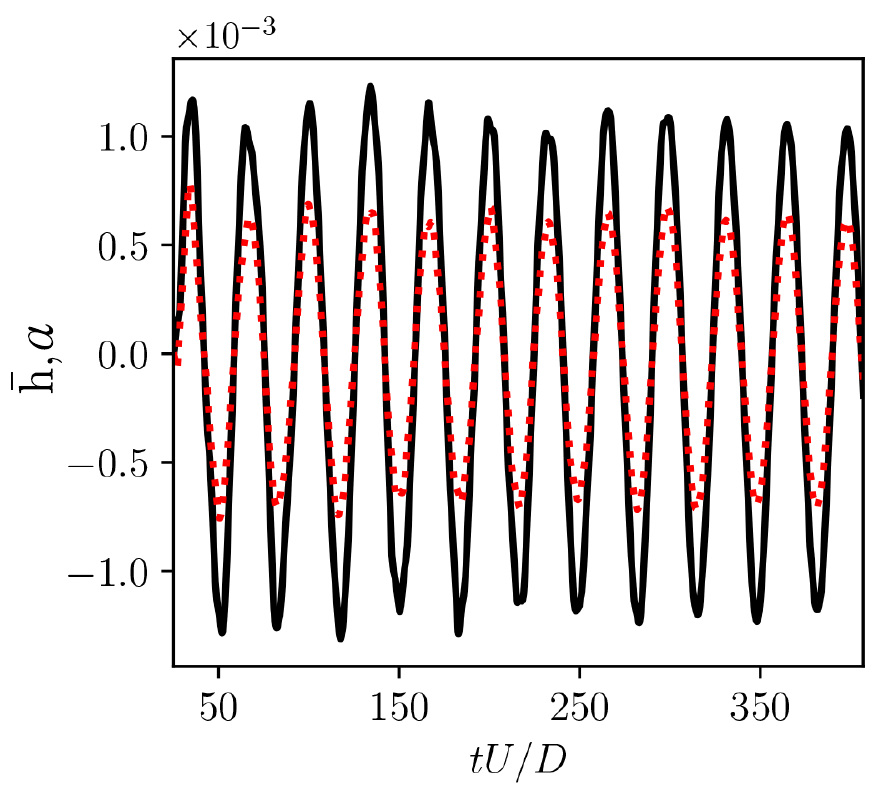}} 
		\subfigure[$ {\rm Re}=1.5\times10^4 $]{\includegraphics[width=0.33\linewidth]{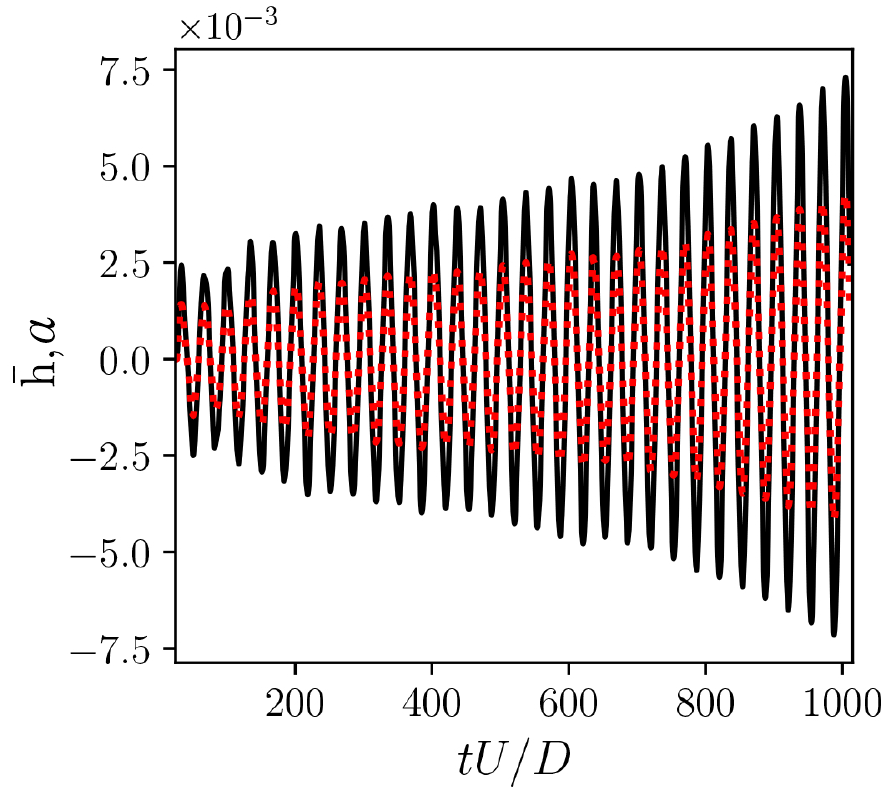}} 
		\caption{\label{fig:2dof_combined} 2DOF system exhibiting different oscillation regimes for transonic flow over NACA64A010 airfoil: damped, limit cycle, and flutter. Normalized plunge displacement (black) and angular displacement (red) over time. 
		}
	\end{figure*}
	\begin{figure*}[t]
		\includegraphics[width=\linewidth]{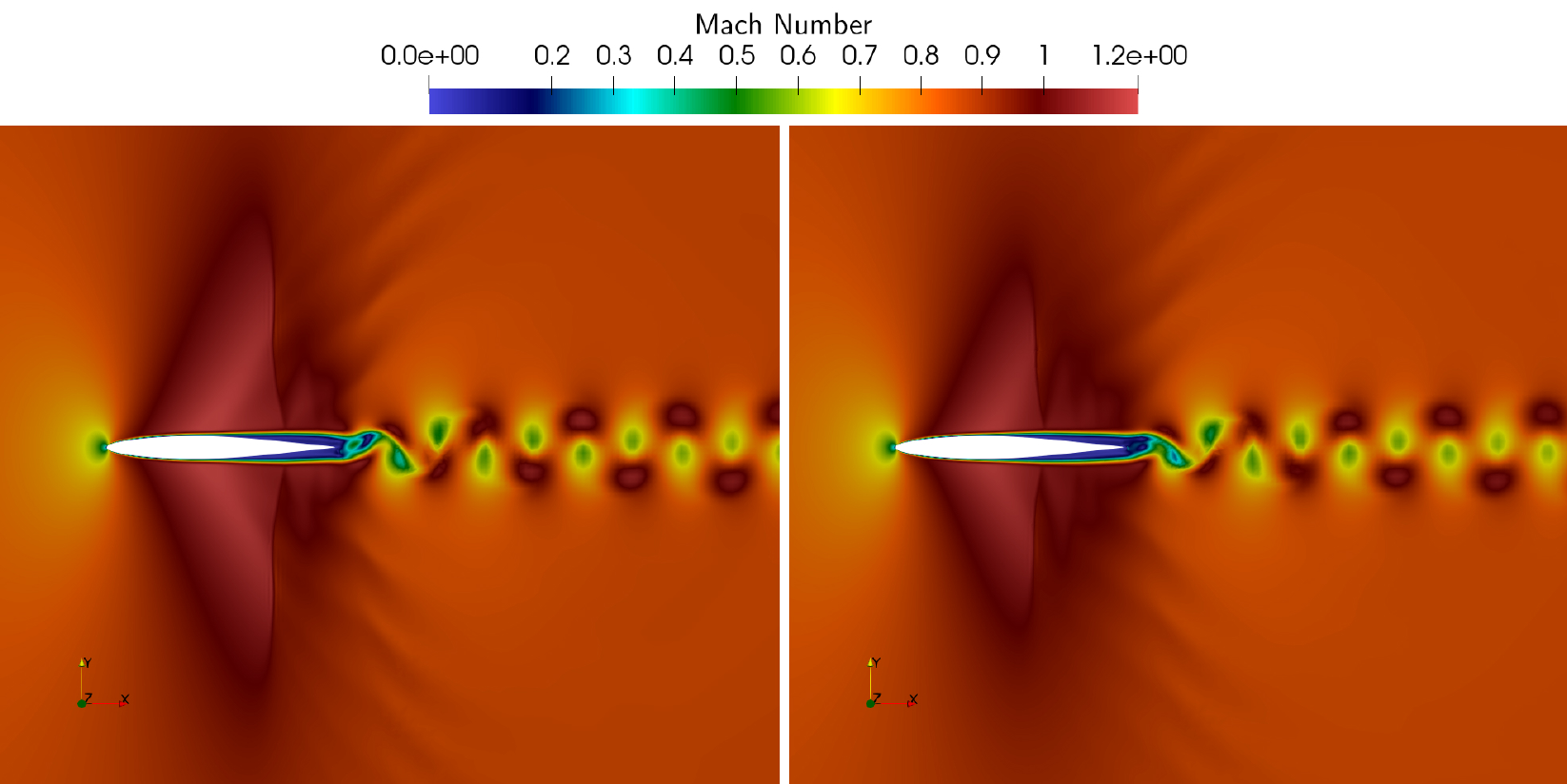}
		\caption{\label{fig:2dof+12p5K} 2DOF system: contours of Mach number for $ {\rm Re}=1.25\times10^4 $ at two time instances showing the movement of the shock over the NACA64A010 airfoil. }
	\end{figure*}

	In order to obtain the non-dimensionalized form of the 2DOF system (Eq.~\eqref{eq:2dofSM}), we introduce the non-dimensionalised plunge, $  {\bar{h}}= h/b $, and  non-dimensionalised time, $ \tau = t\omega_\alpha $, where $\omega_\alpha$ is the pitch frequency. The angular displacement, $\alpha$ is measured in radians. The non-dimensionalised system, can then be written in the form: 
	\begin{align} 				
	\bm{M}\ddot{\bm{y}}+ \bm{K}\bm{y} = \bm{L},
 	\end{align}
	where
	\begin{align} 
	\bm{y} = \begin{bmatrix} {\rm\bar{h}} \\ \alpha \\ \end{bmatrix} , \qquad  \bm{L} = \dfrac{4{{\rm Ma}_{\infty}}^2 \gamma}{\pi m^*} \begin{bmatrix} C_L \\ 2C_M \\ \end{bmatrix}
	\end{align}
	and
	\begin{equation}
 	\begin{aligned} 
        \bm{M} = \begin{bmatrix} 1 & x_\alpha \\ x_\alpha & {r_\alpha}^2\\ \end{bmatrix}, \qquad
		\bm{K} = \dfrac{4{{\rm Ma}_{\infty}}^2 \gamma}{{V_R}^2 m^* }  \begin{bmatrix} {\frac{\omega_h }{\omega_\alpha}}^2 &  0 \\ 0 & {r_\alpha}^2\\ \end{bmatrix}.
		\end{aligned}
	\end{equation}
	Here, we introduce the mass ratio $ m^* $, the reduced velocity $ V_R $ and the uncoupled structural plunge frequency $\omega_h $ respectively. We define the center of mass as the point where the mass of the body can be assumed to be concentrated. The radius of gyration is the distance from the mass center, where the moment of inertia of the concentrated mass equals the moment of inertia of the actual mass distribution of the body. We further denote $ x_\alpha $ as the non-dimensional distance of the center of mass from the elastic axis and $r_\alpha $ is the radius of gyration, which is also measured from the elastic axis. The lift coefficient is defined as $C_L = 2 F_y/ ( \rho U_\infty^2 2b )$ and pitching moment coefficient is given by $C_M = 2 M_{EA}/ ( \rho U_\infty^2 (2b)^2)$.
	
	These non-dimensional quantities and system parameters are defined as:
	
	\begin{align}   m^* = \dfrac{m}{\pi \rho b^2}, \qquad V_R = \dfrac{U_\infty}{b \omega_\alpha \sqrt{\ m^*}},
	\end{align}
	\begin{align} \omega_\alpha = \sqrt{\dfrac{K_\alpha}{I_\alpha}}, \qquad \omega_h =\sqrt{\dfrac{K_h}{m}},
	\end{align}
	\begin{align}  x_\alpha = {\dfrac{S_\alpha}{mb}}, \qquad r_\alpha =\sqrt{\dfrac{I_\alpha}{mb^2}}.
	\end{align}
	
	These parameters are set as follows: $ x_\alpha = 1.8  $ , $ r_\alpha = 1.87 $, $\omega_{h}/\omega_{\alpha} = 1$ , $ m^* = 60 $, and $V_R = 0.5025 $. The elastic axis of the airfoil is located half a chord length in front of the leading edge of the airfoil.
	For our simulations, we use a second-order finite-element mesh with 44,880 elements and 180,192 degrees of freedom, resulting in a near-wall resolution of 0.25 lattice units and 200 lattice units for the airfoil chord.
	The time-step was set to $  \delta t=0.25/3 $. %
	
	Initially, a forced sinusoidal pitching perturbation with $1 \degree $ amplitude was applied for 2 full periods. Subsequently, the airfoil was allowed to oscillate freely under aerodynamic loads.
	
	The simulations are performed at $ {\rm Re} = 1\times10^4 , 1.25\times10^4, \text{ and } 1.5\times10^4 $, where the Reynolds number based on the chord is defined as ${\rm Re} = \rho U_\infty 2b/\mu$. We keep the Reynolds number moderate since the solver is two-dimensional and would produce nonphysical results for larger Reynolds numbers, where the flow would indeed be three-dimensional.
	Nevertheless, even at lower Reynolds numbers, the model can capture the complex dynamic coupling which leads to flutter.
	Fig.~\ref{fig:2dof_combined} shows the normalized transverse displacement and the angular displacement of the airfoil. From left to right, we can see that the airfoil exhibits {damped, limit-cycle, and divergent oscillations, leading to flutter }. The oscillations are damped at ${\rm Re}=1\times10^4$, exhibit LCO at ${\rm Re}=1.25\times10^4$, and flutter starts at ${\rm Re}=1.5\times10^4$. This dependence of flutter oscillations on the Reynolds number has been observed previously, both experimentally and numerically \cite{Nayer2020} \cite{Wood2020}.  
	
	It is important to note that the oscillations are of low amplitude, which is a result of keeping the  Reynolds numbers low in order to maintain two-dimensional flow and warrant the use of a two-dimensional solver.
	This was also reported in the literature, see, e.g., \cite{Nayer2020}, where the authors observed an increase of the oscillation amplitude of two orders of magnitude 
	between $ {\rm Re}=1.6\times10^4$ and ${\rm Re}=2.39\times10^4$. \\
	The contours of Mach number around the airfoil are shown in Fig.~\ref{fig:2dof+12p5K}, where we can observe the expansion of the subsonic flow over the airfoil leading to a locally supersonic region. The two snapshots show the movement of the location of the resulting shock, as it oscillates with time.
	\section{\label{sec:Concl} Conclusions }
	
	In this paper, we have proposed a two-way coupled fluid-structure interaction scheme for compressible flows using the semi-Lagrangian lattice Boltzmann method on unstructured meshes.
	Using an ALE formulation together with blending functions, the dynamics of multiple moving bodies can be described accurately. 
	Thorough validation for vortex induced vibrations of single and tandem cylinders have shown that complex non-linear dynamics can be captured robustly and accurately, in excellent agreement with the literature. 
	Finally, we presented a simulation of transonic flow over an airfoil that can freely pitch and plunge under aerodynamic loads.
	The non-linear coupling between the aerodynamic and structural forces leads to flutter and LCO regimes, which are captured by our model and agrees with what is reported in the literature. 
	While we focused on two-dimensional flows in this paper, the extension to three dimensions is conceptually straightforward and will be focus of future works.

	\begin{acknowledgments}
 This work was supported by European Research Council (ERC) Advanced Grant No. 834763-PonD. Computational resources at the Swiss National SuperComputing Center CSCS were provided under grant No. s1066.	
	\end{acknowledgments}
 
	\nocite{*}
	\bibliography{ref}

\begin{thebibliography}{73}%
\makeatletter
\providecommand \@ifxundefined [1]{%
 \@ifx{#1\undefined}
}%
\providecommand \@ifnum [1]{%
 \ifnum #1\expandafter \@firstoftwo
 \else \expandafter \@secondoftwo
 \fi
}%
\providecommand \@ifx [1]{%
 \ifx #1\expandafter \@firstoftwo
 \else \expandafter \@secondoftwo
 \fi
}%
\providecommand \natexlab [1]{#1}%
\providecommand \enquote  [1]{``#1''}%
\providecommand \bibnamefont  [1]{#1}%
\providecommand \bibfnamefont [1]{#1}%
\providecommand \citenamefont [1]{#1}%
\providecommand \href@noop [0]{\@secondoftwo}%
\providecommand \href [0]{\begingroup \@sanitize@url \@href}%
\providecommand \@href[1]{\@@startlink{#1}\@@href}%
\providecommand \@@href[1]{\endgroup#1\@@endlink}%
\providecommand \@sanitize@url [0]{\catcode `\\12\catcode `\$12\catcode
  `\&12\catcode `\#12\catcode `\^12\catcode `\_12\catcode `\%12\relax}%
\providecommand \@@startlink[1]{}%
\providecommand \@@endlink[0]{}%
\providecommand \url  [0]{\begingroup\@sanitize@url \@url }%
\providecommand \@url [1]{\endgroup\@href {#1}{\urlprefix }}%
\providecommand \urlprefix  [0]{URL }%
\providecommand \Eprint [0]{\href }%
\providecommand \doibase [0]{https://doi.org/}%
\providecommand \selectlanguage [0]{\@gobble}%
\providecommand \bibinfo  [0]{\@secondoftwo}%
\providecommand \bibfield  [0]{\@secondoftwo}%
\providecommand \translation [1]{[#1]}%
\providecommand \BibitemOpen [0]{}%
\providecommand \bibitemStop [0]{}%
\providecommand \bibitemNoStop [0]{.\EOS\space}%
\providecommand \EOS [0]{\spacefactor3000\relax}%
\providecommand \BibitemShut  [1]{\csname bibitem#1\endcsname}%
\let\auto@bib@innerbib\@empty
\bibitem [{\citenamefont {Anand}\ and\ \citenamefont
  {Christov}(2020)}]{Anand_2020}%
  \BibitemOpen
  \bibfield  {author} {\bibinfo {author} {\bibfnamefont {V.}~\bibnamefont
  {Anand}}\ and\ \bibinfo {author} {\bibfnamefont {I.~C.}\ \bibnamefont
  {Christov}},\ }\bibfield  {title} {\bibinfo {title} {Transient compressible
  flow in a compliant viscoelastic tube},\ }\href
  {https://doi.org/10.1063/5.0022406} {\bibfield  {journal} {\bibinfo
  {journal} {Physics of Fluids}\ }\textbf {\bibinfo {volume} {32}},\ \bibinfo
  {pages} {112014} (\bibinfo {year} {2020})}\BibitemShut {NoStop}%
\bibitem [{\citenamefont {Hsiao}\ and\ \citenamefont
  {Chahine}(2015)}]{Hsiao_2015}%
  \BibitemOpen
  \bibfield  {author} {\bibinfo {author} {\bibfnamefont {C.-T.}\ \bibnamefont
  {Hsiao}}\ and\ \bibinfo {author} {\bibfnamefont {G.~L.}\ \bibnamefont
  {Chahine}},\ }\bibfield  {title} {\bibinfo {title} {Dynamic response of a
  composite propeller blade subjected to shock and bubble pressure loading},\
  }\href {https://doi.org/10.1016/j.jfluidstructs.2015.01.012} {\bibfield
  {journal} {\bibinfo  {journal} {Journal of Fluids and Structures}\ }\textbf
  {\bibinfo {volume} {54}},\ \bibinfo {pages} {760} (\bibinfo {year}
  {2015})}\BibitemShut {NoStop}%
\bibitem [{\citenamefont {Chahine}\ and\ \citenamefont
  {Hsiao}(2015)}]{Chahine_2015}%
  \BibitemOpen
  \bibfield  {author} {\bibinfo {author} {\bibfnamefont {G.~L.}\ \bibnamefont
  {Chahine}}\ and\ \bibinfo {author} {\bibfnamefont {C.-T.}\ \bibnamefont
  {Hsiao}},\ }\bibfield  {title} {\bibinfo {title} {Modelling cavitation
  erosion using fluid{\textendash}material interaction simulations},\ }\href
  {https://doi.org/10.1098/rsfs.2015.0016} {\bibfield  {journal} {\bibinfo
  {journal} {Interface Focus}\ }\textbf {\bibinfo {volume} {5}},\ \bibinfo
  {pages} {20150016} (\bibinfo {year} {2015})}\BibitemShut {NoStop}%
\bibitem [{\citenamefont {Subramaniam}\ \emph {et~al.}(2009)\citenamefont
  {Subramaniam}, \citenamefont {Nian},\ and\ \citenamefont
  {Andreopoulos}}]{Subramaniam_2009}%
  \BibitemOpen
  \bibfield  {author} {\bibinfo {author} {\bibfnamefont {K.~V.}\ \bibnamefont
  {Subramaniam}}, \bibinfo {author} {\bibfnamefont {W.}~\bibnamefont {Nian}},\
  and\ \bibinfo {author} {\bibfnamefont {Y.}~\bibnamefont {Andreopoulos}},\
  }\bibfield  {title} {\bibinfo {title} {Blast response simulation of an
  elastic structure: Evaluation of the fluid{\textendash}structure interaction
  effect},\ }\href {https://doi.org/10.1016/j.ijimpeng.2009.01.001} {\bibfield
  {journal} {\bibinfo  {journal} {International Journal of Impact Engineering}\
  }\textbf {\bibinfo {volume} {36}},\ \bibinfo {pages} {965} (\bibinfo {year}
  {2009})}\BibitemShut {NoStop}%
\bibitem [{\citenamefont {Aune}\ \emph {et~al.}(2021)\citenamefont {Aune},
  \citenamefont {Valsamos}, \citenamefont {Casadei}, \citenamefont {Langseth},\
  and\ \citenamefont {B{\o}rvik}}]{Aune_2021}%
  \BibitemOpen
  \bibfield  {author} {\bibinfo {author} {\bibfnamefont {V.}~\bibnamefont
  {Aune}}, \bibinfo {author} {\bibfnamefont {G.}~\bibnamefont {Valsamos}},
  \bibinfo {author} {\bibfnamefont {F.}~\bibnamefont {Casadei}}, \bibinfo
  {author} {\bibfnamefont {M.}~\bibnamefont {Langseth}},\ and\ \bibinfo
  {author} {\bibfnamefont {T.}~\bibnamefont {B{\o}rvik}},\ }\bibfield  {title}
  {\bibinfo {title} {Fluid-structure interaction effects during the dynamic
  response of clamped thin steel plates exposed to blast loading},\ }\href
  {https://doi.org/10.1016/j.ijmecsci.2020.106263} {\bibfield  {journal}
  {\bibinfo  {journal} {International Journal of Mechanical Sciences}\ }\textbf
  {\bibinfo {volume} {195}},\ \bibinfo {pages} {106263} (\bibinfo {year}
  {2021})}\BibitemShut {NoStop}%
\bibitem [{\citenamefont {Thomas}\ \emph {et~al.}(2002)\citenamefont {Thomas},
  \citenamefont {Dowell},\ and\ \citenamefont {Hall}}]{Thomas_2002}%
  \BibitemOpen
  \bibfield  {author} {\bibinfo {author} {\bibfnamefont {J.~P.}\ \bibnamefont
  {Thomas}}, \bibinfo {author} {\bibfnamefont {E.~H.}\ \bibnamefont {Dowell}},\
  and\ \bibinfo {author} {\bibfnamefont {K.~C.}\ \bibnamefont {Hall}},\
  }\bibfield  {title} {\bibinfo {title} {Nonlinear inviscid aerodynamic effects
  on transonic divergence, flutter, and limit-cycle oscillations},\ }\href
  {https://doi.org/10.2514/2.1720} {\bibfield  {journal} {\bibinfo  {journal}
  {{AIAA} Journal}\ }\textbf {\bibinfo {volume} {40}},\ \bibinfo {pages} {638}
  (\bibinfo {year} {2002})}\BibitemShut {NoStop}%
\bibitem [{\citenamefont {Liu}\ \emph {et~al.}(2001)\citenamefont {Liu},
  \citenamefont {Cai}, \citenamefont {Zhu}, \citenamefont {Tsai},\ and\
  \citenamefont {Wong}}]{Liu_2001}%
  \BibitemOpen
  \bibfield  {author} {\bibinfo {author} {\bibfnamefont {F.}~\bibnamefont
  {Liu}}, \bibinfo {author} {\bibfnamefont {J.}~\bibnamefont {Cai}}, \bibinfo
  {author} {\bibfnamefont {Y.}~\bibnamefont {Zhu}}, \bibinfo {author}
  {\bibfnamefont {H.~M.}\ \bibnamefont {Tsai}},\ and\ \bibinfo {author}
  {\bibfnamefont {A.~S.~F.}\ \bibnamefont {Wong}},\ }\bibfield  {title}
  {\bibinfo {title} {Calculation of wing flutter by a coupled fluid-structure
  method},\ }\href {https://doi.org/10.2514/2.2766} {\bibfield  {journal}
  {\bibinfo  {journal} {Journal of Aircraft}\ }\textbf {\bibinfo {volume}
  {38}},\ \bibinfo {pages} {334} (\bibinfo {year} {2001})}\BibitemShut
  {NoStop}%
\bibitem [{\citenamefont {Yuan}\ \emph {et~al.}(2021)\citenamefont {Yuan},
  \citenamefont {Sandhu},\ and\ \citenamefont {Poirel}}]{Yuan2021}%
  \BibitemOpen
  \bibfield  {author} {\bibinfo {author} {\bibfnamefont {W.}~\bibnamefont
  {Yuan}}, \bibinfo {author} {\bibfnamefont {R.}~\bibnamefont {Sandhu}},\ and\
  \bibinfo {author} {\bibfnamefont {D.}~\bibnamefont {Poirel}},\ }\bibfield
  {title} {\bibinfo {title} {Fully coupled aeroelastic analyses of wing flutter
  towards application to complex aircraft configurations},\ }\href
  {https://doi.org/10.1061/(asce)as.1943-5525.0001232} {\bibfield  {journal}
  {\bibinfo  {journal} {Journal of Aerospace Engineering}\ }\textbf {\bibinfo
  {volume} {34}},\ \bibinfo {pages} {04020117} (\bibinfo {year}
  {2021})}\BibitemShut {NoStop}%
\bibitem [{\citenamefont {Frapolli}\ \emph {et~al.}(2018)\citenamefont
  {Frapolli}, \citenamefont {Chikatamarla},\ and\ \citenamefont
  {Karlin}}]{FRAPOLLI20182}%
  \BibitemOpen
  \bibfield  {author} {\bibinfo {author} {\bibfnamefont {N.}~\bibnamefont
  {Frapolli}}, \bibinfo {author} {\bibfnamefont {S.}~\bibnamefont
  {Chikatamarla}},\ and\ \bibinfo {author} {\bibfnamefont {I.}~\bibnamefont
  {Karlin}},\ }\bibfield  {title} {\bibinfo {title} {Entropic lattice boltzmann
  simulation of thermal convective turbulence},\ }\href
  {https://doi.org/https://doi.org/10.1016/j.compfluid.2018.08.021} {\bibfield
  {journal} {\bibinfo  {journal} {Computers \& Fluids}\ }\textbf {\bibinfo
  {volume} {175}},\ \bibinfo {pages} {2 } (\bibinfo {year} {2018})}\BibitemShut
  {NoStop}%
\bibitem [{\citenamefont {Mazloomi}\ \emph {et~al.}(2015)\citenamefont
  {Mazloomi}, \citenamefont {Chikatamarla},\ and\ \citenamefont
  {Karlin}}]{A.Mazloomi2015}%
  \BibitemOpen
  \bibfield  {author} {\bibinfo {author} {\bibfnamefont {A.}~\bibnamefont
  {Mazloomi}}, \bibinfo {author} {\bibfnamefont {S.}~\bibnamefont
  {Chikatamarla}},\ and\ \bibinfo {author} {\bibfnamefont {I.}~\bibnamefont
  {Karlin}},\ }\bibfield  {title} {\bibinfo {title} {Entropic lattice boltzmann
  method for multiphase flows},\ }\href@noop {} {\bibfield  {journal} {\bibinfo
   {journal} {Physical Review Letters}\ }\textbf {\bibinfo {volume} {114}}
  (\bibinfo {year} {2015})}\BibitemShut {NoStop}%
\bibitem [{\citenamefont {Sawant}\ \emph {et~al.}(2020)\citenamefont {Sawant},
  \citenamefont {Dorschner},\ and\ \citenamefont {Karlin}}]{Sawant2020}%
  \BibitemOpen
  \bibfield  {author} {\bibinfo {author} {\bibfnamefont {N.}~\bibnamefont
  {Sawant}}, \bibinfo {author} {\bibfnamefont {B.}~\bibnamefont {Dorschner}},\
  and\ \bibinfo {author} {\bibfnamefont {I.~V.}\ \bibnamefont {Karlin}},\
  }\bibfield  {title} {\bibinfo {title} {Consistent lattice boltzmann model for
  multicomponent mixtures},\ }\href@noop {} {\bibfield  {journal} {\bibinfo
  {journal} {Journal of Fluid Mechanics}\ }\textbf {\bibinfo {volume} {909}}
  (\bibinfo {year} {2020})}\BibitemShut {NoStop}%
\bibitem [{\citenamefont {Staso}\ \emph {et~al.}(2016)\citenamefont {Staso},
  \citenamefont {Clercx}, \citenamefont {Succi},\ and\ \citenamefont
  {Toschi}}]{Staso2016}%
  \BibitemOpen
  \bibfield  {author} {\bibinfo {author} {\bibfnamefont {G.~D.}\ \bibnamefont
  {Staso}}, \bibinfo {author} {\bibfnamefont {H.}~\bibnamefont {Clercx}},
  \bibinfo {author} {\bibfnamefont {S.}~\bibnamefont {Succi}},\ and\ \bibinfo
  {author} {\bibfnamefont {F.}~\bibnamefont {Toschi}},\ }\bibfield  {title}
  {\bibinfo {title} {{DSMC}{\textendash}{LBM} mapping scheme for rarefied and
  non-rarefied gas flows},\ }\href {https://doi.org/10.1016/j.jocs.2016.04.011}
  {\bibfield  {journal} {\bibinfo  {journal} {Journal of Computational
  Science}\ }\textbf {\bibinfo {volume} {17}},\ \bibinfo {pages} {357}
  (\bibinfo {year} {2016})}\BibitemShut {NoStop}%
\bibitem [{\citenamefont {Mendoza}\ \emph {et~al.}(2010)\citenamefont
  {Mendoza}, \citenamefont {Boghosian}, \citenamefont {Herrmann},\ and\
  \citenamefont {Succi}}]{Mendoza2010}%
  \BibitemOpen
  \bibfield  {author} {\bibinfo {author} {\bibfnamefont {M.}~\bibnamefont
  {Mendoza}}, \bibinfo {author} {\bibfnamefont {B.~M.}\ \bibnamefont
  {Boghosian}}, \bibinfo {author} {\bibfnamefont {H.~J.}\ \bibnamefont
  {Herrmann}},\ and\ \bibinfo {author} {\bibfnamefont {S.}~\bibnamefont
  {Succi}},\ }\bibfield  {title} {\bibinfo {title} {Derivation of the lattice
  boltzmann model for relativistic hydrodynamics},\ }\href@noop {} {\bibfield
  {journal} {\bibinfo  {journal} {Physical Review D}\ }\textbf {\bibinfo
  {volume} {82}} (\bibinfo {year} {2010})}\BibitemShut {NoStop}%
\bibitem [{\citenamefont {Chikatamarla}\ and\ \citenamefont
  {Karlin}(2006)}]{Chikatamarla2006}%
  \BibitemOpen
  \bibfield  {author} {\bibinfo {author} {\bibfnamefont {S.~S.}\ \bibnamefont
  {Chikatamarla}}\ and\ \bibinfo {author} {\bibfnamefont {I.~V.}\ \bibnamefont
  {Karlin}},\ }\bibfield  {title} {\bibinfo {title} {Entropy and galilean
  invariance of lattice boltzmann theories},\ }\href@noop {} {\bibfield
  {journal} {\bibinfo  {journal} {Physical Review Letters}\ }\textbf {\bibinfo
  {volume} {97}} (\bibinfo {year} {2006})}\BibitemShut {NoStop}%
\bibitem [{\citenamefont {Shan}\ \emph {et~al.}(2006)\citenamefont {Shan},
  \citenamefont {Yuan},\ and\ \citenamefont {Chen}}]{XIAOWEN2006}%
  \BibitemOpen
  \bibfield  {author} {\bibinfo {author} {\bibfnamefont {X.}~\bibnamefont
  {Shan}}, \bibinfo {author} {\bibfnamefont {X.-F.}\ \bibnamefont {Yuan}},\
  and\ \bibinfo {author} {\bibfnamefont {H.}~\bibnamefont {Chen}},\ }\bibfield
  {title} {\bibinfo {title} {Kinetic theory representation of hydrodynamics: a
  way beyond the navier{\textendash}stokes equation},\ }\href
  {https://doi.org/10.1017/s0022112005008153} {\bibfield  {journal} {\bibinfo
  {journal} {Journal of Fluid Mechanics}\ }\textbf {\bibinfo {volume} {550}},\
  \bibinfo {pages} {413} (\bibinfo {year} {2006})}\BibitemShut {NoStop}%
\bibitem [{\citenamefont {Chikatamarla}\ and\ \citenamefont
  {Karlin}(2009)}]{Chikatamarla2009}%
  \BibitemOpen
  \bibfield  {author} {\bibinfo {author} {\bibfnamefont {S.~S.}\ \bibnamefont
  {Chikatamarla}}\ and\ \bibinfo {author} {\bibfnamefont {I.~V.}\ \bibnamefont
  {Karlin}},\ }\bibfield  {title} {\bibinfo {title} {Lattices for the lattice
  boltzmann method},\ }\href@noop {} {\bibfield  {journal} {\bibinfo  {journal}
  {Physical Review E}\ }\textbf {\bibinfo {volume} {79}} (\bibinfo {year}
  {2009})}\BibitemShut {NoStop}%
\bibitem [{\citenamefont {Siebert}\ \emph {et~al.}(2008)\citenamefont
  {Siebert}, \citenamefont {Hegele},\ and\ \citenamefont
  {Philippi}}]{Siebert2008}%
  \BibitemOpen
  \bibfield  {author} {\bibinfo {author} {\bibfnamefont {D.~N.}\ \bibnamefont
  {Siebert}}, \bibinfo {author} {\bibfnamefont {L.~A.}\ \bibnamefont
  {Hegele}},\ and\ \bibinfo {author} {\bibfnamefont {P.~C.}\ \bibnamefont
  {Philippi}},\ }\bibfield  {title} {\bibinfo {title} {Lattice boltzmann
  equation linear stability analysis: Thermal and athermal models},\
  }\href@noop {} {\bibfield  {journal} {\bibinfo  {journal} {Physical Review
  E}\ }\textbf {\bibinfo {volume} {77}} (\bibinfo {year} {2008})}\BibitemShut
  {NoStop}%
\bibitem [{\citenamefont {Chikatamarla}\ \emph {et~al.}(2010)\citenamefont
  {Chikatamarla}, \citenamefont {Frouzakis}, \citenamefont {Karlin},
  \citenamefont {Tomboulides},\ and\ \citenamefont
  {Boulouchos}}]{CHIKATAMARLA2010}%
  \BibitemOpen
  \bibfield  {author} {\bibinfo {author} {\bibfnamefont {S.~S.}\ \bibnamefont
  {Chikatamarla}}, \bibinfo {author} {\bibfnamefont {C.~E.}\ \bibnamefont
  {Frouzakis}}, \bibinfo {author} {\bibfnamefont {I.~V.}\ \bibnamefont
  {Karlin}}, \bibinfo {author} {\bibfnamefont {A.~G.}\ \bibnamefont
  {Tomboulides}},\ and\ \bibinfo {author} {\bibfnamefont {K.~B.}\ \bibnamefont
  {Boulouchos}},\ }\bibfield  {title} {\bibinfo {title} {Lattice boltzmann
  method for direct numerical simulation of turbulent flows},\ }\href
  {https://doi.org/10.1017/s0022112010002740} {\bibfield  {journal} {\bibinfo
  {journal} {Journal of Fluid Mechanics}\ }\textbf {\bibinfo {volume} {656}},\
  \bibinfo {pages} {298} (\bibinfo {year} {2010})}\BibitemShut {NoStop}%
\bibitem [{\citenamefont {Prasianakis}\ and\ \citenamefont
  {Karlin}(2007)}]{Parsianakis2007}%
  \BibitemOpen
  \bibfield  {author} {\bibinfo {author} {\bibfnamefont {N.~I.}\ \bibnamefont
  {Prasianakis}}\ and\ \bibinfo {author} {\bibfnamefont {I.~V.}\ \bibnamefont
  {Karlin}},\ }\bibfield  {title} {\bibinfo {title} {Lattice boltzmann method
  for thermal flow simulation on standard lattices},\ }\href@noop {} {\bibfield
   {journal} {\bibinfo  {journal} {Physical Review E}\ }\textbf {\bibinfo
  {volume} {76}} (\bibinfo {year} {2007})}\BibitemShut {NoStop}%
\bibitem [{\citenamefont {Prasianakis}\ and\ \citenamefont
  {Karlin}(2008)}]{Prasianakis2008}%
  \BibitemOpen
  \bibfield  {author} {\bibinfo {author} {\bibfnamefont {N.~I.}\ \bibnamefont
  {Prasianakis}}\ and\ \bibinfo {author} {\bibfnamefont {I.~V.}\ \bibnamefont
  {Karlin}},\ }\bibfield  {title} {\bibinfo {title} {Lattice boltzmann method
  for simulation of compressible flows on standard lattices},\ }\href@noop {}
  {\bibfield  {journal} {\bibinfo  {journal} {Phys. Rev. E}\ }\textbf {\bibinfo
  {volume} {78}},\ \bibinfo {pages} {016704} (\bibinfo {year}
  {2008})}\BibitemShut {NoStop}%
\bibitem [{\citenamefont {Feng}\ \emph {et~al.}(2015)\citenamefont {Feng},
  \citenamefont {Sagaut},\ and\ \citenamefont {Tao}}]{Feng_2015}%
  \BibitemOpen
  \bibfield  {author} {\bibinfo {author} {\bibfnamefont {Y.}~\bibnamefont
  {Feng}}, \bibinfo {author} {\bibfnamefont {P.}~\bibnamefont {Sagaut}},\ and\
  \bibinfo {author} {\bibfnamefont {W.}~\bibnamefont {Tao}},\ }\bibfield
  {title} {\bibinfo {title} {A three dimensional lattice model for thermal
  compressible flow on standard lattices},\ }\href
  {https://doi.org/10.1016/j.jcp.2015.09.011} {\bibfield  {journal} {\bibinfo
  {journal} {Journal of Computational Physics}\ }\textbf {\bibinfo {volume}
  {303}},\ \bibinfo {pages} {514} (\bibinfo {year} {2015})}\BibitemShut
  {NoStop}%
\bibitem [{\citenamefont {Huang}\ \emph {et~al.}(2019)\citenamefont {Huang},
  \citenamefont {Wu},\ and\ \citenamefont {Adams}}]{Huang_2019}%
  \BibitemOpen
  \bibfield  {author} {\bibinfo {author} {\bibfnamefont {R.}~\bibnamefont
  {Huang}}, \bibinfo {author} {\bibfnamefont {H.}~\bibnamefont {Wu}},\ and\
  \bibinfo {author} {\bibfnamefont {N.~A.}\ \bibnamefont {Adams}},\ }\bibfield
  {title} {\bibinfo {title} {Lattice boltzmann model with adjustable equation
  of state for coupled thermo-hydrodynamic flows},\ }\href
  {https://doi.org/10.1016/j.jcp.2019.04.044} {\bibfield  {journal} {\bibinfo
  {journal} {Journal of Computational Physics}\ }\textbf {\bibinfo {volume}
  {392}},\ \bibinfo {pages} {227} (\bibinfo {year} {2019})}\BibitemShut
  {NoStop}%
\bibitem [{\citenamefont {Hosseini}\ \emph {et~al.}(2020)\citenamefont
  {Hosseini}, \citenamefont {Darabiha},\ and\ \citenamefont
  {Th{\'{e}}venin}}]{Hosseini_2020}%
  \BibitemOpen
  \bibfield  {author} {\bibinfo {author} {\bibfnamefont {S.~A.}\ \bibnamefont
  {Hosseini}}, \bibinfo {author} {\bibfnamefont {N.}~\bibnamefont {Darabiha}},\
  and\ \bibinfo {author} {\bibfnamefont {D.}~\bibnamefont {Th{\'{e}}venin}},\
  }\bibfield  {title} {\bibinfo {title} {Compressibility in lattice boltzmann
  on standard stencils: effects of deviation from reference temperature},\
  }\href {https://doi.org/10.1098/rsta.2019.0399} {\bibfield  {journal}
  {\bibinfo  {journal} {Philosophical Transactions of the Royal Society A:
  Mathematical, Physical and Engineering Sciences}\ }\textbf {\bibinfo {volume}
  {378}},\ \bibinfo {pages} {20190399} (\bibinfo {year} {2020})}\BibitemShut
  {NoStop}%
\bibitem [{\citenamefont {Saadat}\ \emph {et~al.}(2019)\citenamefont {Saadat},
  \citenamefont {Bösch},\ and\ \citenamefont {Karlin}}]{Saadat2019}%
  \BibitemOpen
  \bibfield  {author} {\bibinfo {author} {\bibfnamefont {M.~H.}\ \bibnamefont
  {Saadat}}, \bibinfo {author} {\bibfnamefont {F.}~\bibnamefont {Bösch}},\
  and\ \bibinfo {author} {\bibfnamefont {I.~V.}\ \bibnamefont {Karlin}},\
  }\bibfield  {title} {\bibinfo {title} {Lattice boltzmann model for
  compressible flows on standard lattices: Variable prandtl number and
  adiabatic exponent},\ }\href@noop {} {\bibfield  {journal} {\bibinfo
  {journal} {Physical Review E}\ }\textbf {\bibinfo {volume} {99}} (\bibinfo
  {year} {2019})}\BibitemShut {NoStop}%
\bibitem [{\citenamefont {Saadat}\ \emph {et~al.}(2021)\citenamefont {Saadat},
  \citenamefont {Hosseini}, \citenamefont {Dorschner},\ and\ \citenamefont
  {Karlin}}]{saadat2021extendedGas}%
  \BibitemOpen
  \bibfield  {author} {\bibinfo {author} {\bibfnamefont {M.~H.}\ \bibnamefont
  {Saadat}}, \bibinfo {author} {\bibfnamefont {S.~A.}\ \bibnamefont
  {Hosseini}}, \bibinfo {author} {\bibfnamefont {B.}~\bibnamefont
  {Dorschner}},\ and\ \bibinfo {author} {\bibfnamefont {I.~V.}\ \bibnamefont
  {Karlin}},\ }\bibfield  {title} {\bibinfo {title} {Extended lattice boltzmann
  model for gas dynamics},\ }\href {https://doi.org/10.1063/5.0048029}
  {\bibfield  {journal} {\bibinfo  {journal} {Physics of Fluids}\ }\textbf
  {\bibinfo {volume} {33}},\ \bibinfo {pages} {046104} (\bibinfo {year}
  {2021})}\BibitemShut {NoStop}%
\bibitem [{\citenamefont {Nannelli}\ and\ \citenamefont
  {Succi}(1992)}]{Nannelli1992}%
  \BibitemOpen
  \bibfield  {author} {\bibinfo {author} {\bibfnamefont {F.}~\bibnamefont
  {Nannelli}}\ and\ \bibinfo {author} {\bibfnamefont {S.}~\bibnamefont
  {Succi}},\ }\bibfield  {title} {\bibinfo {title} {The lattice boltzmann
  equation on irregular lattices},\ }\href {https://doi.org/10.1007/bf01341755}
  {\bibfield  {journal} {\bibinfo  {journal} {Journal of Statistical Physics}\
  }\textbf {\bibinfo {volume} {68}},\ \bibinfo {pages} {401} (\bibinfo {year}
  {1992})}\BibitemShut {NoStop}%
\bibitem [{\citenamefont {Xi}\ \emph {et~al.}(1999)\citenamefont {Xi},
  \citenamefont {Peng},\ and\ \citenamefont {Chou}}]{Xi1999}%
  \BibitemOpen
  \bibfield  {author} {\bibinfo {author} {\bibfnamefont {H.}~\bibnamefont
  {Xi}}, \bibinfo {author} {\bibfnamefont {G.}~\bibnamefont {Peng}},\ and\
  \bibinfo {author} {\bibfnamefont {S.-H.}\ \bibnamefont {Chou}},\ }\bibfield
  {title} {\bibinfo {title} {Finite-volume lattice boltzmann method},\ }\href
  {https://doi.org/10.1103/physreve.59.6202} {\bibfield  {journal} {\bibinfo
  {journal} {Physical Review E}\ }\textbf {\bibinfo {volume} {59}},\ \bibinfo
  {pages} {6202} (\bibinfo {year} {1999})}\BibitemShut {NoStop}%
\bibitem [{\citenamefont {Patil}\ and\ \citenamefont
  {Lakshmisha}(2009)}]{Patil2009}%
  \BibitemOpen
  \bibfield  {author} {\bibinfo {author} {\bibfnamefont {D.}~\bibnamefont
  {Patil}}\ and\ \bibinfo {author} {\bibfnamefont {K.}~\bibnamefont
  {Lakshmisha}},\ }\bibfield  {title} {\bibinfo {title} {Finite volume {TVD}
  formulation of lattice boltzmann simulation on unstructured mesh},\ }\href
  {https://doi.org/10.1016/j.jcp.2009.04.008} {\bibfield  {journal} {\bibinfo
  {journal} {Journal of Computational Physics}\ }\textbf {\bibinfo {volume}
  {228}},\ \bibinfo {pages} {5262} (\bibinfo {year} {2009})}\BibitemShut
  {NoStop}%
\bibitem [{\citenamefont {Fakhari}\ and\ \citenamefont
  {Lee}(2015)}]{Fakhari2015}%
  \BibitemOpen
  \bibfield  {author} {\bibinfo {author} {\bibfnamefont {A.}~\bibnamefont
  {Fakhari}}\ and\ \bibinfo {author} {\bibfnamefont {T.}~\bibnamefont {Lee}},\
  }\bibfield  {title} {\bibinfo {title} {Numerics of the lattice boltzmann
  method on nonuniform grids: Standard {LBM} and finite-difference {LBM}},\
  }\href {https://doi.org/10.1016/j.compfluid.2014.11.013} {\bibfield
  {journal} {\bibinfo  {journal} {Computers {\&} Fluids}\ }\textbf {\bibinfo
  {volume} {107}},\ \bibinfo {pages} {205} (\bibinfo {year}
  {2015})}\BibitemShut {NoStop}%
\bibitem [{\citenamefont {Hejranfar}\ and\ \citenamefont
  {Ezzatneshan}(2014)}]{Hejranfar2014}%
  \BibitemOpen
  \bibfield  {author} {\bibinfo {author} {\bibfnamefont {K.}~\bibnamefont
  {Hejranfar}}\ and\ \bibinfo {author} {\bibfnamefont {E.}~\bibnamefont
  {Ezzatneshan}},\ }\bibfield  {title} {\bibinfo {title} {Implementation of a
  high-order compact finite-difference lattice boltzmann method in generalized
  curvilinear coordinates},\ }\href {https://doi.org/10.1016/j.jcp.2014.02.030}
  {\bibfield  {journal} {\bibinfo  {journal} {Journal of Computational
  Physics}\ }\textbf {\bibinfo {volume} {267}},\ \bibinfo {pages} {28}
  (\bibinfo {year} {2014})}\BibitemShut {NoStop}%
\bibitem [{\citenamefont {Düster}\ \emph {et~al.}(2006)\citenamefont
  {Düster}, \citenamefont {Demkowicz},\ and\ \citenamefont
  {Rank}}]{Duester2006}%
  \BibitemOpen
  \bibfield  {author} {\bibinfo {author} {\bibfnamefont {A.}~\bibnamefont
  {Düster}}, \bibinfo {author} {\bibfnamefont {L.}~\bibnamefont {Demkowicz}},\
  and\ \bibinfo {author} {\bibfnamefont {E.}~\bibnamefont {Rank}},\ }\bibfield
  {title} {\bibinfo {title} {High-order finite elements applied to the discrete
  boltzmann equation},\ }\href {https://doi.org/10.1002/nme.1657} {\bibfield
  {journal} {\bibinfo  {journal} {International Journal for Numerical Methods
  in Engineering}\ }\textbf {\bibinfo {volume} {67}},\ \bibinfo {pages} {1094}
  (\bibinfo {year} {2006})}\BibitemShut {NoStop}%
\bibitem [{\citenamefont {Li}\ \emph {et~al.}(2005)\citenamefont {Li},
  \citenamefont {LeBoeuf},\ and\ \citenamefont {Basu}}]{Li2005}%
  \BibitemOpen
  \bibfield  {author} {\bibinfo {author} {\bibfnamefont {Y.}~\bibnamefont
  {Li}}, \bibinfo {author} {\bibfnamefont {E.~J.}\ \bibnamefont {LeBoeuf}},\
  and\ \bibinfo {author} {\bibfnamefont {P.~K.}\ \bibnamefont {Basu}},\
  }\bibfield  {title} {\bibinfo {title} {Least-squares finite-element scheme
  for the lattice boltzmann method on an unstructured mesh},\ }\href@noop {}
  {\bibfield  {journal} {\bibinfo  {journal} {Physical Review E}\ }\textbf
  {\bibinfo {volume} {72}} (\bibinfo {year} {2005})}\BibitemShut {NoStop}%
\bibitem [{\citenamefont {Krämer}\ \emph {et~al.}(2017)\citenamefont
  {Krämer}, \citenamefont {Küllmer}, \citenamefont {Reith}, \citenamefont
  {Joppich},\ and\ \citenamefont {Foysi}}]{Kraemer2017}%
  \BibitemOpen
  \bibfield  {author} {\bibinfo {author} {\bibfnamefont {A.}~\bibnamefont
  {Krämer}}, \bibinfo {author} {\bibfnamefont {K.}~\bibnamefont {Küllmer}},
  \bibinfo {author} {\bibfnamefont {D.}~\bibnamefont {Reith}}, \bibinfo
  {author} {\bibfnamefont {W.}~\bibnamefont {Joppich}},\ and\ \bibinfo {author}
  {\bibfnamefont {H.}~\bibnamefont {Foysi}},\ }\bibfield  {title} {\bibinfo
  {title} {Semi-lagrangian off-lattice boltzmann method for weakly compressible
  flows},\ }\href@noop {} {\bibfield  {journal} {\bibinfo  {journal} {Physical
  Review E}\ }\textbf {\bibinfo {volume} {95}} (\bibinfo {year}
  {2017})}\BibitemShut {NoStop}%
\bibitem [{\citenamefont {Shu}\ \emph {et~al.}(2002)\citenamefont {Shu},
  \citenamefont {Niu},\ and\ \citenamefont {Chew}}]{shu2002taylor}%
  \BibitemOpen
  \bibfield  {author} {\bibinfo {author} {\bibfnamefont {C.}~\bibnamefont
  {Shu}}, \bibinfo {author} {\bibfnamefont {X.}~\bibnamefont {Niu}},\ and\
  \bibinfo {author} {\bibfnamefont {Y.}~\bibnamefont {Chew}},\ }\bibfield
  {title} {\bibinfo {title} {Taylor-series expansion and least-squares-based
  lattice boltzmann method: two-dimensional formulation and its applications},\
  }\href@noop {} {\bibfield  {journal} {\bibinfo  {journal} {Physical Review
  E}\ }\textbf {\bibinfo {volume} {65}},\ \bibinfo {pages} {036708} (\bibinfo
  {year} {2002})}\BibitemShut {NoStop}%
\bibitem [{\citenamefont {Cheng}\ and\ \citenamefont
  {Hung}(2004)}]{cheng2004lattice}%
  \BibitemOpen
  \bibfield  {author} {\bibinfo {author} {\bibfnamefont {M.}~\bibnamefont
  {Cheng}}\ and\ \bibinfo {author} {\bibfnamefont {K.~C.}\ \bibnamefont
  {Hung}},\ }\bibfield  {title} {\bibinfo {title} {Lattice boltzmann method on
  nonuniform mesh},\ }\href@noop {} {\bibfield  {journal} {\bibinfo  {journal}
  {International Journal of Computational Engineering Science}\ }\textbf
  {\bibinfo {volume} {5}},\ \bibinfo {pages} {291} (\bibinfo {year}
  {2004})}\BibitemShut {NoStop}%
\bibitem [{\citenamefont {Bardow}\ \emph {et~al.}(2006)\citenamefont {Bardow},
  \citenamefont {Karlin},\ and\ \citenamefont {Gusev}}]{bardow2006general}%
  \BibitemOpen
  \bibfield  {author} {\bibinfo {author} {\bibfnamefont {A.}~\bibnamefont
  {Bardow}}, \bibinfo {author} {\bibfnamefont {I.~V.}\ \bibnamefont {Karlin}},\
  and\ \bibinfo {author} {\bibfnamefont {A.~A.}\ \bibnamefont {Gusev}},\
  }\bibfield  {title} {\bibinfo {title} {General characteristic-based algorithm
  for off-lattice boltzmann simulations},\ }\href@noop {} {\bibfield  {journal}
  {\bibinfo  {journal} {EPL (Europhysics Letters)}\ }\textbf {\bibinfo {volume}
  {75}},\ \bibinfo {pages} {434} (\bibinfo {year} {2006})}\BibitemShut
  {NoStop}%
\bibitem [{\citenamefont {Wilde}\ \emph {et~al.}(2020)\citenamefont {Wilde},
  \citenamefont {Kr{\"a}mer}, \citenamefont {Reith},\ and\ \citenamefont
  {Foysi}}]{wilde2020semi}%
  \BibitemOpen
  \bibfield  {author} {\bibinfo {author} {\bibfnamefont {D.}~\bibnamefont
  {Wilde}}, \bibinfo {author} {\bibfnamefont {A.}~\bibnamefont {Kr{\"a}mer}},
  \bibinfo {author} {\bibfnamefont {D.}~\bibnamefont {Reith}},\ and\ \bibinfo
  {author} {\bibfnamefont {H.}~\bibnamefont {Foysi}},\ }\bibfield  {title}
  {\bibinfo {title} {Semi-lagrangian lattice boltzmann method for compressible
  flows},\ }\href@noop {} {\bibfield  {journal} {\bibinfo  {journal} {Physical
  Review E}\ }\textbf {\bibinfo {volume} {101}},\ \bibinfo {pages} {053306}
  (\bibinfo {year} {2020})}\BibitemShut {NoStop}%
\bibitem [{\citenamefont {Di~Ilio}\ \emph {et~al.}(2018)\citenamefont
  {Di~Ilio}, \citenamefont {Dorschner}, \citenamefont {Bella}, \citenamefont
  {Succi},\ and\ \citenamefont {Karlin}}]{DIILIO_2018}%
  \BibitemOpen
  \bibfield  {author} {\bibinfo {author} {\bibfnamefont {G.}~\bibnamefont
  {Di~Ilio}}, \bibinfo {author} {\bibfnamefont {B.}~\bibnamefont {Dorschner}},
  \bibinfo {author} {\bibfnamefont {G.}~\bibnamefont {Bella}}, \bibinfo
  {author} {\bibfnamefont {S.}~\bibnamefont {Succi}},\ and\ \bibinfo {author}
  {\bibfnamefont {I.~V.}\ \bibnamefont {Karlin}},\ }\bibfield  {title}
  {\bibinfo {title} {Simulation of turbulent flows with the entropic
  multirelaxation time lattice boltzmann method on body-fitted meshes},\ }\href
  {https://doi.org/10.1017/jfm.2018.413} {\bibfield  {journal} {\bibinfo
  {journal} {Journal of Fluid Mechanics}\ }\textbf {\bibinfo {volume} {849}},\
  \bibinfo {pages} {35–56} (\bibinfo {year} {2018})}\BibitemShut {NoStop}%
\bibitem [{\citenamefont {Saadat}\ \emph {et~al.}(2020)\citenamefont {Saadat},
  \citenamefont {Bösch},\ and\ \citenamefont {Karlin}}]{Saadat2020}%
  \BibitemOpen
  \bibfield  {author} {\bibinfo {author} {\bibfnamefont {M.~H.}\ \bibnamefont
  {Saadat}}, \bibinfo {author} {\bibfnamefont {F.}~\bibnamefont {Bösch}},\
  and\ \bibinfo {author} {\bibfnamefont {I.~V.}\ \bibnamefont {Karlin}},\
  }\bibfield  {title} {\bibinfo {title} {Semi-lagrangian lattice boltzmann
  model for compressible flows on unstructured meshes},\ }\href@noop {}
  {\bibfield  {journal} {\bibinfo  {journal} {Physical Review E}\ }\textbf
  {\bibinfo {volume} {101}} (\bibinfo {year} {2020})}\BibitemShut {NoStop}%
\bibitem [{\citenamefont {Saadat}\ and\ \citenamefont
  {Karlin}(2020)}]{Saadat2020a}%
  \BibitemOpen
  \bibfield  {author} {\bibinfo {author} {\bibfnamefont {M.~H.}\ \bibnamefont
  {Saadat}}\ and\ \bibinfo {author} {\bibfnamefont {I.~V.}\ \bibnamefont
  {Karlin}},\ }\bibfield  {title} {\bibinfo {title} {Arbitrary
  lagrangian{\textendash}eulerian formulation of lattice boltzmann model for
  compressible flows on unstructured moving meshes},\ }\href
  {https://doi.org/10.1063/5.0004024} {\bibfield  {journal} {\bibinfo
  {journal} {Physics of Fluids}\ }\textbf {\bibinfo {volume} {32}},\ \bibinfo
  {pages} {046105} (\bibinfo {year} {2020})}\BibitemShut {NoStop}%
\bibitem [{\citenamefont {Dorschner}\ \emph {et~al.}(2018)\citenamefont
  {Dorschner}, \citenamefont {Chikatamarla},\ and\ \citenamefont
  {Karlin}}]{Dorschner2018}%
  \BibitemOpen
  \bibfield  {author} {\bibinfo {author} {\bibfnamefont {B.}~\bibnamefont
  {Dorschner}}, \bibinfo {author} {\bibfnamefont {S.~S.}\ \bibnamefont
  {Chikatamarla}},\ and\ \bibinfo {author} {\bibfnamefont {I.~V.}\ \bibnamefont
  {Karlin}},\ }\bibfield  {title} {\bibinfo {title} {Fluid-structure
  interaction with the entropic lattice boltzmann method},\ }\href@noop {}
  {\bibfield  {journal} {\bibinfo  {journal} {Physical Review E}\ }\textbf
  {\bibinfo {volume} {97}} (\bibinfo {year} {2018})}\BibitemShut {NoStop}%
\bibitem [{\citenamefont {Dorschner}(2018)}]{Dorschner2018a}%
  \BibitemOpen
  \bibfield  {author} {\bibinfo {author} {\bibfnamefont {B.}~\bibnamefont
  {Dorschner}},\ }\emph {\bibinfo {title} {Entropic Lattice Boltzmann Method
  for Complex Flows}},\ \href {https://doi.org/10.3929/ethz-b-000278644} {Ph.D.
  thesis},\ \bibinfo  {school} {ETH Zurich} (\bibinfo {year}
  {2018})\BibitemShut {NoStop}%
\bibitem [{\citenamefont {Feng}\ and\ \citenamefont
  {Michaelides}(2004)}]{Feng_2004}%
  \BibitemOpen
  \bibfield  {author} {\bibinfo {author} {\bibfnamefont {Z.-G.}\ \bibnamefont
  {Feng}}\ and\ \bibinfo {author} {\bibfnamefont {E.~E.}\ \bibnamefont
  {Michaelides}},\ }\bibfield  {title} {\bibinfo {title} {The immersed
  boundary-lattice boltzmann method for solving fluid{\textendash}particles
  interaction problems},\ }\href {https://doi.org/10.1016/j.jcp.2003.10.013}
  {\bibfield  {journal} {\bibinfo  {journal} {Journal of Computational
  Physics}\ }\textbf {\bibinfo {volume} {195}},\ \bibinfo {pages} {602}
  (\bibinfo {year} {2004})}\BibitemShut {NoStop}%
\bibitem [{\citenamefont {Zhu}\ \emph {et~al.}(2011)\citenamefont {Zhu},
  \citenamefont {He}, \citenamefont {Wang}, \citenamefont {Miller},
  \citenamefont {Zhang}, \citenamefont {You},\ and\ \citenamefont
  {Fang}}]{Zhu_2011}%
  \BibitemOpen
  \bibfield  {author} {\bibinfo {author} {\bibfnamefont {L.}~\bibnamefont
  {Zhu}}, \bibinfo {author} {\bibfnamefont {G.}~\bibnamefont {He}}, \bibinfo
  {author} {\bibfnamefont {S.}~\bibnamefont {Wang}}, \bibinfo {author}
  {\bibfnamefont {L.}~\bibnamefont {Miller}}, \bibinfo {author} {\bibfnamefont
  {X.}~\bibnamefont {Zhang}}, \bibinfo {author} {\bibfnamefont
  {Q.}~\bibnamefont {You}},\ and\ \bibinfo {author} {\bibfnamefont
  {S.}~\bibnamefont {Fang}},\ }\bibfield  {title} {\bibinfo {title} {An
  immersed boundary method based on the lattice boltzmann approach in three
  dimensions, with application},\ }\href
  {https://doi.org/10.1016/j.camwa.2010.03.022} {\bibfield  {journal} {\bibinfo
   {journal} {Computers {\&} Mathematics with Applications}\ }\textbf {\bibinfo
  {volume} {61}},\ \bibinfo {pages} {3506} (\bibinfo {year}
  {2011})}\BibitemShut {NoStop}%
\bibitem [{\citenamefont {Nagar}\ \emph {et~al.}(2015)\citenamefont {Nagar},
  \citenamefont {Song}, \citenamefont {Zhu},\ and\ \citenamefont
  {Lin}}]{Nagar_2015}%
  \BibitemOpen
  \bibfield  {author} {\bibinfo {author} {\bibfnamefont {P.}~\bibnamefont
  {Nagar}}, \bibinfo {author} {\bibfnamefont {F.}~\bibnamefont {Song}},
  \bibinfo {author} {\bibfnamefont {L.}~\bibnamefont {Zhu}},\ and\ \bibinfo
  {author} {\bibfnamefont {L.}~\bibnamefont {Lin}},\ }\bibfield  {title}
  {\bibinfo {title} {{LBM}-{IB}: A parallel library to solve 3d fluid-structure
  interaction problems on manycore systems},\ }in\ \href
  {https://doi.org/10.1109/icpp.2015.14} {\emph {\bibinfo {booktitle} {2015
  44th International Conference on Parallel Processing}}}\ (\bibinfo
  {publisher} {{IEEE}},\ \bibinfo {year} {2015})\BibitemShut {NoStop}%
\bibitem [{\citenamefont {Donea}\ \emph {et~al.}(2004)\citenamefont {Donea},
  \citenamefont {Huerta}, \citenamefont {Ponthot},\ and\ \citenamefont
  {Rodríguez-Ferran}}]{Donea2004}%
  \BibitemOpen
  \bibfield  {author} {\bibinfo {author} {\bibfnamefont {J.}~\bibnamefont
  {Donea}}, \bibinfo {author} {\bibfnamefont {A.}~\bibnamefont {Huerta}},
  \bibinfo {author} {\bibfnamefont {J.-P.}\ \bibnamefont {Ponthot}},\ and\
  \bibinfo {author} {\bibfnamefont {A.}~\bibnamefont {Rodríguez-Ferran}},\
  }\bibinfo {title} {Arbitrary lagrangian–eulerian methods},\ in\ \href
  {https://doi.org/10.1002/0470091355.ecm009} {\emph {\bibinfo {booktitle}
  {Encyclopedia of Computational Mechanics}}}\ (\bibinfo  {publisher} {American
  Cancer Society},\ \bibinfo {year} {2004})\ Chap.~\bibinfo {chapter}
  {14}\BibitemShut {NoStop}%
\bibitem [{\citenamefont {Persson}\ \emph {et~al.}(2009)\citenamefont
  {Persson}, \citenamefont {Bonet},\ and\ \citenamefont
  {Peraire}}]{Persson2009}%
  \BibitemOpen
  \bibfield  {author} {\bibinfo {author} {\bibfnamefont {P.-O.}\ \bibnamefont
  {Persson}}, \bibinfo {author} {\bibfnamefont {J.}~\bibnamefont {Bonet}},\
  and\ \bibinfo {author} {\bibfnamefont {J.}~\bibnamefont {Peraire}},\
  }\bibfield  {title} {\bibinfo {title} {Discontinuous galerkin solution of the
  navier{\textendash}stokes equations on deformable domains},\ }\href
  {https://doi.org/10.1016/j.cma.2009.01.012} {\bibfield  {journal} {\bibinfo
  {journal} {Computer Methods in Applied Mechanics and Engineering}\ }\textbf
  {\bibinfo {volume} {198}},\ \bibinfo {pages} {1585} (\bibinfo {year}
  {2009})}\BibitemShut {NoStop}%
\bibitem [{\citenamefont {Akhtar}\ \emph {et~al.}(2007)\citenamefont {Akhtar},
  \citenamefont {Mittal}, \citenamefont {Lauder},\ and\ \citenamefont
  {Drucker}}]{Akhtar2007}%
  \BibitemOpen
  \bibfield  {author} {\bibinfo {author} {\bibfnamefont {I.}~\bibnamefont
  {Akhtar}}, \bibinfo {author} {\bibfnamefont {R.}~\bibnamefont {Mittal}},
  \bibinfo {author} {\bibfnamefont {G.~V.}\ \bibnamefont {Lauder}},\ and\
  \bibinfo {author} {\bibfnamefont {E.}~\bibnamefont {Drucker}},\ }\bibfield
  {title} {\bibinfo {title} {Hydrodynamics of a biologically inspired tandem
  flapping foil configuration},\ }\href
  {https://doi.org/10.1007/s00162-007-0045-2} {\bibfield  {journal} {\bibinfo
  {journal} {Theoretical and Computational Fluid Dynamics}\ }\textbf {\bibinfo
  {volume} {21}},\ \bibinfo {pages} {155} (\bibinfo {year} {2007})}\BibitemShut
  {NoStop}%
\bibitem [{\citenamefont {Wick}(2011)}]{WICK20111456}%
  \BibitemOpen
  \bibfield  {author} {\bibinfo {author} {\bibfnamefont {T.}~\bibnamefont
  {Wick}},\ }\bibfield  {title} {\bibinfo {title} {Fluid-structure interactions
  using different mesh motion techniques},\ }\href@noop {} {\bibfield
  {journal} {\bibinfo  {journal} {Computers \& Structures}\ }\textbf {\bibinfo
  {volume} {89}},\ \bibinfo {pages} {1456 } (\bibinfo {year}
  {2011})}\BibitemShut {NoStop}%
\bibitem [{\citenamefont {Wick}(2013)}]{Wick2013}%
  \BibitemOpen
  \bibfield  {author} {\bibinfo {author} {\bibfnamefont {T.}~\bibnamefont
  {Wick}},\ }\bibfield  {title} {\bibinfo {title} {Solving monolithic
  fluid-structure interaction problems in arbitrary lagrangian eulerian
  coordinates with the deal.ii library},\ }\href@noop {} {\bibfield  {journal}
  {\bibinfo  {journal} {Archive of Numerical Software}\ }\textbf {\bibinfo
  {volume} {Vol 1}} (\bibinfo {year} {2013})}\BibitemShut {NoStop}%
\bibitem [{\citenamefont {Karlin}\ \emph {et~al.}(2013)\citenamefont {Karlin},
  \citenamefont {Sichau},\ and\ \citenamefont {Chikatamarla}}]{KARLIN2013}%
  \BibitemOpen
  \bibfield  {author} {\bibinfo {author} {\bibfnamefont {I.~V.}\ \bibnamefont
  {Karlin}}, \bibinfo {author} {\bibfnamefont {D.}~\bibnamefont {Sichau}},\
  and\ \bibinfo {author} {\bibfnamefont {S.~S.}\ \bibnamefont {Chikatamarla}},\
  }\bibfield  {title} {\bibinfo {title} {Consistent two-population lattice
  boltzmann model for thermal flows},\ }\href
  {https://doi.org/10.1103/PhysRevE.88.063310} {\bibfield  {journal} {\bibinfo
  {journal} {Phys. Rev. E}\ }\textbf {\bibinfo {volume} {88}},\ \bibinfo
  {pages} {063310} (\bibinfo {year} {2013})}\BibitemShut {NoStop}%
\bibitem [{\citenamefont {Piperno}\ and\ \citenamefont
  {Farhat}(2001)}]{Piperno_2001}%
  \BibitemOpen
  \bibfield  {author} {\bibinfo {author} {\bibfnamefont {S.}~\bibnamefont
  {Piperno}}\ and\ \bibinfo {author} {\bibfnamefont {C.}~\bibnamefont
  {Farhat}},\ }\bibfield  {title} {\bibinfo {title} {Partitioned procedures for
  the transient solution of coupled aeroelastic problems {\textendash} part
  {II}: energy transfer analysis and three-dimensional applications},\ }\href
  {https://doi.org/10.1016/s0045-7825(00)00386-8} {\bibfield  {journal}
  {\bibinfo  {journal} {Computer Methods in Applied Mechanics and Engineering}\
  }\textbf {\bibinfo {volume} {190}},\ \bibinfo {pages} {3147} (\bibinfo {year}
  {2001})}\BibitemShut {NoStop}%
\bibitem [{\citenamefont {Kollmannsberger}\ \emph {et~al.}(2009)\citenamefont
  {Kollmannsberger}, \citenamefont {Geller}, \citenamefont {Düster},
  \citenamefont {Tölke}, \citenamefont {Sorger}, \citenamefont {Krafczyk},\
  and\ \citenamefont {Rank}}]{Kollmannsberger_2009}%
  \BibitemOpen
  \bibfield  {author} {\bibinfo {author} {\bibfnamefont {S.}~\bibnamefont
  {Kollmannsberger}}, \bibinfo {author} {\bibfnamefont {S.}~\bibnamefont
  {Geller}}, \bibinfo {author} {\bibfnamefont {A.}~\bibnamefont {Düster}},
  \bibinfo {author} {\bibfnamefont {J.}~\bibnamefont {Tölke}}, \bibinfo
  {author} {\bibfnamefont {C.}~\bibnamefont {Sorger}}, \bibinfo {author}
  {\bibfnamefont {M.}~\bibnamefont {Krafczyk}},\ and\ \bibinfo {author}
  {\bibfnamefont {E.}~\bibnamefont {Rank}},\ }\bibfield  {title} {\bibinfo
  {title} {Fixed-grid fluid-structure interaction in two dimensions based on a
  partitioned lattice boltzmann andp-{FEM} approach},\ }\href
  {https://doi.org/10.1002/nme.2581} {\bibfield  {journal} {\bibinfo  {journal}
  {International Journal for Numerical Methods in Engineering}\ }\textbf
  {\bibinfo {volume} {79}},\ \bibinfo {pages} {817} (\bibinfo {year}
  {2009})}\BibitemShut {NoStop}%
\bibitem [{\citenamefont {Dorschner}\ \emph {et~al.}(2015)\citenamefont
  {Dorschner}, \citenamefont {Chikatamarla}, \citenamefont {Bösch},\ and\
  \citenamefont {Karlin}}]{dorschner2015_grad}%
  \BibitemOpen
  \bibfield  {author} {\bibinfo {author} {\bibfnamefont {B.}~\bibnamefont
  {Dorschner}}, \bibinfo {author} {\bibfnamefont {S.~S.}\ \bibnamefont
  {Chikatamarla}}, \bibinfo {author} {\bibfnamefont {F.}~\bibnamefont
  {Bösch}},\ and\ \bibinfo {author} {\bibfnamefont {I.~V.}\ \bibnamefont
  {Karlin}},\ }\bibfield  {title} {\bibinfo {title} {Grad's approximation for
  moving and stationary walls in entropic lattice boltzmann simulations},\
  }\href {https://doi.org/https://doi.org/10.1016/j.jcp.2015.04.017} {\bibfield
   {journal} {\bibinfo  {journal} {Journal of Computational Physics}\ }\textbf
  {\bibinfo {volume} {295}},\ \bibinfo {pages} {340 } (\bibinfo {year}
  {2015})}\BibitemShut {NoStop}%
\bibitem [{\citenamefont {Qiu}\ \emph {et~al.}(2019)\citenamefont {Qiu},
  \citenamefont {Zhang}, \citenamefont {Liang},\ and\ \citenamefont
  {Xu}}]{Qiu2019}%
  \BibitemOpen
  \bibfield  {author} {\bibinfo {author} {\bibfnamefont {Z.}~\bibnamefont
  {Qiu}}, \bibinfo {author} {\bibfnamefont {B.}~\bibnamefont {Zhang}}, \bibinfo
  {author} {\bibfnamefont {C.}~\bibnamefont {Liang}},\ and\ \bibinfo {author}
  {\bibfnamefont {M.}~\bibnamefont {Xu}},\ }\bibfield  {title} {\bibinfo
  {title} {A high-order solver for simulating vortex-induced vibrations using
  the sliding-mesh spectral difference method and hybrid grids},\ }\href
  {https://doi.org/10.1002/fld.4717} {\bibfield  {journal} {\bibinfo  {journal}
  {International Journal for Numerical Methods in Fluids}\ }\textbf {\bibinfo
  {volume} {90}},\ \bibinfo {pages} {171} (\bibinfo {year} {2019})}\BibitemShut
  {NoStop}%
\bibitem [{\citenamefont {Zhao}(2013{\natexlab{a}})}]{Zhao2013a}%
  \BibitemOpen
  \bibfield  {author} {\bibinfo {author} {\bibfnamefont {M.}~\bibnamefont
  {Zhao}},\ }\bibfield  {title} {\bibinfo {title} {Flow induced vibration of
  two rigidly coupled circular cylinders in tandem and side-by-side
  arrangements at a low reynolds number of 150},\ }\href
  {https://doi.org/10.1063/1.4832956} {\bibfield  {journal} {\bibinfo
  {journal} {Physics of Fluids}\ }\textbf {\bibinfo {volume} {25}},\ \bibinfo
  {pages} {123601} (\bibinfo {year} {2013}{\natexlab{a}})}\BibitemShut
  {NoStop}%
\bibitem [{\citenamefont {Prasanth}\ and\ \citenamefont
  {Mittal}(2008)}]{prasanth_mittal_2008}%
  \BibitemOpen
  \bibfield  {author} {\bibinfo {author} {\bibfnamefont {T.~K.}\ \bibnamefont
  {Prasanth}}\ and\ \bibinfo {author} {\bibfnamefont {S.}~\bibnamefont
  {Mittal}},\ }\bibfield  {title} {\bibinfo {title} {Vortex-induced vibrations
  of a circular cylinder at low reynolds numbers},\ }\href
  {https://doi.org/10.1017/S0022112007009202} {\bibfield  {journal} {\bibinfo
  {journal} {Journal of Fluid Mechanics}\ }\textbf {\bibinfo {volume} {594}},\
  \bibinfo {pages} {463–491} (\bibinfo {year} {2008})}\BibitemShut {NoStop}%
\bibitem [{\citenamefont {Zdravkovich}(1988)}]{Zdravkovich1988}%
  \BibitemOpen
  \bibfield  {author} {\bibinfo {author} {\bibfnamefont {M.}~\bibnamefont
  {Zdravkovich}},\ }\bibfield  {title} {\bibinfo {title} {Review of
  interference-induced oscillations in flow past two parallel circular
  cylinders in various arrangements},\ }\href
  {https://doi.org/10.1016/0167-6105(88)90115-8} {\bibfield  {journal}
  {\bibinfo  {journal} {Journal of Wind Engineering and Industrial
  Aerodynamics}\ }\textbf {\bibinfo {volume} {28}},\ \bibinfo {pages} {183}
  (\bibinfo {year} {1988})}\BibitemShut {NoStop}%
\bibitem [{\citenamefont {Zhao}(2013{\natexlab{b}})}]{Zhao2013}%
  \BibitemOpen
  \bibfield  {author} {\bibinfo {author} {\bibfnamefont {M.}~\bibnamefont
  {Zhao}},\ }\bibfield  {title} {\bibinfo {title} {Flow induced vibration of
  two rigidly coupled circular cylinders in tandem and side-by-side
  arrangements at a low reynolds number of 150},\ }\href
  {https://doi.org/10.1063/1.4832956} {\bibfield  {journal} {\bibinfo
  {journal} {Physics of Fluids}\ }\textbf {\bibinfo {volume} {25}},\ \bibinfo
  {pages} {123601} (\bibinfo {year} {2013}{\natexlab{b}})}\BibitemShut
  {NoStop}%
\bibitem [{\citenamefont {Jester}\ and\ \citenamefont
  {Kallinderis}(2004)}]{Jester2004}%
  \BibitemOpen
  \bibfield  {author} {\bibinfo {author} {\bibfnamefont {W.}~\bibnamefont
  {Jester}}\ and\ \bibinfo {author} {\bibfnamefont {Y.}~\bibnamefont
  {Kallinderis}},\ }\bibfield  {title} {\bibinfo {title} {Numerical study of
  incompressible flow about transversely oscillating cylinder pairs},\ }\href
  {https://doi.org/10.1115/1.1834618} {\bibfield  {journal} {\bibinfo
  {journal} {Journal of Offshore Mechanics and Arctic Engineering}\ }\textbf
  {\bibinfo {volume} {126}},\ \bibinfo {pages} {310} (\bibinfo {year}
  {2004})}\BibitemShut {NoStop}%
\bibitem [{\citenamefont {King}\ and\ \citenamefont {Johns}(1976)}]{King1976}%
  \BibitemOpen
  \bibfield  {author} {\bibinfo {author} {\bibfnamefont {R.}~\bibnamefont
  {King}}\ and\ \bibinfo {author} {\bibfnamefont {D.}~\bibnamefont {Johns}},\
  }\bibfield  {title} {\bibinfo {title} {Wake interaction experiments with two
  flexible circular cylinders in flowing water},\ }\href
  {https://doi.org/10.1016/0022-460x(76)90601-5} {\bibfield  {journal}
  {\bibinfo  {journal} {Journal of Sound and Vibration}\ }\textbf {\bibinfo
  {volume} {45}},\ \bibinfo {pages} {259} (\bibinfo {year} {1976})}\BibitemShut
  {NoStop}%
\bibitem [{\citenamefont {Nayer}\ \emph {et~al.}(2020)\citenamefont {Nayer},
  \citenamefont {Breuer},\ and\ \citenamefont {Wood}}]{Nayer2020}%
  \BibitemOpen
  \bibfield  {author} {\bibinfo {author} {\bibfnamefont {G.~D.}\ \bibnamefont
  {Nayer}}, \bibinfo {author} {\bibfnamefont {M.}~\bibnamefont {Breuer}},\ and\
  \bibinfo {author} {\bibfnamefont {J.}~\bibnamefont {Wood}},\ }\bibfield
  {title} {\bibinfo {title} {Numerical investigations on the dynamic behavior
  of a 2-{DOF} airfoil in the transitional re number regime based on fully
  coupled simulations relying on an eddy-resolving technique},\ }\href
  {https://doi.org/10.1016/j.ijheatfluidflow.2020.108631} {\bibfield  {journal}
  {\bibinfo  {journal} {International Journal of Heat and Fluid Flow}\ }\textbf
  {\bibinfo {volume} {85}},\ \bibinfo {pages} {108631} (\bibinfo {year}
  {2020})}\BibitemShut {NoStop}%
\bibitem [{\citenamefont {Wood}\ \emph {et~al.}(2020)\citenamefont {Wood},
  \citenamefont {Breuer},\ and\ \citenamefont {Nayer}}]{Wood2020}%
  \BibitemOpen
  \bibfield  {author} {\bibinfo {author} {\bibfnamefont {J.}~\bibnamefont
  {Wood}}, \bibinfo {author} {\bibfnamefont {M.}~\bibnamefont {Breuer}},\ and\
  \bibinfo {author} {\bibfnamefont {G.~D.}\ \bibnamefont {Nayer}},\ }\bibfield
  {title} {\bibinfo {title} {Experimental investigations on the dynamic
  behavior of a 2-{DOF} airfoil in the transitional re number regime based on
  digital-image correlation measurements},\ }\href
  {https://doi.org/10.1016/j.jfluidstructs.2020.103052} {\bibfield  {journal}
  {\bibinfo  {journal} {Journal of Fluids and Structures}\ }\textbf {\bibinfo
  {volume} {96}},\ \bibinfo {pages} {103052} (\bibinfo {year}
  {2020})}\BibitemShut {NoStop}%
\bibitem [{\citenamefont {Toomey}\ and\ \citenamefont
  {Eldredge}(2008)}]{Toomey2008}%
  \BibitemOpen
  \bibfield  {author} {\bibinfo {author} {\bibfnamefont {J.}~\bibnamefont
  {Toomey}}\ and\ \bibinfo {author} {\bibfnamefont {J.~D.}\ \bibnamefont
  {Eldredge}},\ }\bibfield  {title} {\bibinfo {title} {Numerical and
  experimental study of the fluid dynamics of a flapping wing with low order
  flexibility},\ }\href {https://doi.org/10.1063/1.2956372} {\bibfield
  {journal} {\bibinfo  {journal} {Physics of Fluids}\ }\textbf {\bibinfo
  {volume} {20}},\ \bibinfo {pages} {073603} (\bibinfo {year}
  {2008})}\BibitemShut {NoStop}%
\bibitem [{\citenamefont {Borazjani}\ and\ \citenamefont
  {Sotiropoulos}(2009)}]{BORAZJANI2009}%
  \BibitemOpen
  \bibfield  {author} {\bibinfo {author} {\bibfnamefont {I.}~\bibnamefont
  {Borazjani}}\ and\ \bibinfo {author} {\bibfnamefont {F.}~\bibnamefont
  {Sotiropoulos}},\ }\bibfield  {title} {\bibinfo {title} {Vortex-induced
  vibrations of two cylinders in tandem arrangement in the
  proximity{\textendash}wake interference region},\ }\href
  {https://doi.org/10.1017/s0022112008004850} {\bibfield  {journal} {\bibinfo
  {journal} {Journal of Fluid Mechanics}\ }\textbf {\bibinfo {volume} {621}},\
  \bibinfo {pages} {321} (\bibinfo {year} {2009})}\BibitemShut {NoStop}%
\bibitem [{\citenamefont {Mittal}\ and\ \citenamefont
  {Kumar}(2001)}]{MITTAL2001}%
  \BibitemOpen
  \bibfield  {author} {\bibinfo {author} {\bibfnamefont {S.}~\bibnamefont
  {Mittal}}\ and\ \bibinfo {author} {\bibfnamefont {V.}~\bibnamefont {Kumar}},\
  }\bibfield  {title} {\bibinfo {title} {{Flow}-{Induced} {Osclillations} {Of}
  {Two} {Cylinders} {In} {Tandem} {And} {Staggered} {Arrangement}},\ }\href
  {https://doi.org/10.1006/jfls.2000.0376} {\bibfield  {journal} {\bibinfo
  {journal} {Journal of Fluids and Structures}\ }\textbf {\bibinfo {volume}
  {15}},\ \bibinfo {pages} {717} (\bibinfo {year} {2001})}\BibitemShut
  {NoStop}%
\bibitem [{\citenamefont {Zhang}\ \emph {et~al.}(2015)\citenamefont {Zhang},
  \citenamefont {Gao}, \citenamefont {Liu}, \citenamefont {Ye},\ and\
  \citenamefont {Jiang}}]{Zhang2015}%
  \BibitemOpen
  \bibfield  {author} {\bibinfo {author} {\bibfnamefont {W.}~\bibnamefont
  {Zhang}}, \bibinfo {author} {\bibfnamefont {C.}~\bibnamefont {Gao}}, \bibinfo
  {author} {\bibfnamefont {Y.}~\bibnamefont {Liu}}, \bibinfo {author}
  {\bibfnamefont {Z.}~\bibnamefont {Ye}},\ and\ \bibinfo {author}
  {\bibfnamefont {Y.}~\bibnamefont {Jiang}},\ }\bibfield  {title} {\bibinfo
  {title} {The interaction between flutter and buffet in transonic flow},\
  }\href {https://doi.org/10.1007/s11071-015-2282-z} {\bibfield  {journal}
  {\bibinfo  {journal} {Nonlinear Dynamics}\ }\textbf {\bibinfo {volume}
  {82}},\ \bibinfo {pages} {1851} (\bibinfo {year} {2015})}\BibitemShut
  {NoStop}%
\bibitem [{\citenamefont {Gao}\ \emph {et~al.}(2015)\citenamefont {Gao},
  \citenamefont {Zhang}, \citenamefont {Liu}, \citenamefont {Ye},\ and\
  \citenamefont {Jiang}}]{Gao2015}%
  \BibitemOpen
  \bibfield  {author} {\bibinfo {author} {\bibfnamefont {C.}~\bibnamefont
  {Gao}}, \bibinfo {author} {\bibfnamefont {W.}~\bibnamefont {Zhang}}, \bibinfo
  {author} {\bibfnamefont {Y.}~\bibnamefont {Liu}}, \bibinfo {author}
  {\bibfnamefont {Z.}~\bibnamefont {Ye}},\ and\ \bibinfo {author}
  {\bibfnamefont {Y.}~\bibnamefont {Jiang}},\ }\bibfield  {title} {\bibinfo
  {title} {Numerical study on the correlation of transonic
  single-degree-of-freedom flutter and buffet},\ }\href@noop {} {\bibfield
  {journal} {\bibinfo  {journal} {Science China Physics, Mechanics {\&}
  Astronomy}\ }\textbf {\bibinfo {volume} {58}} (\bibinfo {year}
  {2015})}\BibitemShut {NoStop}%
\bibitem [{\citenamefont {He}\ \emph {et~al.}(1998)\citenamefont {He},
  \citenamefont {Chen},\ and\ \citenamefont {Doolen}}]{HE1998282}%
  \BibitemOpen
  \bibfield  {author} {\bibinfo {author} {\bibfnamefont {X.}~\bibnamefont
  {He}}, \bibinfo {author} {\bibfnamefont {S.}~\bibnamefont {Chen}},\ and\
  \bibinfo {author} {\bibfnamefont {G.~D.}\ \bibnamefont {Doolen}},\ }\bibfield
   {title} {\bibinfo {title} {A novel thermal model for the lattice boltzmann
  method in incompressible limit},\ }\href
  {https://doi.org/https://doi.org/10.1006/jcph.1998.6057} {\bibfield
  {journal} {\bibinfo  {journal} {Journal of Computational Physics}\ }\textbf
  {\bibinfo {volume} {146}},\ \bibinfo {pages} {282 } (\bibinfo {year}
  {1998})}\BibitemShut {NoStop}%
\bibitem [{\citenamefont {Karlin}\ \emph {et~al.}(2014)\citenamefont {Karlin},
  \citenamefont {Bösch},\ and\ \citenamefont {Chikatamarla}}]{Karlin2014}%
  \BibitemOpen
  \bibfield  {author} {\bibinfo {author} {\bibfnamefont {I.~V.}\ \bibnamefont
  {Karlin}}, \bibinfo {author} {\bibfnamefont {F.}~\bibnamefont {Bösch}},\
  and\ \bibinfo {author} {\bibfnamefont {S.~S.}\ \bibnamefont {Chikatamarla}},\
  }\bibfield  {title} {\bibinfo {title} {Gibb's principle for the
  lattice-kinetic theory of fluid dynamics},\ }\href@noop {} {\bibfield
  {journal} {\bibinfo  {journal} {Physical Review E}\ }\textbf {\bibinfo
  {volume} {90}} (\bibinfo {year} {2014})}\BibitemShut {NoStop}%
\bibitem [{\citenamefont {Karlin}\ and\ \citenamefont
  {Succi}(1998)}]{karlin1999}%
  \BibitemOpen
  \bibfield  {author} {\bibinfo {author} {\bibfnamefont {I.~V.}\ \bibnamefont
  {Karlin}}\ and\ \bibinfo {author} {\bibfnamefont {S.}~\bibnamefont {Succi}},\
  }\bibfield  {title} {\bibinfo {title} {Equilibria for discrete kinetic
  equations},\ }\href {https://doi.org/10.1103/PhysRevE.58.R4053} {\bibfield
  {journal} {\bibinfo  {journal} {Phys. Rev. E}\ }\textbf {\bibinfo {volume}
  {58}},\ \bibinfo {pages} {R4053} (\bibinfo {year} {1998})}\BibitemShut
  {NoStop}%
\bibitem [{\citenamefont {Ansumali}\ and\ \citenamefont
  {Karlin}(2005)}]{Ansumali2005}%
  \BibitemOpen
  \bibfield  {author} {\bibinfo {author} {\bibfnamefont {S.}~\bibnamefont
  {Ansumali}}\ and\ \bibinfo {author} {\bibfnamefont {I.~V.}\ \bibnamefont
  {Karlin}},\ }\bibfield  {title} {\bibinfo {title} {Consistent lattice
  boltzmann method},\ }\href@noop {} {\bibfield  {journal} {\bibinfo  {journal}
  {Physical Review Letters}\ }\textbf {\bibinfo {volume} {95}} (\bibinfo {year}
  {2005})}\BibitemShut {NoStop}%
\bibitem [{\citenamefont {Prasanth}\ and\ \citenamefont
  {Mittal}(2009)}]{Prasanth2009}%
  \BibitemOpen
  \bibfield  {author} {\bibinfo {author} {\bibfnamefont {T.}~\bibnamefont
  {Prasanth}}\ and\ \bibinfo {author} {\bibfnamefont {S.}~\bibnamefont
  {Mittal}},\ }\bibfield  {title} {\bibinfo {title} {Flow-induced oscillation
  of two circular cylinders in tandem arrangement at low re},\ }\href
  {https://doi.org/10.1016/j.jfluidstructs.2009.04.001} {\bibfield  {journal}
  {\bibinfo  {journal} {Journal of Fluids and Structures}\ }\textbf {\bibinfo
  {volume} {25}},\ \bibinfo {pages} {1029} (\bibinfo {year}
  {2009})}\BibitemShut {NoStop}%
\end{thebibliography}%


%
	
\end{document}